\documentclass[twocolumn]{aastex62}

\usepackage{graphicx}
\usepackage[caption=false]{subfig}	

\usepackage{longtable}
\usepackage{latexsym}
\usepackage{lineno}
\usepackage{amsmath}
\usepackage{color}

\newcommand{\HI}{H\,{\sc i} }

\newcommand{\arcs}{\arcsec }
\newcommand{\wise}{{\it WISE} }

\accepted{June 12, 2020}
\submitjournal{ApJ}

\shorttitle{GAMA-\wise $z<0.1$ Groups}
\shortauthors{Cluver et al.}

\begin{document}

\title{Galaxy and Mass Assembly (GAMA): Demonstrating the power of \wise in the study of Galaxy Groups to $z<0.1$}

\correspondingauthor{Michelle Cluver}
\email{michelle.cluver@gmail.com}

\author{M.E. Cluver}
\affiliation{Centre for Astrophysics and Supercomputing, Swinburne University of Technology, John Street, Hawthorn 3122, Victoria, Australia}
\affiliation{Department of Physics and Astronomy, University of the Western Cape,
Robert Sobukwe Road, Bellville, South Africa}

\author{T.H. Jarrett}
\affiliation{Department of Astronomy, University of Cape Town, Rondebosch,
 South Africa}

\author{E.N. Taylor}
\affiliation{Centre for Astrophysics and Supercomputing, Swinburne University of Technology, John Street, Hawthorn 3122, Victoria, Australia}

\author{A.M. Hopkins}
\affil{Australian Astronomical Optics, Macquarie University, 105 Delhi Rd, North Ryde, NSW 2113, Australia}

\author{S. Brough}
\affil{School of Physics, University of New South Wales, NSW 2052, Australia}

\author{S. Casura}
\affil{Hamburger Sternwarte, Universit\"{a}t Hamburg, Gojenbergsweg 112, 21029 Hamburg, Germany}

\author{B.W. Holwerda}
\affil{Department of Physics and Astronomy, 102 Natural Science Building, University of Louisville, Louisville KY 40292, USA}

\author{J. Liske}
\affil{Hamburger Sternwarte, Universit\"{a}t Hamburg, Gojenbergsweg 112, 21029 Hamburg, Germany}

\author{K.A. Pimbblet}
\affil{E.A.Milne Centre for Astrophysics, Department of Physics, University of Hull, Cottingham Road, Kingston-upon-Hull, HU6 7RX, UK}

\author{A.H. Wright}
\affil{Ruhr-Universit\"{a}t Bochum, Astronomisches Institut, German Centre for Cosmological Lensing (GCCL), Universit\"{a}tsstr. 150, 44801 Bochum, Germany}

\begin{abstract}

Combining high-fidelity group characterisation from the Galaxy and Mass Assembly (GAMA) survey and source-tailored $z<0.1$ photometry from the \wise survey, we present a comprehensive study of the properties of ungrouped galaxies, compared to 497 galaxy groups (4$\leq$ N$_{\rm FoF}$ $\leq$ 20) as a function of stellar and halo mass. Ungrouped galaxies are largely unimodal in \wise color, the result of being dominated by star-forming, late-type galaxies. Grouped galaxies, however, show a clear bimodality in \wise color, which correlates strongly with stellar mass and morphology. We find evidence for an increasing early-type fraction, in stellar mass bins between $10^{10}\lesssim$M$_{\rm stellar} \lesssim10^{11}$\,M$_\odot$, with increasing halo mass. Using ungrouped, late-type galaxies with star-forming colors (W2$-$W3$>$3), we define a star-forming main-sequence (SFMS), which we use to delineate systems that have moved below the sequence (``quenched" for the purposes of this work). We find that with increasing halo mass, the relative number of late-type systems on the SFMS decreases, with a corresponding increase in early-type, quenched systems at high stellar mass (M$_{\rm stellar}>{10}^{10.5}$\,M$_\odot$), consistent with mass quenching. Group galaxies with masses M$_{\rm stellar}<{10}^{10.5}$\,M$_\odot$ show evidence of quenching consistent with environmentally-driven processes. The stellar mass distribution of late-type, quenched galaxies suggests they may be an intermediate population as systems transition from being star-forming and late-type to the ``red sequence". Finally, we use the projected area of groups on the sky to extract groups that are (relatively) compact for their halo mass. Although these show a marginal increase in their proportion of high mass and early-type galaxies compared to nominal groups, a clear increase in quenched fraction is not evident. 

\end{abstract}


\section{Introduction} \label{sec:intro}

The local ($z<0.1$) Universe offers us the clearest and most complete view for studying the feeding, feedback, and quenching processes that drive and regulate star formation within cosmic structures of varying density. The formation of large-scale structure and sub-structure in the universe lies at the heart of the hierarchical paradigm of $\Lambda$-CDM \citep[e.g.][]{Davis85}, where groups of galaxies merge into clusters, filaments, walls, and superclusters, creating the cosmic web. 

In the nearby universe, we observe the so-called ``morphology-density" relation \citep[e.g.][]{Dres80, PG84, Got03, Bl09}, the suppression of star formation in high density environments \citep[e.g.][]{Bal98, Couch01, Lew02}, and the bimodality of the local galaxy population as a ``blue cloud" of star-forming galaxies and a ``red sequence" of quenched, passively evolving systems  \citep[e.g.][]{Strat01, Bl03, Bald04, Bal04, Tay15}. However, the pathways that lead to these observed trends, and how they are connected, remain unclear. This is in part due to the challenge of disentangling a number of possible mechanisms, acting as a function of either stellar mass \citep[e.g.][]{Bald06, Peng10}, environment \citep[e.g.][]{Peng10, Peng12, Bluck16}, or morphology \citep[e.g.][]{Mar09, Bluck14}. The complication in distinguishing between these is that it is clear that {\it both} mass {\it and} environment play a role, which means that we have to carefully measure and control for mass-dependent effects in order to isolate and characterise the effects of environment. 

``Mass" or secular quenching \citep[][]{Driv06, Peng10} translates to more massive galaxies quenching independent of environment, i.e. due to internal processes such as AGN feedback. Alternatively, ``environmental quenching" applies to galaxies quenched due to external processes, i.e. their environment, independent of stellar mass \citep{Peng10}. It is worth noting that ``halo quenching" \citep[e.g.][]{Birn03, Dek06}, the virial shock heating of intergalactic gas falling into a galactic dark matter halo, has been proposed as driving both processes \citep[e.g.][]{Gab15}. 

A further complication, however, is the timescales on which the physical quenching mechanisms operate, e.g. the starvation (strangulation) of the gas supply \citep[e.g.][]{Lar80, Peng15} due to dense environments \citep[e.g.][]{Voort17}, or the heating of galactic halos by, e.g.  large-scale AGN jets \citep{Cro06} and shocks \citep[e.g.][]{Birn03, Dek06}. Here the former would be an example of environmental quenching and the latter, mass quenching. The proposed framework of cosmic web detachment suggests that the starvation process due to the disruption of filaments that supply gas to galaxies, encompasses the role of external processes and is able to reproduce observations such as the dependence of the red fraction on mass and local density \citep{Ar19}.

In dense environments, particularly, processes such as galaxy harassment \citep[e.g.][]{Far81, Moore96}, tidal encounters \citep[e.g.][]{TT72, Bar92}, and various mechanisms of gas stripping, such as ram-pressure stripping \citep[e.g.][]{GG72}, tidal stripping \citep[e.g.][]{Mih04}, and viscous stripping \citep[e.g.][]{Nul82, Ras06}, can contribute to the evolutionary pathways (morphological transformation and quenching) of individual galaxies. It is therefore prudent, albeit challenging, to control for environment when investigating the pathways of galaxy evolution.

Galaxy evolution in the group environment is of particular interest given that 40-50\% of galaxies in the local ($z\sim0$) Universe, reside in groups \citep{Eke04,Robot11}. It is the most common environment in which galaxies are found, especially compared to clusters which are rare by comparison ($\sim$5\%). In addition, it has been suggested that the dominance of early-type galaxies in rich clusters is the result of galaxies being ``pre-processed" in low to medium density environments \citep[][]{Zab98, Fuj04} before being assimilated into larger structures \citep[e.g.][]{Bian18, Just19, Hain15}. Studies examining the infall regions of clusters suggest that pre-processing may be occurring in overdensities (groups) located in filamentary structures, as part of the hierarchical growth of structure \citep[e.g.][]{Port08, Kral18, Sarr19}.

Within groups themselves, evidence of pre-processing can be seen as changes in morphology and star formation, particularly in low-mass galaxies infalling into high-mass group haloes \citep[e.g.][]{Rob17, Bar18}. Pre-processing is also expected to impact the gas content of galaxies, and provides an opportunity to study transformation in situ \citep[e.g.][]{Cor06, Dudz19}. In the study of \citet{Hess13}, combining group information from the SDSS survey \citep{York00} and \HI from the ALFALFA survey \citep{Gio05}, the processing of galaxies in the group environment is observed as an increasing deficiency of \HI-rich galaxies at the centers of groups, with increasing optical group membership.

It is therefore evident that galaxy groups play a key role in understanding galaxy transformations and identifying the mechanisms that dominate in these environments is key. In this work, we aim to provide a benchmark view of the mid-infrared properties of galaxy groups, focusing on groups with membership between 4 and 20 (halo mass between $10^{10.5}$ to $10^{14}$ M$_\odot$/h), representative of the most typical overdensities in the local  ($z<0.1$) universe, excluding pair and triple systems \citep[see][]{Robot11}. We make use of the {\it Wide-Field Infrared Explorer} \citep[\wise,][]{Wr10} to investigate the diversity of mass (halo and stellar), morphology, and star formation properties within this population. 

\wise surveyed the entire sky at wavelengths of 3.4\micron\ (W1), 4.6\micron\ (W2), 12\micron\ (W3), and 23\micron\ \citep[W4;][]{Br14a} and hence traces both dust-free stellar mass and dust-reprocessed star formation \citep[e.g.][]{Jar12, Jar13, Clu14}, particularly in the local universe where its sensitivity to both is the most uniform. It is therefore a valuable resource for studying galaxy populations in wide-area surveys and particularly useful for studying the global measurements of interacting systems potentially generating excess dust through triggered star formation \citep[e.g.][]{Mar07}.
The mid-infrared photometry used in this study has been carefully tailored to suit both galaxies that are resolved and unresolved by \wise (see Section 2.2. for details), enabling a detailed mid-infrared study of the color, stellar mass and star formation properties of grouped and ungrouped galaxies, and exploring the roles of morphology, halo mass, and compactness. This can therefore be extended in a straightforward way to larger areas, with improved uniformity and statistics, provided highly complete redshifts are available.

In this study we exploit the robust identification of group galaxies in Galaxy and Mass Assembly survey \citep{Driv11}, which was constructed with an emphasis on high completeness, making it an ideal dataset for group galaxy science. In the equatorial regions of G09, G12, and G15, covering $\sim 180$\,degrees$^2$ \citep{Hop13, Bald18}, the survey achieves 98.48\% completeness to a limiting magnitude of $r_{\rm AB}=19.8$\,mag \citep{Lisk15},  i.e. 2\, mag deeper than SDSS \citep{York00}. The GroupFinding DMU \citep{Robot11} is considered a key data product of the survey \citep{Lisk15}.

The high fidelity characterisation of environment in GAMA has been used to investigate several aspects of environment-driven evolution, for example, the impact of pairs, mergers, and close interactions \citep[e.g.][]{Robot14, Prop14, Dav15}, the influence of group, cluster, local, and large-scale environment on galaxy properties \citep[e.g.][]{Wij12, Br13, Al15, Dav16, Sch17, Groot17, Groot18, Bar18, WangL18, Sch19, Dav19a}, and the quenching of centrals and satellites in groups \citep{Dav19b}. Summarising some of these results related to the group environment, \citet{Al15} investigated the impact of environment on several quantities (optical color, luminosity, morphology), finding that a mass-controlled sample does not show an environmental dependence, whereas removing mass-matching amplifies trends as a function of environment/density, suggesting that  stellar mass is the dominant driver of galaxy properties. As part of their study they show that the characteristic mass, i.e. the knee of the mass function, increases with group halo mass. \citet{Bar18} find increased star formation for galaxies located on the outskirts of groups, compared to those in the central regions, and \citet{WangL18} find evidence for a decrease in the fraction of star-forming satellites, with increasing halo mass. The study of \citet{Dav19b} find that irrespective of whether a galaxy is a central or satellite, more massive galaxies are more likely to be passive. They find that with increasing halo mass, both centrals and satellites have an increasing passive fraction, with centrals having a higher passive fraction, likely due to being more massive than satellites. These works reflect the ongoing progress in looking at different aspects of galaxy evolution, but our understanding of the baryon cycle as a function of stellar mass and environment remains murky.

This paper is organized as follows: in Section 2 we provide details of the data and samples that form the basis of our study, in Section 3 we present our analysis, including \wise colors (3.1), stellar mass and morphology (3.2), the star-forming main sequence (3.3), and compactness (3.4). A summary of our main results and their significance is given in Section \ref{sec:disc}, and conclusions in Section 5.

The cosmology adopted throughout this paper is $H_0 =70$ km s$^{-1}$ Mpc$^{-1}$, $h = H_0/100$, $\Omega_M = 0.27$ and $\Omega_\Lambda = 0.73$. All magnitudes are in the Vega system, as adopted by the \wise survey \citep[as described in][]{Jar11}. All linear fits are performed using the Hyper-Fit package \citep{hyper}.

\section{Data and Sample Selection}

Our primary dataset is drawn from the three equatorial regions (G09, G12, and G15) of the GAMA II spectroscopic survey which cover an area of 180 degree$^2$ to a limiting magnitude of $r_{\rm AB}=19.8$\,mag \citep[][]{Lisk15, Bald18}. 

\subsection{The GAMA Group Catalog}

The GAMA Galaxy Group Catalog (G$^3$C) is constructed using an iterative friends-of-friends algorithm, making use of mock GAMA light cones in order to refine the group-finding algorithm \citep[full details are provided in][]{Robot11}. The G$^3$C assigns $\sim 40\%$ of galaxies to groups with multiplicity N$>1$, i.e. pairs and groups \citep{Robot11}. For this work, we make use of the most recent version of the G$^3$C, G3Cv10, constructed after the completion of the equatorial fields. The catalog has been constructed from the main survey catalog, TilingCatv46, extracting galaxies with an AUTOZ redshift quality flag of NQ $\geq 2$ \citep[i.e. science quality; see][for further details]{Bald14} and CMB-corrected redshift limits of $0.003 < {\rm Z}\_{\rm CMB} < 0.6$ -- this sample is included in the G3Cv10 DMU (Data Management Unit) as G3CGalv10. 

The group catalog extracted from G3CGalv10 is provided as G3CFoFGroupv10 and we impose the following selection criteria: 

\begin{enumerate}

\item{median redshift of all groups is $z<0.1$ (i.e. Zfof $<0.1$) -- although this greatly reduces the size of our sample, it is most suited to the angular resolution (6\arcs) and sensitivity of \wise}

\item{group membership (multiplicity) between 4 and 20 galaxies, i.e. 4$\leq$ Nfof $\leq$ 20, corresponding to derived halo masses between $10^{10}$ and $10^{14}$ M$_\odot$/h (see Figure \ref{fig:A1} of the Appendix)}

\item{select only groups that are entirely contained within the survey volume (GroupEdge$=$1), i.e no partial groups due to the edges of the survey are included}

\end{enumerate}

These criteria reflect our focus on the nearby universe (with higher sensitivity and fewer observational biases) and our aim of investigating the properties of systems dominated by multi-member interactions in the absence of virialised halos associated with a pervasive hot intracluster medium.

We find 498 groups that satisfy our criteria, consisting of 3195 galaxies. Galaxies that are left ungrouped in G3CGalv10 are designated ``non-G3C galaxies" and we consider these to be the ``least grouped" galaxies within GAMA. In addition to isolated field galaxies, they are most likely the central galaxies of groups where the other member galaxies were too faint to be detected. In the study of \citet{Bar18}, they showed that many of these galaxies can be associated with groups using an alternative prescription, such as projected phase space.  However, since we are interested in examining the impact of group environment specifically, we postulate any effects will be strongest within the FoF-defined groups and weakest in the non-G3C galaxies. To this end we select least-grouped, non-G3C galaxies with $z<0.1$ from G3CGalv10, resulting in 12\,594 galaxies as a control sample.

In the G3C, group halo masses are calculated by matching to simulated light cones using several different methods \citep[see][]{Robot11}. In this work, we make use of the ``MassAFunc" quantity in the G3CFoFGroupv10 catalog; this corresponds to the MassProxy ($\sim R_{50} \sigma^2$, where $R_{50}$ is the radius containing 50\% of the group members and $\sigma$ is the velocity dispersion) multiplied by a scaling factor, ``A", required to get a median-unbaised halo mass estimate (for Nfof $\geq4$). For our purposes, the A factor is a function of Nfof (the number of galaxies in a group) and the IterCenZ (the redshift of the iterative central galaxy); see \citet{Robot11} for full details. The distribution of dynamical (halo) mass and group membership within our sample is included as Figure \ref{fig:A1} in the Appendix. Groups of membership 4 and 5 dominate our sample and show the broadest range in dynamical (halo) mass. As discussed in section \ref{sec:A2} of the Appendix, groups with the lowest membership have the largest uncertainty in their velocity dispersion estimation, and therefore derived dynamical masses. 

\begin{figure*}[!thb]
\gridline{\fig{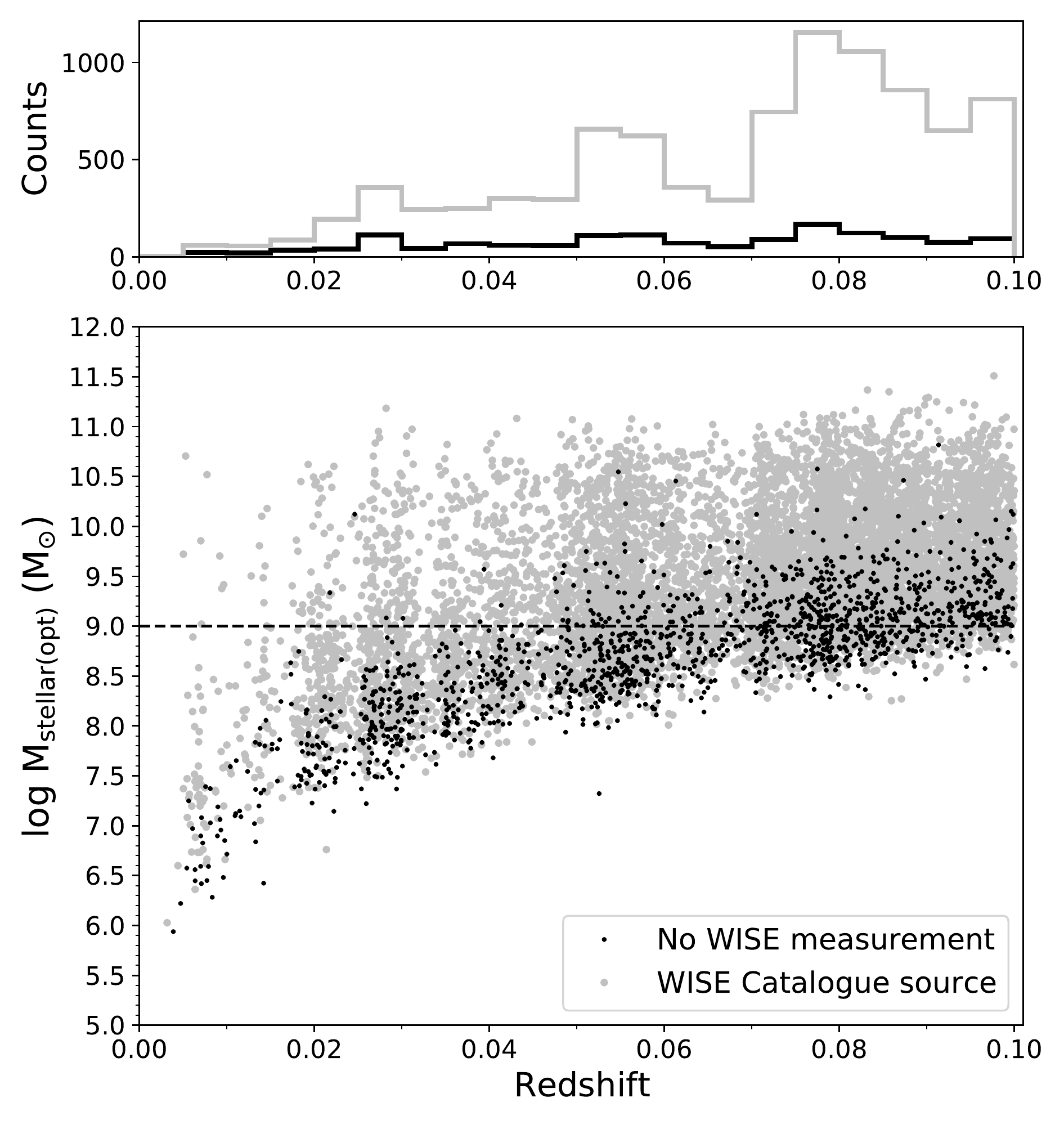}{0.5\textwidth}{(b) non-G3C Galaxies (ungrouped)}
              \fig{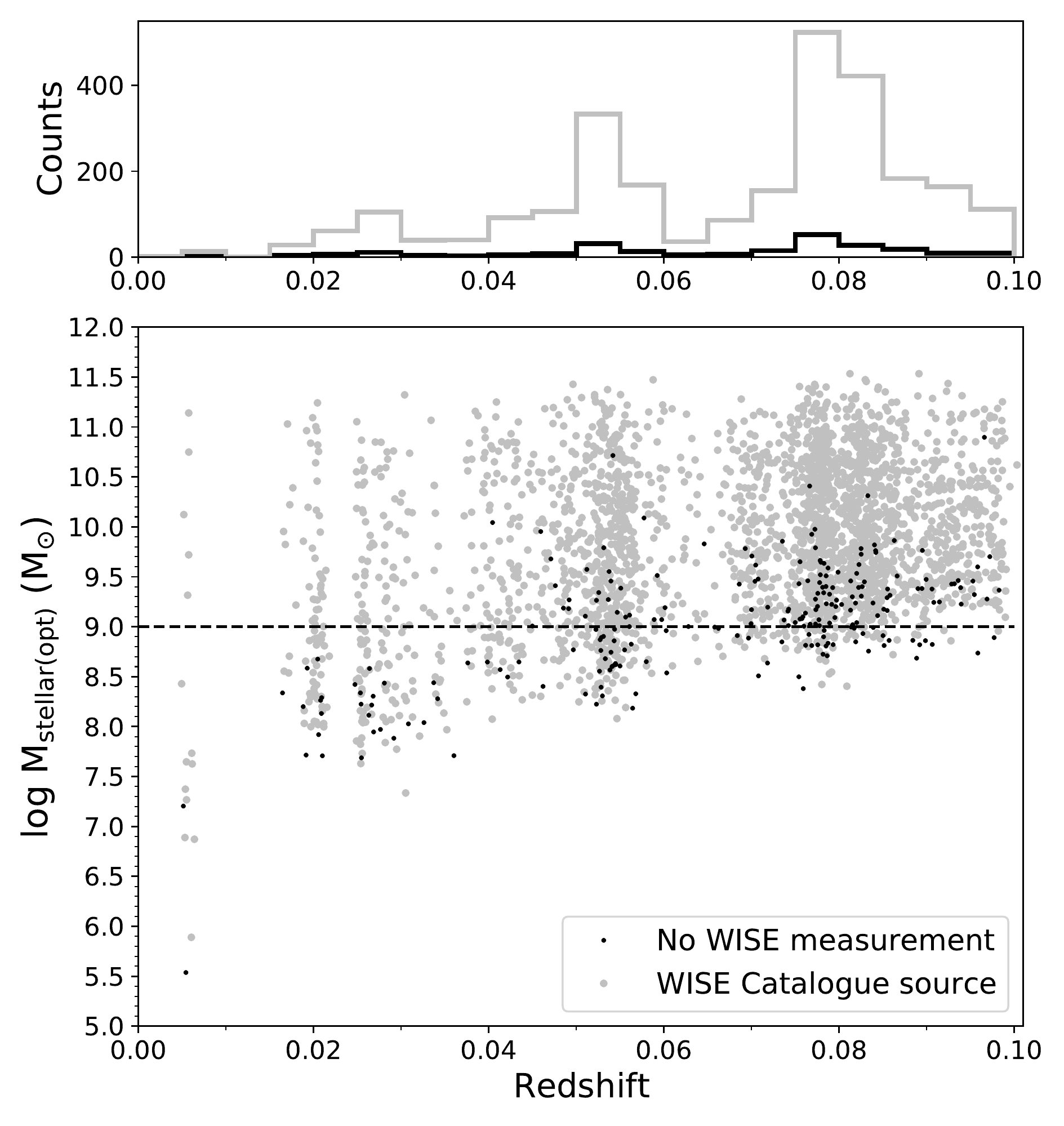}{0.5\textwidth}{(a) G3C Group Catalog Galaxies}}
\caption{Optically-derived stellar masses for a) the G3C galaxies in our sample, and b) the non-G3C galaxies indicate the WISE cross-match sample to be highly complete when imposing a mass cut of $10^{9}$ M$_\sun$. \label{fig:ff1}}
\end{figure*}

\subsection{\wise Photometry}

Our base catalog is the WISECatv02 table available from the GAMA database (www.gama-survey.org). A detailed description of the construction of this catalog can be found in \citet{Clu14}, with an update included in \citet{Dri16} as part of the GAMA panchromatic data release. We summarise here the main features of the catalog:

\begin{enumerate}

\item The ALLWISE Data Release catalog\footnote{https://irsa.ipac.caltech.edu/data/download/wise-allwise/} is the primary source of \wise counterparts to GAMA sources.

\item The WISECatv02 catalog covers the G09, G12 and G15 fields.

\item Sources that are potentially resolved by \wise \citep[see][]{Clu14}, are measured on reconstructed `drizzle' images \citep{Jar12}, with native \wise resolution and 1\arcsec\ pixels, as described in \citet{Clu14}. In bands where the source is resolved, the isophotal (integrated flux) photometry is reported, 
which captures better than 90\% of the total flux for the source \citep{Jar19}.

\item Due to the sensitivity of the W1 and W2 bands, unresolved extended sources (i.e. galaxies resolved by optical imaging) are not well-measured by profile-fit photometry. Here the standard aperture photometry (w1mag, w2mag), corresponding to a circular aperture of 8.25\arcsec, is reported.  

\item For sources unresolved in the W3 and W4 bands, the profile-fit photometry (w3mpro, w4mpro) from the ALLWISE catalog is reported, 
providing the best sensitivity for these cases \citep{Clu14}.

\end{enumerate}

We modify this catalog by replacing non-detection, upper limits (from ALLWISE) with forced photometry provided by the LAMDAR code DMU (LamdarCatv01) as detailed in \citet{Wr16}. This provides useful constraints when shifting to restframe photometry, and enables the propagation of meaningful upper limits. The G3C and non-G3C galaxies of our sample are then crossmatched to the modified \wise photometry catalog using the CATAID identifier; the statistics are listed in Table \ref{tab:t1}. 

\begin{deluxetable*}{lccC}[th!]
\tablecaption{Crossmatch statistics \label{tab:t1}}
\tablecolumns{4}
\tablenum{1}
\tablewidth{0pt}
\tablehead{
\colhead{} &
\colhead{No. of Galaxies} &
\colhead{\wise Matches} &
\colhead{Completeness} 
}
\startdata
G3C & 3195 &  2871 &  90\% \\
G3C$_\textrm{optical stellar mass cut}$\tablenotemark{a} & 2454  & 2324 & 97\% \\
G3C$_\textrm{WISE stellar mass cut}$\tablenotemark{b} &  2583  &   &   \\
\tableline
non-G3C  & 12 594 &  9461 &  75\% \\
non-G3C$_\textrm{optical stellar mass cut}$\tablenotemark{a} & 7042  & 6531  & 93\%\\
non-G3C$_\textrm{WISE stellar mass cut}$\tablenotemark{b} & 7534   &   &   \\
\enddata

\tablenotetext{a}{Using optically-derived stellar masses where available}
\tablenotetext{b}{Using \wise-derived stellar masses}
\end{deluxetable*}

The magnitude limit used for galaxy selection in the GAMA survey leads to a sample whose mass completeness is a function of redshift. We make use of the optically-derived StellarMasses DMU (StellarMassesv19) as detailed in \citet{Tay11}, using $h=0.7$ and removing galaxies with uncertain fluxscale corrections, to identify sources that do not have a \wise counterpart (and therefore do not have a \wise stellar mass), as a function of redshift. The W1 (3.4\micron) band of \wise is its most sensitive, but Figure \ref{fig:ff1} indicates that sources with low stellar mass can be missed by \wise as they lack a substantial old stellar population. As shown, a M$_{\rm stellar} \geq 10^{9.0}$ M$_\odot$ selection creates an approximately mass complete sample to $z<0.1$ \citep[see also][]{Al15}. This improves the completeness of the matched sample since it removes many of the low-mass systems that \wise is less suited to detect (see Table \ref{tab:t1}).

\subsection{Derived Quantities}

\begin{figure}[!thb]
\begin{center}
\includegraphics[width=8cm]{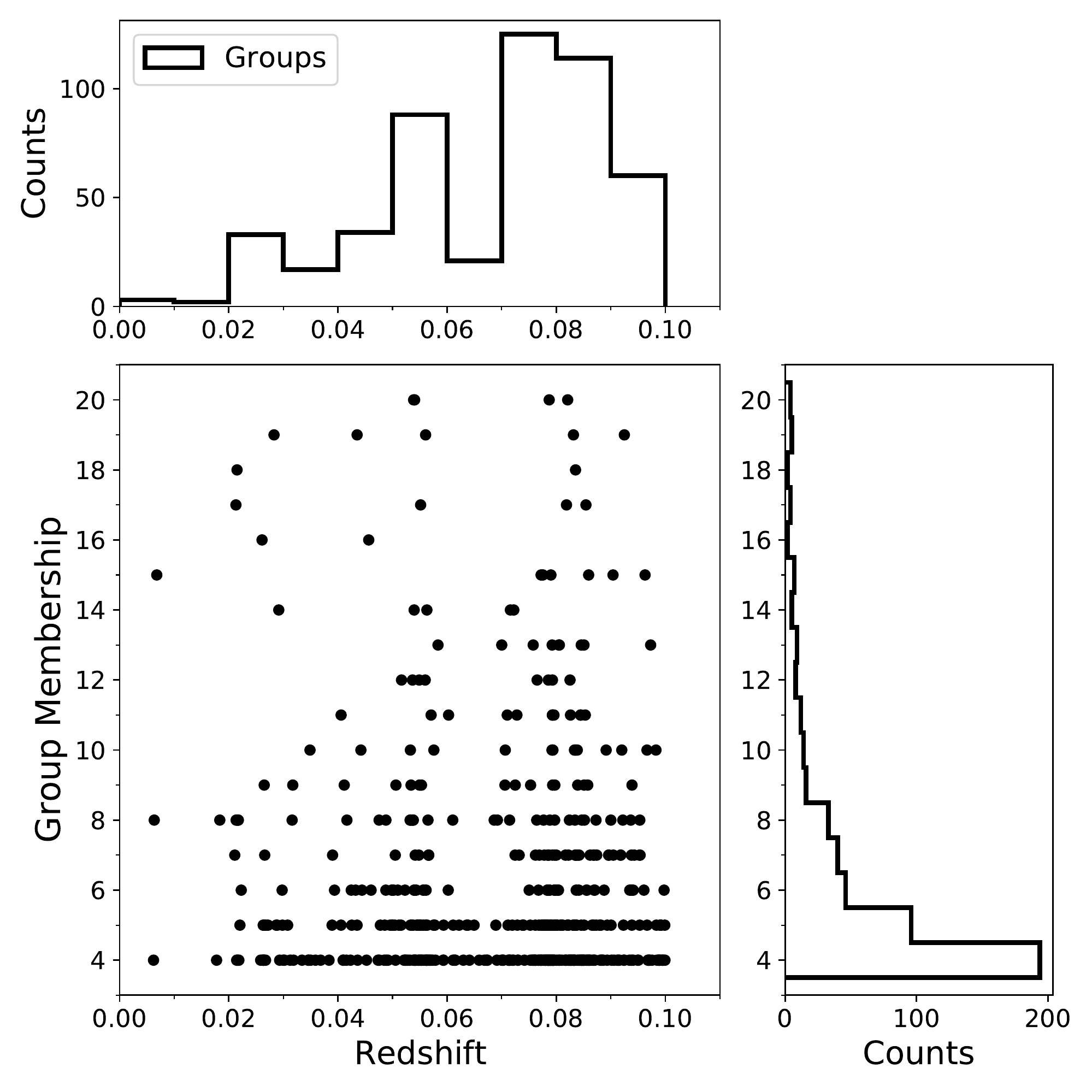}
\caption{Group membership distribution, as a function of redshift, of the G3C groups in the sample.}
\label{fig:f1}
\end{center}
\end{figure}

Multiwavelength optical and near-infrared photometry from LamdarCatv01 \citep{Wr16}, corrected for Galactic foreground dust extinction using \citet{Sch98} (GalacticExtinctionv03), is used in combination with the \wise mid-infrared (mid-IR) catalog described above to determine rest-frame photometry by fitting to the empirical template library of \citet{Br14b}.

The aim of our study is to use \wise as our primary data source; given its all-sky coverage, a study of this kind can be extended in a uniform way. In a sense we are therefore simulating regions of the sky that do not share the extensive multiwavelength coverage in GAMA, and hence we use \wise to determine both the stellar mass and SFRs of the galaxies in our sample.
The stellar masses are derived using equation (1) of \citet{Clu14}. This relation was determined using \wise-resolved galaxies calibrated to the GAMA stellar masses of \citet{Tay11}, derived from stellar population synthesis modelling, and assuming a \citet{Chab03} IMF. For convenience we reproduce it here:
\begin{equation}
{\rm log_{10}}\, {\rm M}_{\rm stellar}/L_{\rm W1} = -2.54 ({\rm W1} - {\rm W2}) - 0.17, \\ 
 \end{equation}
with $  L_{\rm W1}\ ( L_{\odot,W1}) = 10^{ -0.4 (M - M_{\rm \odot,W1})} $, where $M$ is the absolute magnitude of the source in W1, $M_{\rm \odot,W1} = 3.24$ mag (the W1 in-band value of the Sun), and ${\rm W}_{3.4\mu m} - {\rm W}_{4.6\mu m}$ is the rest-frame W1$-$W2 color of the source \citep[see][for further details]{Jar13}. This equation is only applied within the limits of the calibration i.e. W1$-$W2 color from $-$0.05 to 0.2 mag (corresponding to a M$/L_{\rm W1}$  of 0.21 to 0.91); for galaxies with only W1 detections (e.g. dwarfs) a constant M$/L_{\rm W1}$ of 0.6 is used \citep[see][]{Jar19}. For consistency with our SFR relation (see below), we convert our stellar masses to a \citet{Kroup01} IMF using the offsets from \citet{Zahid12} corresponding to 0.03\,dex.

After calculating the \wise-derived stellar masses, we impose a stellar mass cut of log M$_{\rm stellar} \geq 9.0$ M$_\odot$ as discussed above (see Table \ref{tab:t1}). After the cut, one group of 4 galaxies (GroupID: 200857) is no longer represented in our sample, leaving 497 groups whose redshift distribution is shown in Figure \ref{fig:f1}. Inspection of the properties of this excluded group (located at $z=0.029$) find that all its members have an optically-derived stellar mass $<10^{9}$ M$_\odot$, reflecting broad consistency between the two measures of stellar mass. This would be an interesting group in its own right, but illustrates the rarity of such groups in our volume. 

SFRs are determined using the W3 (12\micron) band, after removing the contribution from the stellar continuum \citep[see][]{Clu17}. Dust-reprocessed star formation, as traced by the mid-infrared, probes star formation on timescales $\gtrsim 100$ Myr, but compares favourably to optically-derived values that require, sometimes large, dust corrections \citep[see ][]{Clu17}. We make use of equation (4) from \citet{Clu17}, calibrated to the total infrared luminosities of the SINGS/KINGFISH sample \citep{Dale17} and assuming a Kroupa IMF \citep{Kroup01}, reproduced here:
\begin{multline}
 {\rm log\, SFR}\ (\rm M_{\odot}\, {\rm yr}^{-1}) = \\ (0.889\pm 0.018)\, {\rm log}\, L_{12\micron} (L_\odot) - (7.76\pm 0.15), \\
\end{multline}
where $L_{12\micron}$ is the monochromatic 12\micron\ luminosity, $\nu L_{\nu}(12\micron)$, determined from the restframe-corrected W3 band.

We note that some dust-reprocessed emission within the W3 band is due to heating from the old stellar population; this is particularly true of massive ellipticals.  Although this is somewhat ameliorated when removing the contribution of the stellar continuum from the W3 band, it does mean that an artificially elevated SFR can occur, although this effect is not expected to be large enough to impact the analysis presented here \citep[see][]{Clu17}.

\subsection{Photometry Quality Cuts \label{photcut}}

For our primary analysis, we are particularly interested in the stellar mass dependence of our samples which requires controlling for stellar mass in a reliable way. We impose a restriction of signal to noise (S/N) in W1$-$W2 color (S/N$>$5) which corresponds to a stellar mass error less than 0.5 dex; this is necessary to limit contamination across bins. We explore the consequences of this selection further in section \ref{sec:A1} of the Appendix, but note that it chiefly impacts the number of galaxies in our sample at the low mass end (M$_{\rm stellar}<10^{9.5}$ M$_\odot$). The requirements we have imposed on our sample mean that all numbers and figures presented in this work should be compared in relative and not absolute terms.

We further apply a S/N cut in W2$-$W3 (S/N$>2$) to assign reliable colors and SFRs. For systems with a lower S/N W2$-$W3 color, we report a low S/N SFR. Upper limit SFRs to $z<0.1$ can arise when there is a reliable W2$-$W3 color, but a large stellar continuum dominates the W3 band -- such as is the case for high mass elliptical galaxies -- and results in an upper limit after correcting for this. In addition, an upper limit arises when little to no star formation is detected; this occurs at high mass due to systems having low or negligible star formation, but also at the low mass end (M$_{\rm stellar}<10^{10}$ M$_\odot$), where the low surface brightness of these systems impacts the W3 detections of these systems. We note that for the analysis presented in section \ref{sec:MS}, the \wise W3 band sensitivity means we are complete to $z<0.1$ for detecting star formation at the level of the quenching separator for M$_{\rm stellar}\geq 10^{10}$ M$_\odot$. However, for the low levels of star formation of the M$_{\rm stellar}< 10^{10}$ M$_\odot$ population, there is redshift dependence; we therefore test each source against its distance to determine if a W3 flux (and hence SFR) could be detected. This limits the number of sources we can count in our lowest stellar mass bins and therefore reduces our statistical power there.

\subsection{Visual Morphological Classification}

In order to investigate the morphological mix in group environments, we make use of the VisualMorphologyv03 DMU which contains the visual identification \citep[following][]{Driv12} for galaxies in the GAMA II equatorial regions to $z < 0.1$ (i.e. encompassing our entire sample). Here galaxies with ELLIPTICAL$\_$CODE$=$1 are classed as ``Elliptical" and those with ELLIPTICAL$\_$CODE$=$10 are classified as ``NotElliptical". No reliable classification is given ELLIPTICAL$\_$CODE$=$0, and 3\% are classified as Little Blue Spheroids (with ELLIPTICAL$\_$CODE$=$2). The classification was performed on three colour $giH$-band images from the SDSS \citep{York00}, VIKING \citep{Ed13} or UKIDSS \citep{Law07} large area survey data. 

This DMU additionally includes the Hubble Type classifications, following \citet{Kel14b}, for the GAMA II equatorial regions, but which is only available for galaxies to $z<0.06$. Of the galaxies with both an ELLIPTICAL$\_$CODE and HUBBLE$\_$TYPE$\_$CODE classification (i.e. $z<0.06$), we find the following:

\begin{itemize}

\item{ELLIPTICAL$\_$CODE$=$1: 73\% are classified as E and 22\% are classified S0-Sa, with only 1\% classified as either Sab-cd or Sd-Irr.}

\item{ELLIPTICAL$\_$CODE$=$10: 79\% are classified as either Sab-cd or Sd-Irr, with 4\% classified as E, and 10\% classified as S0-Sa.}

\end{itemize}

For the purposes of this study, therefore, we make use of the ELLIPTICAL$\_$CODE classification, where available, and assign galaxies with ELLIPTICAL$\_$CODE$=$1 to be ``Early-type", i.e. bulge-dominated systems, and those with ELLIPTICAL$\_$CODE$=$10 as ``Late-type", i.e. disk-dominated galaxies. After applying our stellar mass cut, we have only 2 galaxies in the non-G3C sample without a classification (ELLIPTICAL$\_$CODE$=$0). Similarly for 11 galaxies in the G3C sample.

\section{Analysis}

\subsection{\wise colours}

\begin{figure*}[!htb]
\gridline{\fig{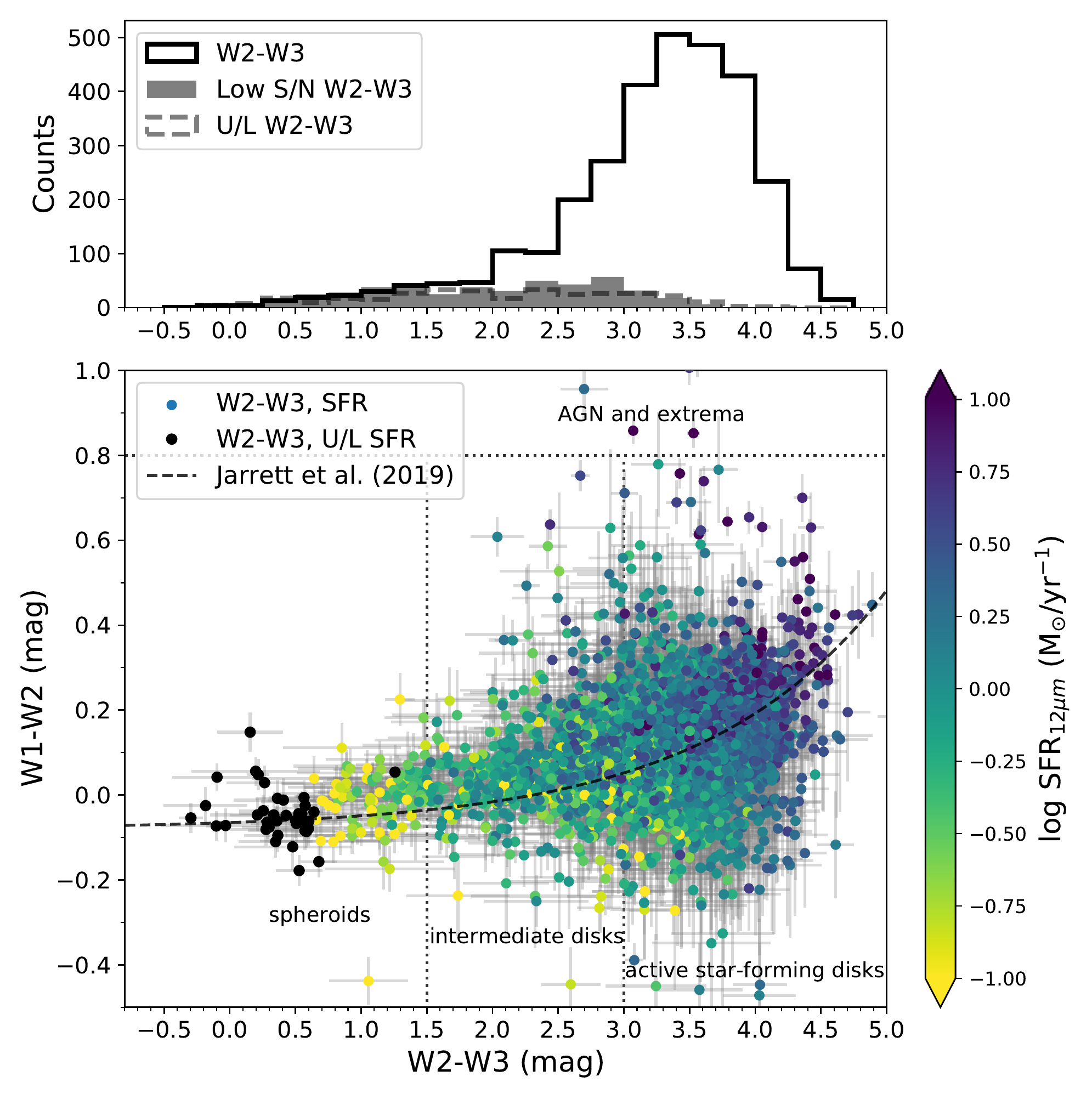}{0.5\textwidth}{(a) Non-G3C sample}
             \fig{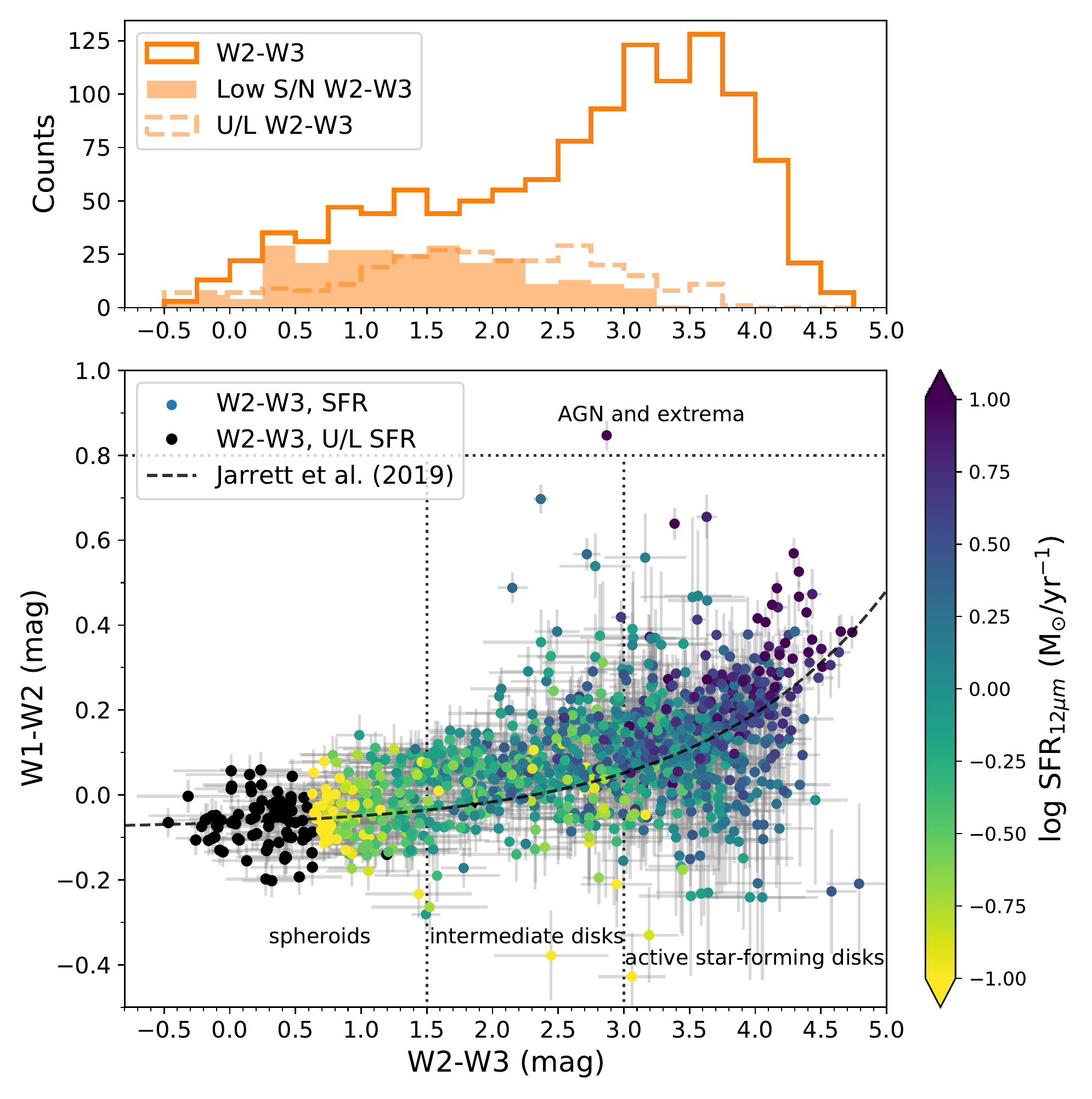}{0.5\textwidth}{(b) G3C sample}}
\caption{\wise color-color diagram (W1$-$W2 vs W2$-$W3), color-coded by SFR, for the (a) non-G3C sample and (b) G3C galaxies, after applying the W2$-$W3 S/N cut. Also shown are galaxies where the requirement of a robust SFR have been removed (black points). The upper panels additionally reflect sources with low S/N and upper limit W2$-$W3 color that are now shown in the main panel. The non-G3C sample is dominated by star-forming systems with large W2$-$W3 color, while the G3C sample shows relatively more sources at low W2$-$W3 color. The color-morphology divisions and color-color ``sequence" are from \citet{Jar19}.\label{fig:f3}}
\end{figure*}

We consider first the \wise color-color diagrams of the non-G3C and G3C galaxies, color-coded by 12\micron\ SFR, as shown in Figure \ref{fig:f3}; we include the ``color sequence" and delimited regions from \citet{Jar19}. In the upper panels we show the distribution of systems with reliable W2$-$W3 colors (S/N$>2$), which are plotted in the main panel. We also include in the upper panels the color distribution of the low S/N and upper limit W2$-$W3 sources.  

As expected, star formation is generally highest at large W2$-$W3 color, decreasing to the left where low star-forming, large stellar mass systems reside \citep{Jar19}. It is apparent from the histogram in the upper panel of Figure \ref{fig:f3}a that the W2$-$W3 color distribution of the non-G3C galaxies is largely unimodal, dominated by mid-IR ``red" colors corresponding to actively star-forming systems; this is consistent with the optical color distribution ($u-r$) found for ungrouped galaxies at $z<0.1$ in \citet{Al15}. By comparison, the G3C galaxies (Figure \ref{fig:f3}b) show a bimodal distribution in W2$-$W3 color. We note that the mid-IR-derived SFRs of galaxies in the ``AGN and extrema" zone \citep[see][]{Jar11, Stern12} are likely contaminated by hot dust from AGN heating, and are considered to be unreliable in their mass and SFR tracers (and therefore excluded from this analysis). 

Included in the main panels of Figure \ref{fig:f3}, as black points, are galaxies with a well-determined W2$-$W3 color, but without a robustly determined SFR. There is an inherent challenge in studying systems that are ceasing to form stars (or have ceased star formation, i.e. passive) using diagnostics that require a well-determined SFR.  In the case of the mid-IR, the galaxy must have dust-obscured SF in order to gauge the activity through ISM heating, wholly separate from the photospheric emission from evolved stars. In the mid-IR, bulge-dominated galaxies with little star formation will still have a reliable, if low, W2$-$W3 color as their Rayleigh-Jeans continuum dominates the W3 band (and not warm dust from star formation). This continuum is removed when calculating the 12\micron-derived SFR, which can result in little to no (reliably) detectable warm dust. Figure \ref{fig:f3} shows these cases as black points and demonstrates that the W2$-$W3 color probes further down by almost an entire magnitude (W2$-$W3 $<0.5$), particularly for the G3C sample, by removing the SFR requirement (but retaining the S/N selection). 

\begin{figure*}[!thb]
\gridline{\fig{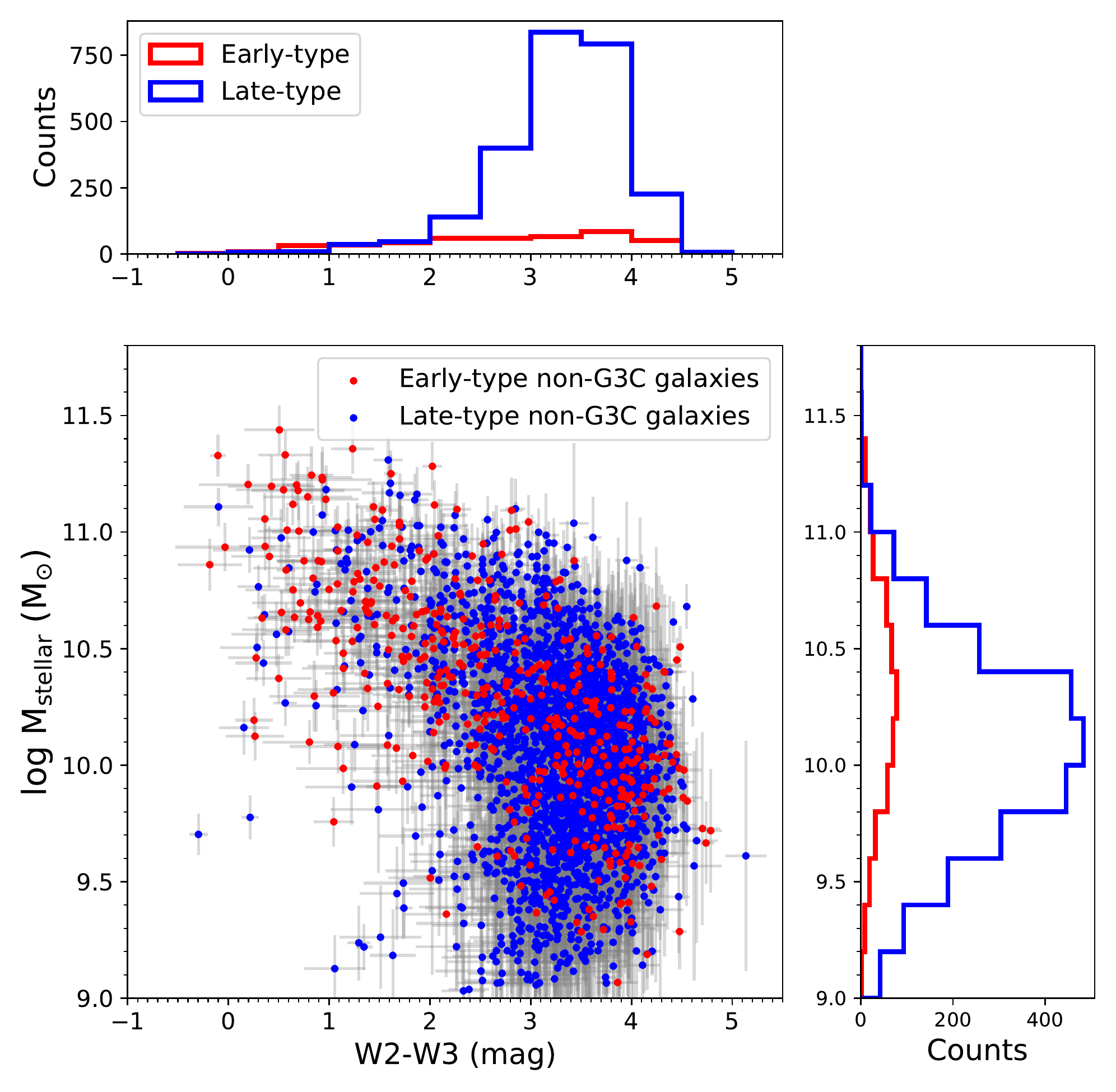}{0.5\textwidth}{(a) non-G3C sample}
              \fig{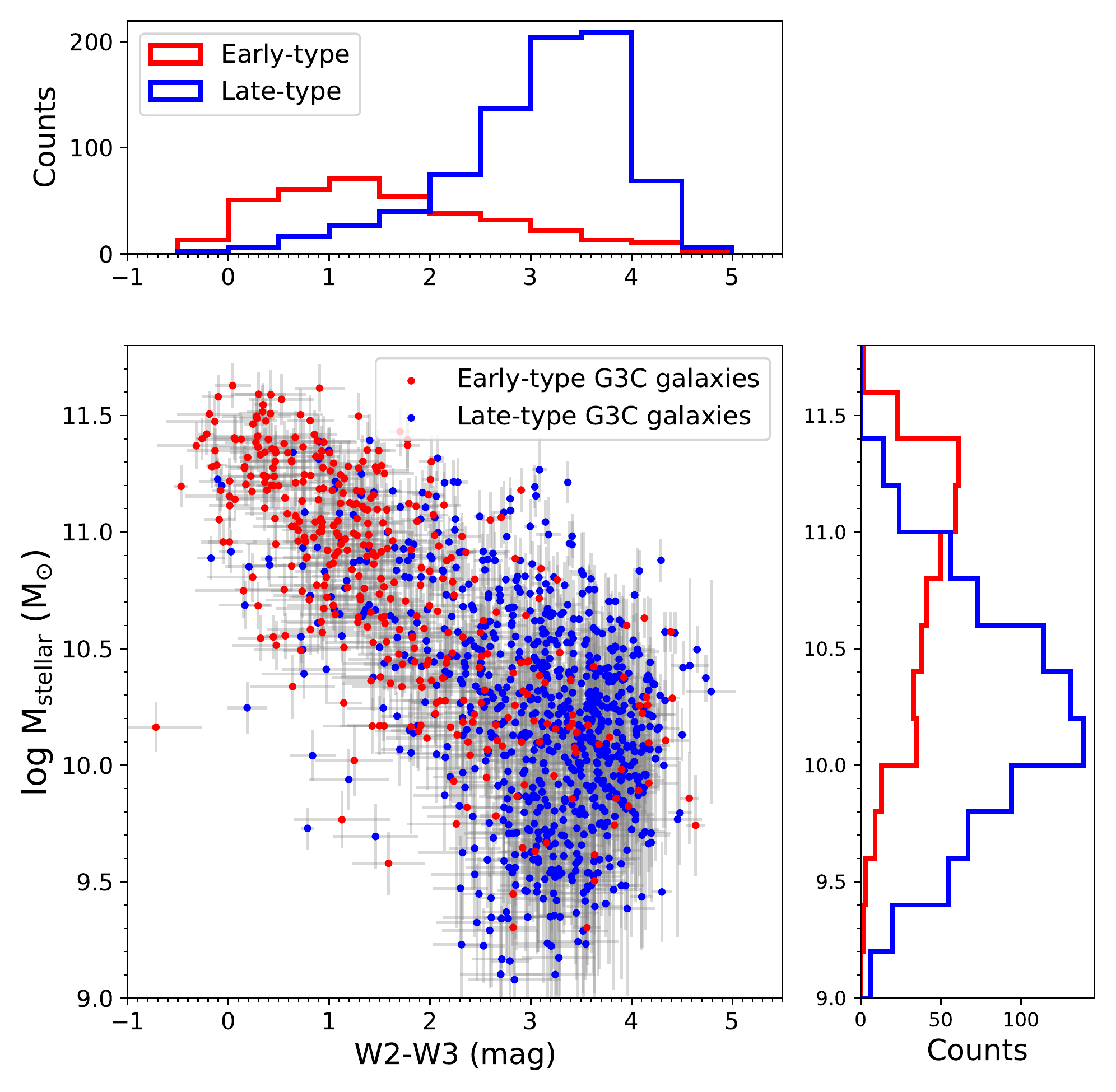}{0.5\textwidth}{(b) G3C sample}}
\caption{Stellar mass as a function of W2$-$W3 color, color-coded by morphological classification (early-type$=$red, late-type$=$blue). The (a) non-G3C sample and (b) G3C sample have very different distributions in this phase space. \label{fig:f4}}
\end{figure*}

\subsection{Stellar Mass and Morphology}

We next explore a more physically-informed diagnostic, stellar mass (M$_{\rm stellar}$) versus W2$-$W3 color. In Figure \ref{fig:f4} we divide the non-G3C and G3C samples into early-type (i.e. bulge-dominated) and late-type (i.e. disk-dominated), making use of the visual morphology classification outlined previously. This clearly shows that the dominance of star-forming galaxies seen in Figure \ref{fig:f3}a is directly attributed to late-type systems, with little contribution from early-types (only 14\% of non-G3C galaxies in our sample are classified as early-type), even at low W2$-$W3 color/ high stellar mass (M$_{\rm stellar}$ $>10^{11}$\,M$_\odot$). 

In contrast, the G3C sample (Figure \ref{fig:f4}b) shows a clearly bimodal distribution in stellar mass, with early-type systems (33\% of our sample) dominating at the high stellar mass/low W2$-$W3 end. This framing of stellar mass vs W2$-$W3 color also shows that low-mass systems do not have the largest W2$-$W3 color, i.e. the late-type distribution turns back to bluer mid-IR colors at low mass indicating lower dust content and SFR activity. 

The difference in the grouped and ungrouped stellar mass distributions is consistent with what was found by \citet{Al15} in GAMA. We note that the lower mass and late-type nature of the non-G3C galaxies (when considered the central of an undetected group) are in agreement with \citet{Robot13} who find that centrals of lower mass (M$_{\rm stellar}$ $<10^{10.5}$\,M$_\odot$) are more likely to be late-type than centrals of higher mass.

\begin{figure*}[!thb]
\centering
\subfloat[Group Mass 1 ($<10^{12.95}$\,M$_\odot$/h)]{
\includegraphics[width=0.5\textwidth]{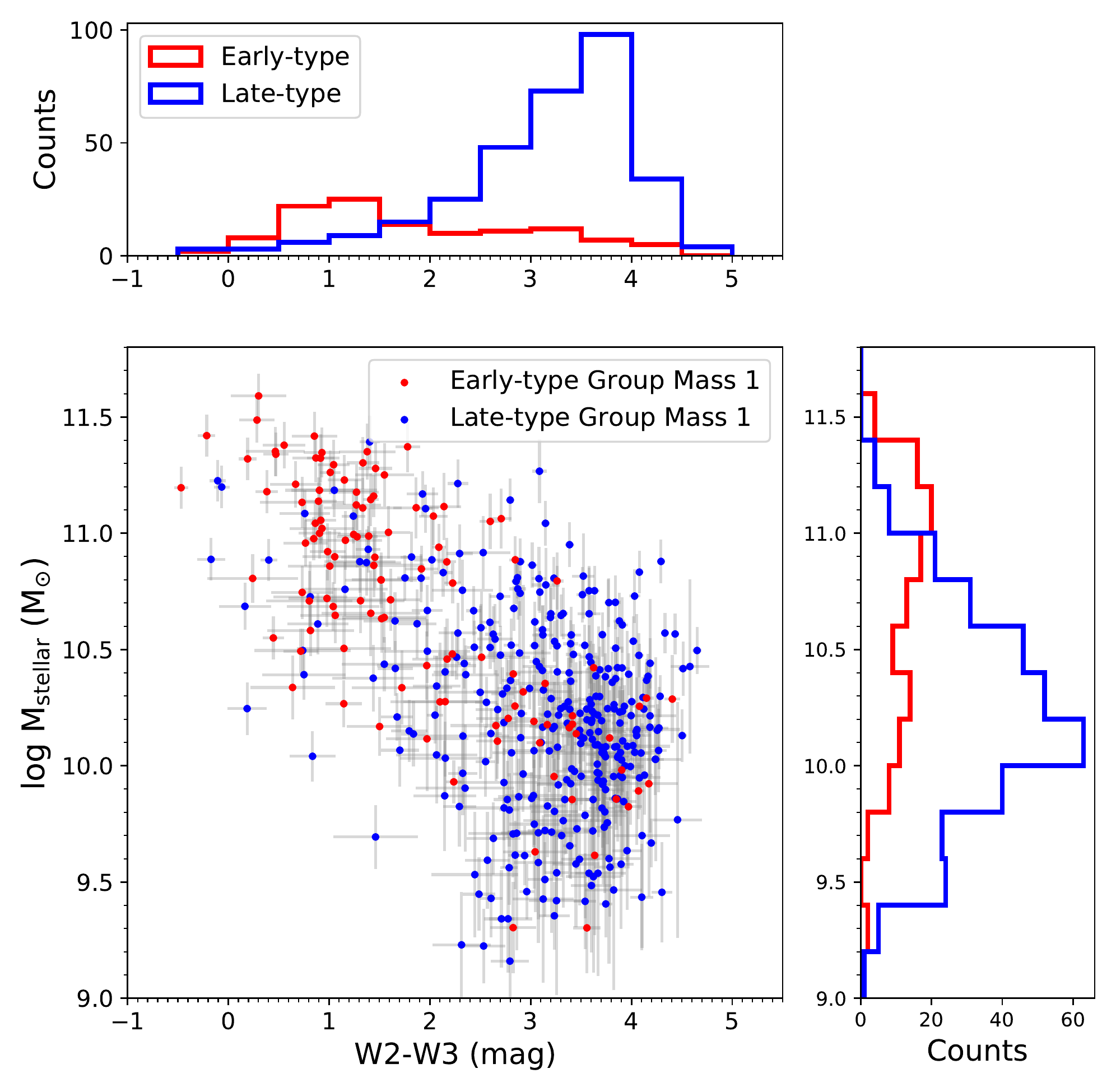}}
\\
\subfloat[Group Mass 2 ($10^{12.95}$ -- $10^{13.4}$\,M$_\odot$/h)]{
\includegraphics[width=0.5\textwidth]{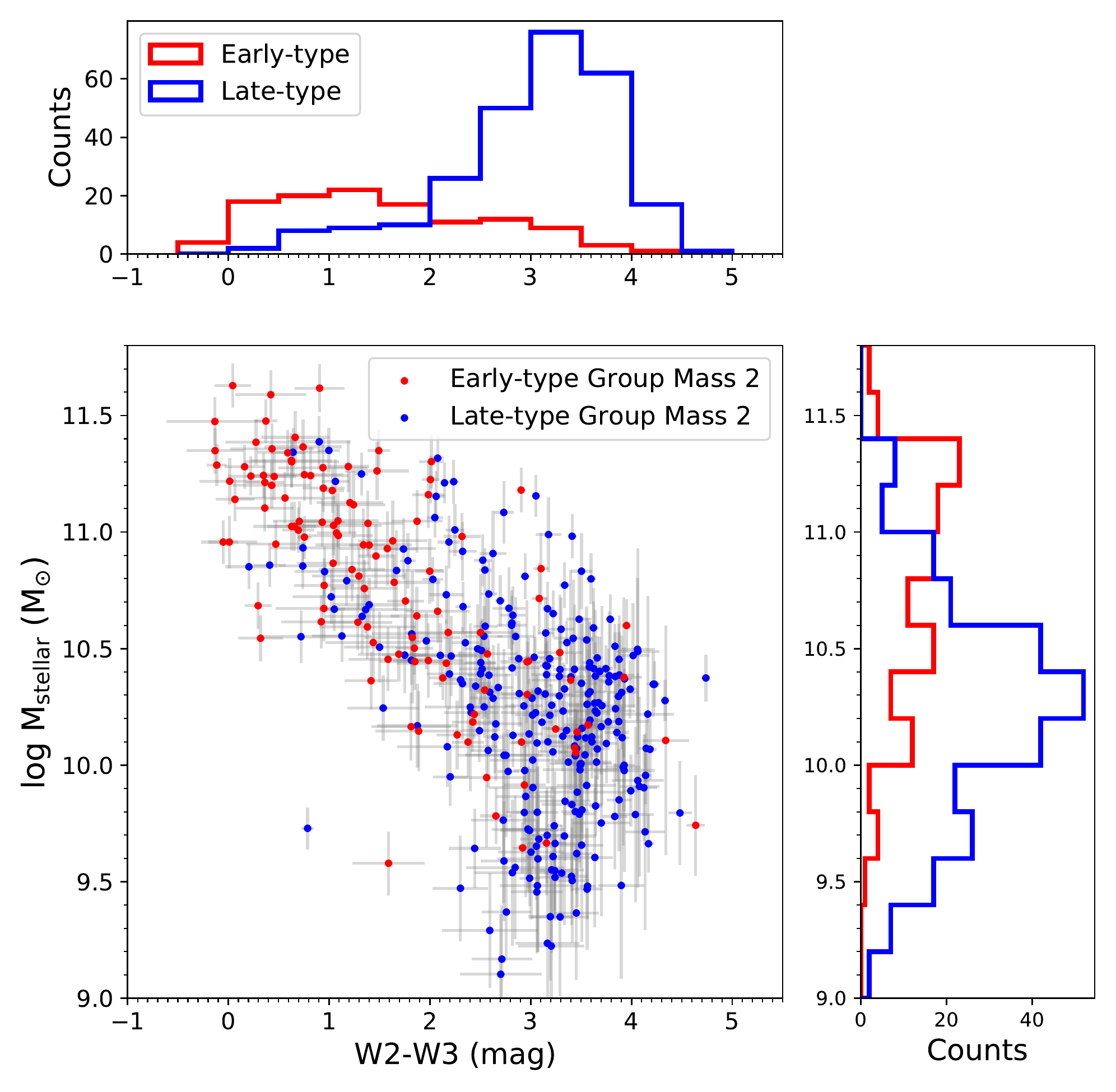}}
\subfloat[Group Mass 3 ($>10^{13.4}$\,M$_\odot$/h)]{
\includegraphics[width=0.5\textwidth]{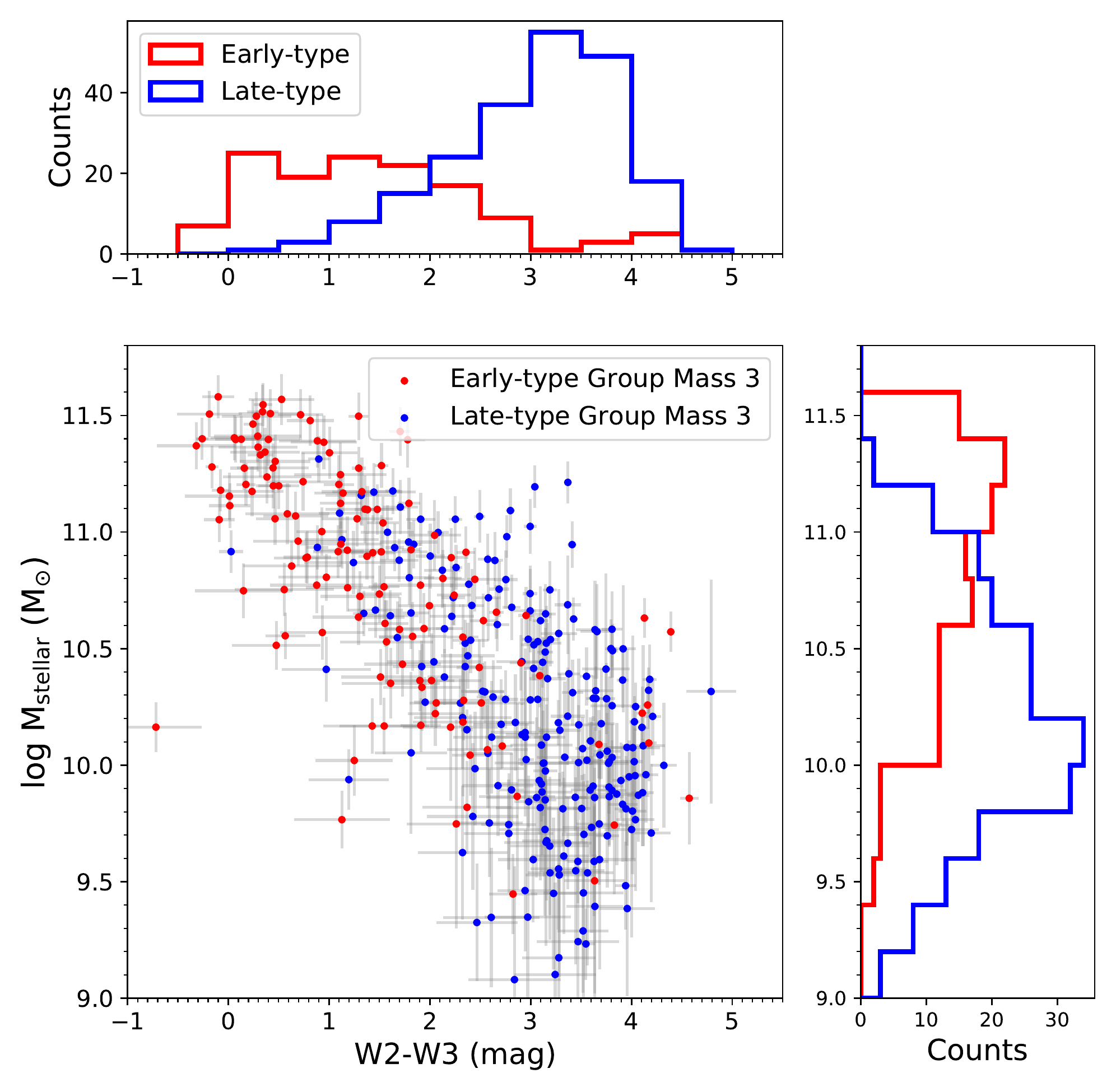}}
\caption{As in Figure \ref{fig:f4}, but for the individual group mass bins listed in Table \ref{tab:t2}. With increasing group mass, the relative proportion of early-type systems increases (corresponding to an increase of systems at high stellar masses): in Group Mass 1, 27\% are early-type, 31\% in Group 2, and and 39\% for the largest halos.
\label{fig:f5}}
\end{figure*}

We see from Figure \ref{fig:f4} that, not only do the source counts of the ungrouped sample drop off rapidly for M$_{\rm stellar}$ $>10^{10.5}$\,M$_\odot$, but that very few are classified as early-type compared to the same stellar mass range in the grouped sample; this makes them very different populations. In the mass range of M$_{\rm stellar}$ $>10^{10.5}$\,M$_\odot$ where galaxies are turning off the star-forming main sequence, it shows that, at fixed mass, the grouped galaxies are more likely to be early-type. i.e. one important consequence of being in a group is the associated increase in the early-type fraction; we investigate this further in the next section. 

Considering the stellar mass distributions (right panels in Figure \ref{fig:f4}), the late-types (blue histograms) in the G3C sample show a tendency to higher mass compared to the non-G3C sample. This implies that the changing stellar mass function also modifies the mass distributions of different morphological types. 

Next we turn to the dynamical (halo) masses of the groups in our sample. We divide the \wise-G3C sample into three halo mass bins, choosing limits that provide roughly equal numbers of galaxies. Three groups that have MassAFunc$=0$ are excluded; the remaining are divided as shown in Table \ref{tab:t2}.

\begin{deluxetable}{lccC}[htb!]
\tablecaption{Dynamical Mass of Group Sample \label{tab:t2}}
\tablecolumns{4}
\tablenum{2}
\tablewidth{0pt}
\tablehead{
\colhead{} &
\colhead{Log Mass Range} &
\colhead{Number of} &
\colhead{Number of} \\
\colhead{} &
\colhead{(M$_\odot$/h)} &
\colhead{Groups} &
\colhead{Galaxies} 
}
\startdata
Group Mass 1 & $\leq12.95$ &  223 &  853 \\
Group Mass 2 & 12.95 -- 13.4  &  159 &  891 \\
Group Mass 3 & 13.4 -- 14.1 &  112 &  824 \\
\enddata

\end{deluxetable}

\begin{figure}[!thp]
\begin{center}
\includegraphics[width=8cm]{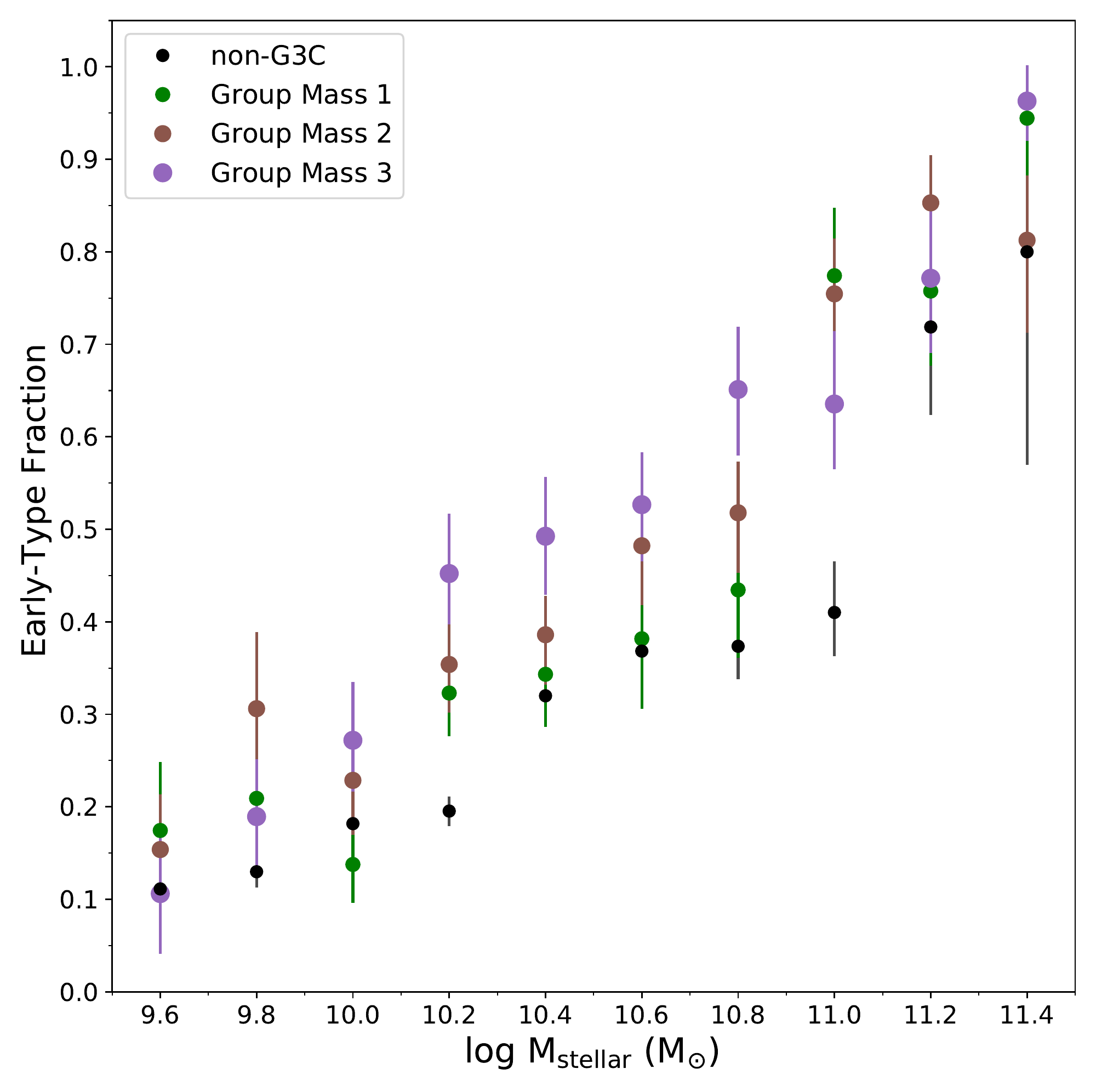}
\caption{The Early-type fraction in bins of stellar mass for the non-G3C (black), Group Mass 1 (green), Group Mass 2 (brown), and Group Mass 3 (purple). Errors are determined from bootstrap resampling. This indicates that in the bins between M$_{\rm stellar}$ $\sim10^{10}$\,M$_\odot$ and M$_{\rm stellar}$ $\sim10^{11}$\,M$_\odot$ we may be seeing an increase in early-type fraction associated with increased halo mass.}
\label{fig:new}
\end{center}
\end{figure}

Splitting into group mass bins (Figure \ref{fig:f5}), we note the increase in systems at high stellar mass, M$_{\rm stellar}$ $>10^{10.5}$\,M$_\odot$, (shown in the right panels) with increasing halo mass. This is consistent with the variation of the galaxy stellar mass function with halo mass \citep[e.g.][]{Yang09, Al15}. In addition, the shift to low W2$-$W3 colors observed in Figure \ref{fig:f3}b appears to be driven by an increased population of early-type systems (with increasing halo mass), which dominates the high stellar mass population. In Figure \ref{fig:f5}, Group Mass 1 is 26.7($\pm$2.8)\% early-type, compared to 31.1($\pm$3.3)\% and 38.8($\pm$3.9)\%, respectively, for Group Mass 2 and 3. However, this includes a W2$-$W3 cut. If we consider our entire sample (i.e. with only a W1$-$W2 selection) and M$_{\rm stellar}$ $>10^{10}$\,M$_\odot$ where our sample is most complete, we find that Group Mass 1 is 29.7($\pm2.5$)\% early-type, with 33.3($\pm$2.6)\% and 36.4($\pm$2.8)\%, respectively, for Group Mass 2 and 3. This suggests an increase in early-type fraction with increasing group mass, contrary to what was found by \citet{Bam09} and also \citet{Al15} who do not see a change in the early-type fraction with group mass. Our samples (e.g. halo mass bins) and selections are, however, quite different. \citet{Bam09} do, though, observe that at fixed stellar mass, the fraction of early-types increases with local density, implying that the morphology–density relation is not simply a product of a morphology–mass relationship and the changing stellar mass function.

We investigate the early-type fraction (per stellar mass bin) for our non-G3C and 3 G3C samples in Figure \ref{fig:new}. Although we lack the large samples needed to make a definitive statement, we do see that between M$_{\rm stellar}$ $\sim10^{10}$\,M$_\odot$ and M$_{\rm stellar}$ $\sim10^{11}$\,M$_\odot$ the early-type fraction appears to increase with group halo mass within the stellar mass bins.


We can see that the transition from ungrouped to our lowest halo mass has changed the stellar mass profile and fraction of early-types in a noticeable way. Galaxies in groups have either built (high) stellar mass more efficiently compared to ungrouped systems, with more high-mass galaxies as halo mass increases, or their formation history means they have had more time to build mass. One may expect increased interactions and merging in group environments, which would be consistent with the accompanied increase in early-types (i.e. bulge growth) with increasing halo mass. Gravitational torques due to tidal interactions can cause gas to flow to the centres of galaxies, leading to centrally-concentrated star formation and corresponding bulge growth. \citet{Sch19} find that galaxies with M$_{\rm stellar}$ $>10^{10}$\,M$_\odot$ in high-mass groups are more likely to experience centrally-concentrated star formation, whilst \citet{Bluck14} find that bulge mass is most strongly correlated with passive fraction, consistent with the inside-out growth paradigm.
This observed increase in high stellar mass galaxies and early-types with increasing group halo mass will likely impact the observed fraction of quenched galaxies \citep[e.g.][]{Peng10,Dav19b}, which we examine in the next section. 

To summarise, we find progressively more high-mass and early-type systems with increasing group halo mass compared to the ungrouped sample which is dominated by late-type galaxies and has relatively few galaxies with M$_{\rm stellar}$ $>10^{10.5}$\,M$_\odot$.

\subsection{The Star-forming Main Sequence}\label{sec:MS}

The correlation between stellar mass and SFR for star-forming galaxies \citep[][]{Noes07, El07, Dad07} has become an indispensable tool for identifying and studying the properties of typical star-forming galaxies to high redshift \citep[e.g.][]{Bris19}. It also provides a natural means to separate samples into star-forming, transitioning and passive galaxies \citep[see, for example,][]{Bluck14, Ren15, Bluck16, Dav19b, WangB20}.

\begin{figure*}[!thp]
\begin{center}
\includegraphics[width=12cm]{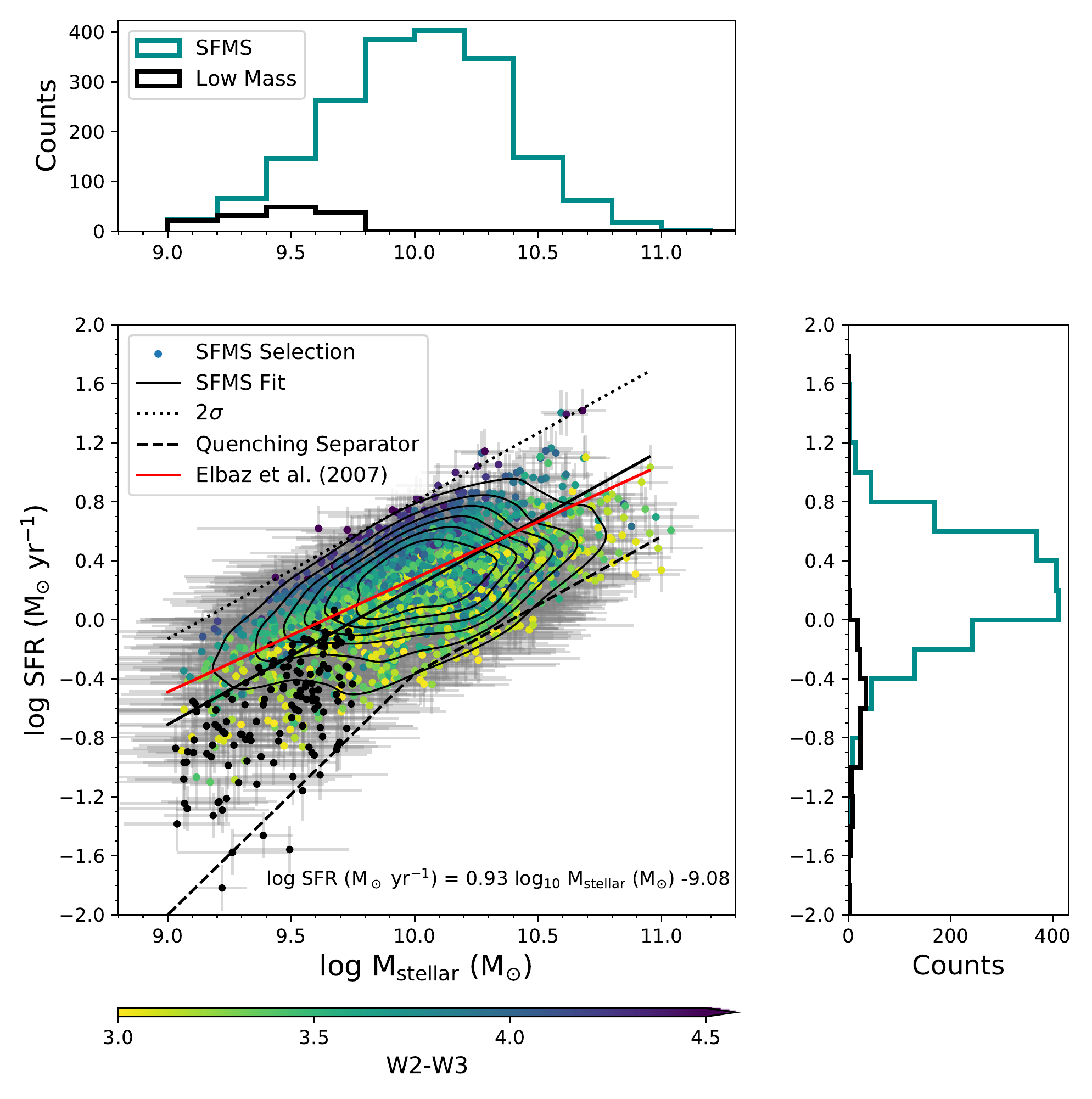}
\caption{The SFMS as determined from the late-type non-G3C sample (colored points), selecting galaxies with W2$-$W3$>3$ \citep[see Figure \ref{fig:f3} and also][]{Jar19}. For comparison, we include the relation from \citet{El07} derived using SDSS for $z<0.1$. Low mass galaxies with high S/N SFRs are shown as black points, with no morphology or color selection imposed. The dotted line shows the upper 2$\sigma$ envelope, while the dashed line represents the quenching separator, which is the lower 2$\sigma$ boundary from the SFMS fit for M$_{\rm stellar}$ $\geq10^{10}$\,M$_\odot$ and a modified selection for M$_{\rm stellar}$ $<10^{10}$\,M$_\odot$  to accommodate lower SFRs at the low mass end. }
\label{fig:ms1}
\end{center}
\end{figure*}

The color-coding used in Figure \ref{fig:f3} showed the connection between SFR and W2$-$W3 color, with high star-formation broadly corresponding to large W2$-$W3 color (due in part to SFR being derived from the W3 luminosities after removing the stellar contribution). We exploit this using the late-type, non-G3C sample and select star-forming galaxies based on their W2$-$W3 color (using W2$-$W3 $>3$). This forms a relatively tight sequence in the log\,SFR--log\,M$_{\rm stellar}$ plane, as shown in Figure \ref{fig:ms1} \citep[corresponding to an assumed Kroupa IMF,][]{Clu17} with a best-fit relation given by:
\begin{multline}
{\rm log_{10}}\, {\rm SFR} ({\rm M}_{\odot}\,{\rm yr^{-1}}) = 0.93\, {\rm log_{10}}\, {\rm M}_{\rm stellar} ({\rm M}_{\odot}) - 9.08,
 \end{multline}
 with $\sigma=0.29$ reflecting the intrinsic spread of the distribution. The distribution is well-contained within $\pm$2-$\sigma$ of the relation, as shown in Figure \ref{fig:ms1}. We find that our relation closely matches the relation of \citet{El07}, derived using SDSS galaxies at $z<0.1$.

The slope of the star-forming main sequence (SFMS) is heavily impacted by the selection of star-forming vs ``mixed" samples, where the inclusion of galaxies that have already ```turned-over" reduces the slope. 
It should be borne in mind that a mid-infrared-derived SFMS makes use of dust-reprocessed star formation and is well-suited to dusty star-formers, but probes star formation on longer timescales ($\sim$100 Myr) compared to, for example, H$\alpha$ sensitive to $\lesssim 20$ Myr \citep[e.g.][]{Ken98}.

We use the lower 2-$\sigma$ (i.e. 0.6 dex) boundary (see Figure \ref{fig:ms1}) to separate systems on the SFMS and those that are below (where M$_{\rm stellar}$ $\geq10^{10}$\,M$_\odot$) -- we refer to these as  ``quenched", but note that these include transitional (or ``quenching") systems, in the process of moving off the SFMS. At the high mass end, our selection of quenched systems, therefore, includes galaxies that would be considered in the ``green valley" by some studies \citep[see e.g.][]{Bluck16, Jan20}.

Our SFMS selection, however, has not taken into account the low mass population in the non-G3C sample, with correspondingly low SFRs; the W2$-$W3$>3$ selection is likely too restrictive at these masses (see Figure \ref{fig:f4}). We therefore show in Figure \ref{fig:ms1} (black points) the high S/N SFRs for galaxies with M$_{\rm stellar}$ $<10^{9.75}$\,M$_\odot$ and W2$-$W3 $\leq 3$. To account for the increased scatter, we bin in stellar mass and determine the values at 3 standard deviations below the SFMS and fit to those points, intersecting the 2$\sigma$ line at $10^{10}$\,M$_\odot$. The equation for the quenching separator at low mass is therefore given by: 
\begin{multline}
{\rm log_{10}}\, {\rm SFR} ({\rm M}_{\odot}\,{\rm yr^{-1}}) = 1.635\, {\rm log_{10}}\, {\rm M}_{\rm stellar} ({\rm M}_{\odot}) - 16.715; \\
M_{\rm stellar} <10^{10}\,M_\odot
 \end{multline}

\begin{figure}[!thp]
\begin{center}
\includegraphics[width=9cm]{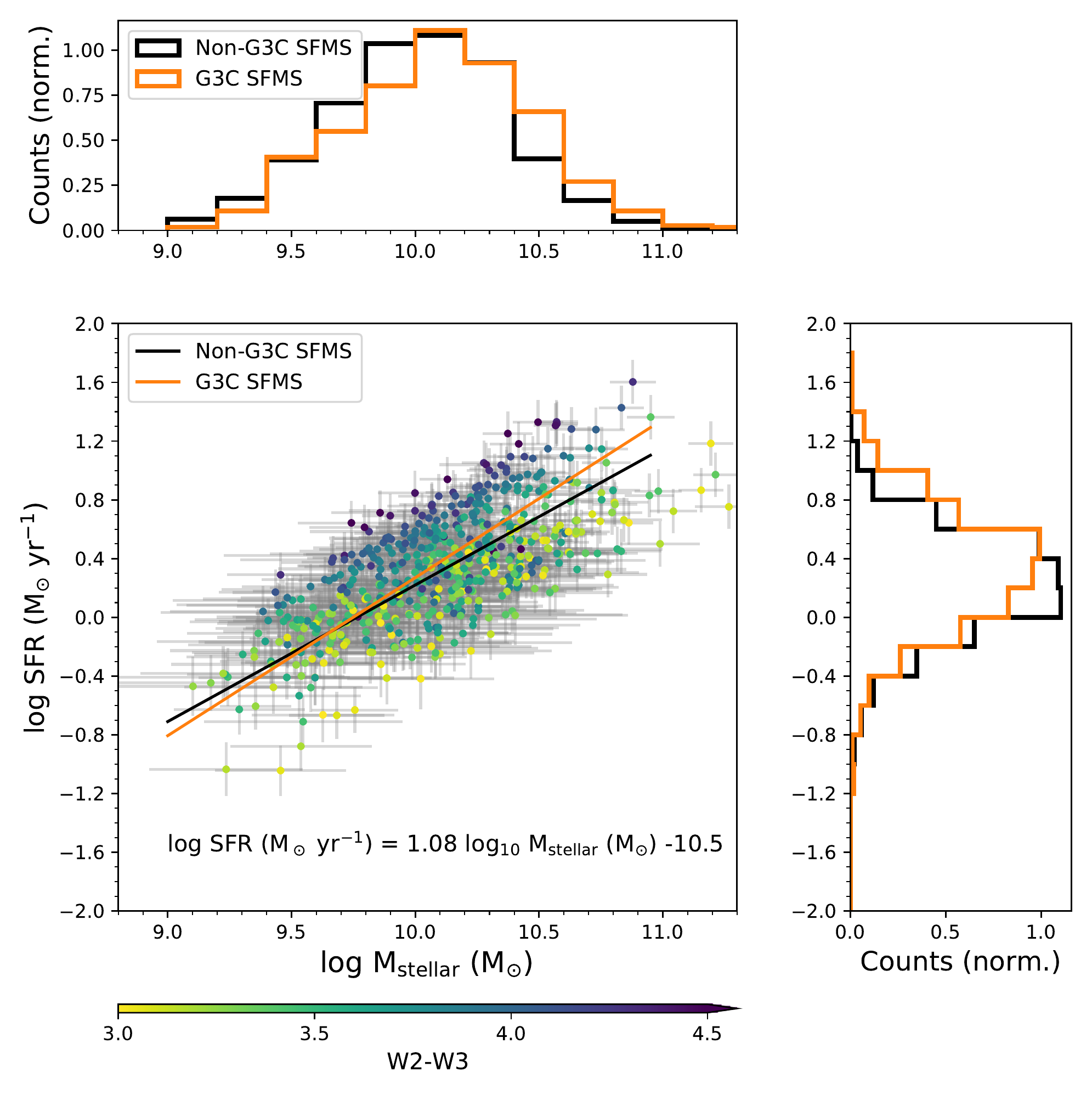}
\caption{Replicating the selection used in Figure \ref{fig:ms1} (late-type galaxies with W2$-$W3$>3$), but using the G3C sample finds a steeper slope (orange line) compared to the non-G3C galaxies. This is reflected in the stellar mass (top panel) and star formation histograms (right panel) and suggests increased star formation in this population relative to the ungrouped galaxies.}
\label{fig:ms2}
\end{center}
\end{figure}

\begin{figure*}[!t]
\gridline{\fig{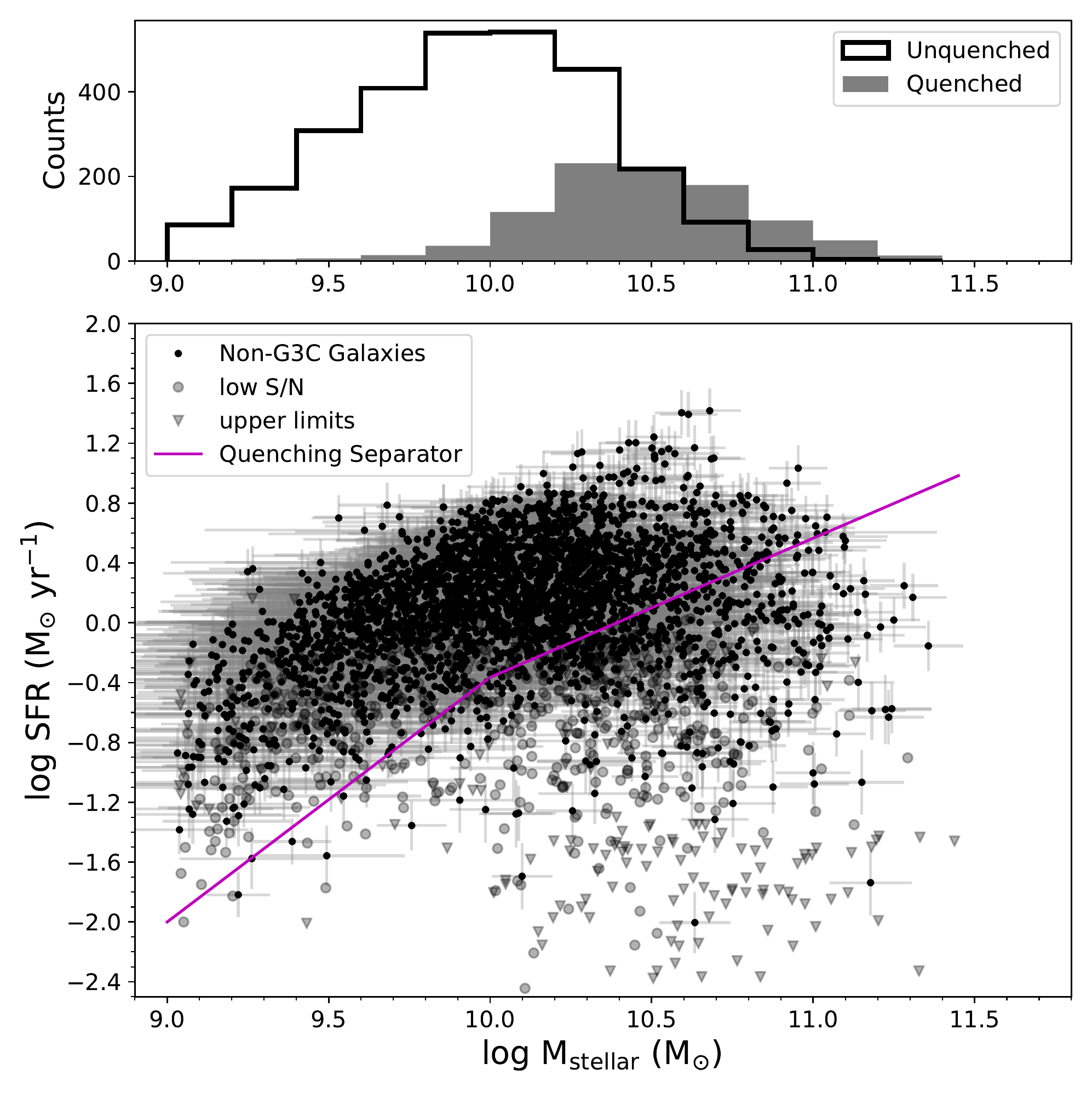}{0.5\textwidth}{(a) non-G3C sample}
               \fig{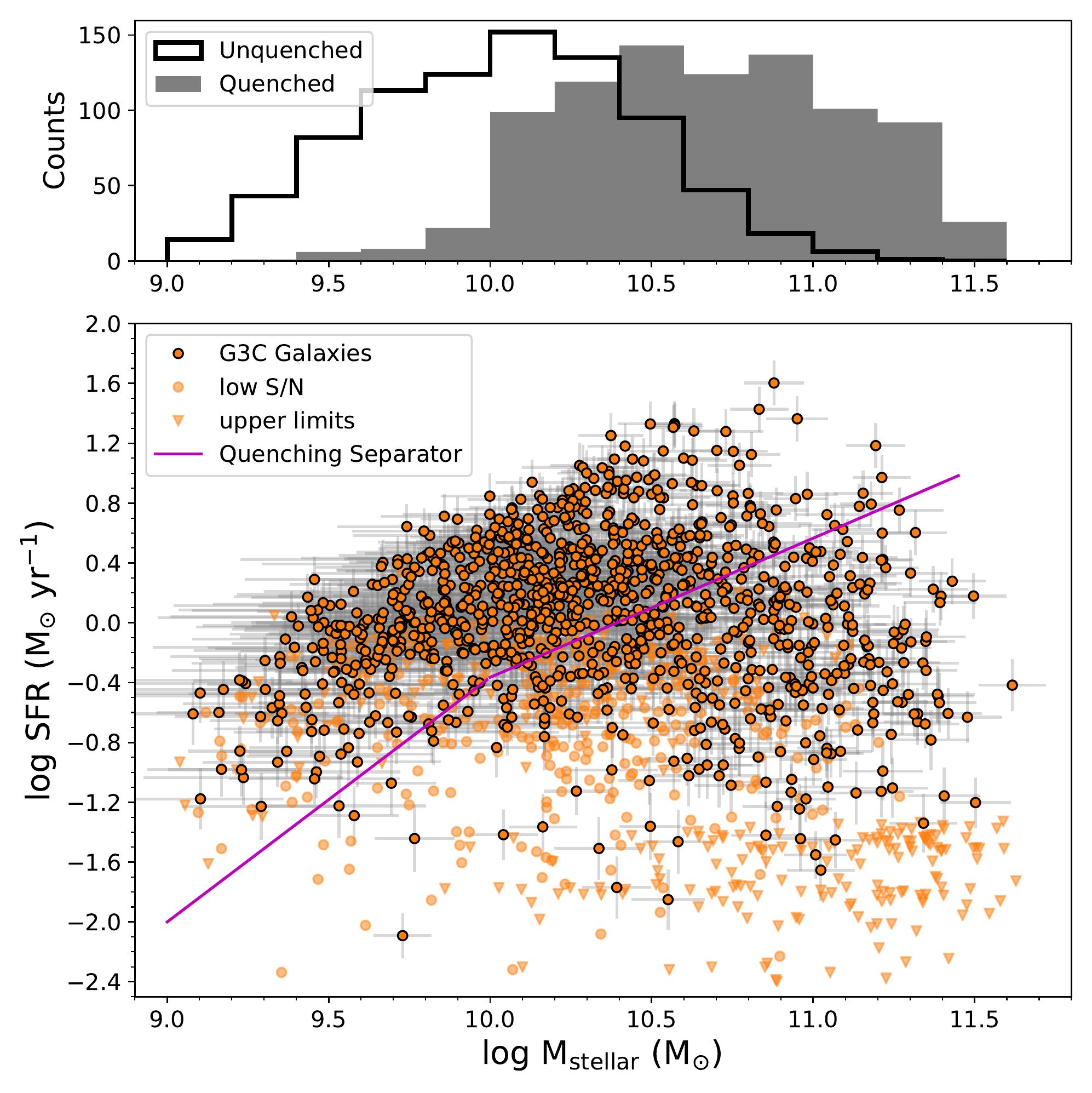}{0.5\textwidth}{(b) G3C sample}}
\caption{The log\,SFR--log\,M$_{\rm stellar}$ distribution for the (a) non-G3C and (b) G3C samples, respectively; the dividing line separates what we consider in this study to be unquenched (above the line) and quenched (below the line) systems. The upper panels reflect the stellar mass distribution of the unquenched (unfilled histogram) and quenched (filled histogram) population. The overall fraction of quenched systems in the G3C sample is 51.5\% compared to just 25\% in the non-G3C sample; see Table \ref{tab:t3}.  
\label{fig:ms3}}
\end{figure*}

Considering now the equivalent SFMS for the G3C sample, we impose the same selection (late-type morphology with W2$-$W3$>3$) and obtain Figure \ref{fig:ms2}, where we see a slightly steeper relation, consistent with a small shift to higher stellar mass and higher SFR (as seen in the normalised histogram comparisons). This would suggest that in the group sample, galaxies on the star-forming sequence are experiencing a ``feast before the famine", tending to higher star formation compared to their ungrouped counterparts. This could be the result of triggered elevated star formation due to increased interactions \citep[e.g. ][]{Mor19}, or minor mergers, occurring in the group environment. It is therefore clear that the selection employed when determining a SFMS selection is important. For the analysis that follows, we will use the quenching separator as determined from the SFMS and low mass population of the non-G3C sample, as a control to test for differences compared to the group environment.

\subsubsection{The SFMS and the Quenching of Star Formation}

In Figure \ref{fig:ms3} we show the log\,SFR--log\,M$_{\rm stellar}$ distribution, for the entire (a) non-G3C and (b) G3C samples. We also show the low S/N and upper limit SFR values, which are included in the stellar mass distributions (upper panels). Using the quenching separator defined in the previous section, the upper panels reflect the quenched and unquenched distributions for both samples. The unquenched distributions of both appear similar as a function of stellar mass. This suggests that the overall SFRs of galaxies within the SFMS is largely agnostic to being in a grouped or ungrouped environment, although we have shown in Figure \ref{fig:ms2} that the slope of the relation is somewhat steeper, suggesting a slightly different mass dependence.

\begin{figure*}[!thp]
\gridline{\fig{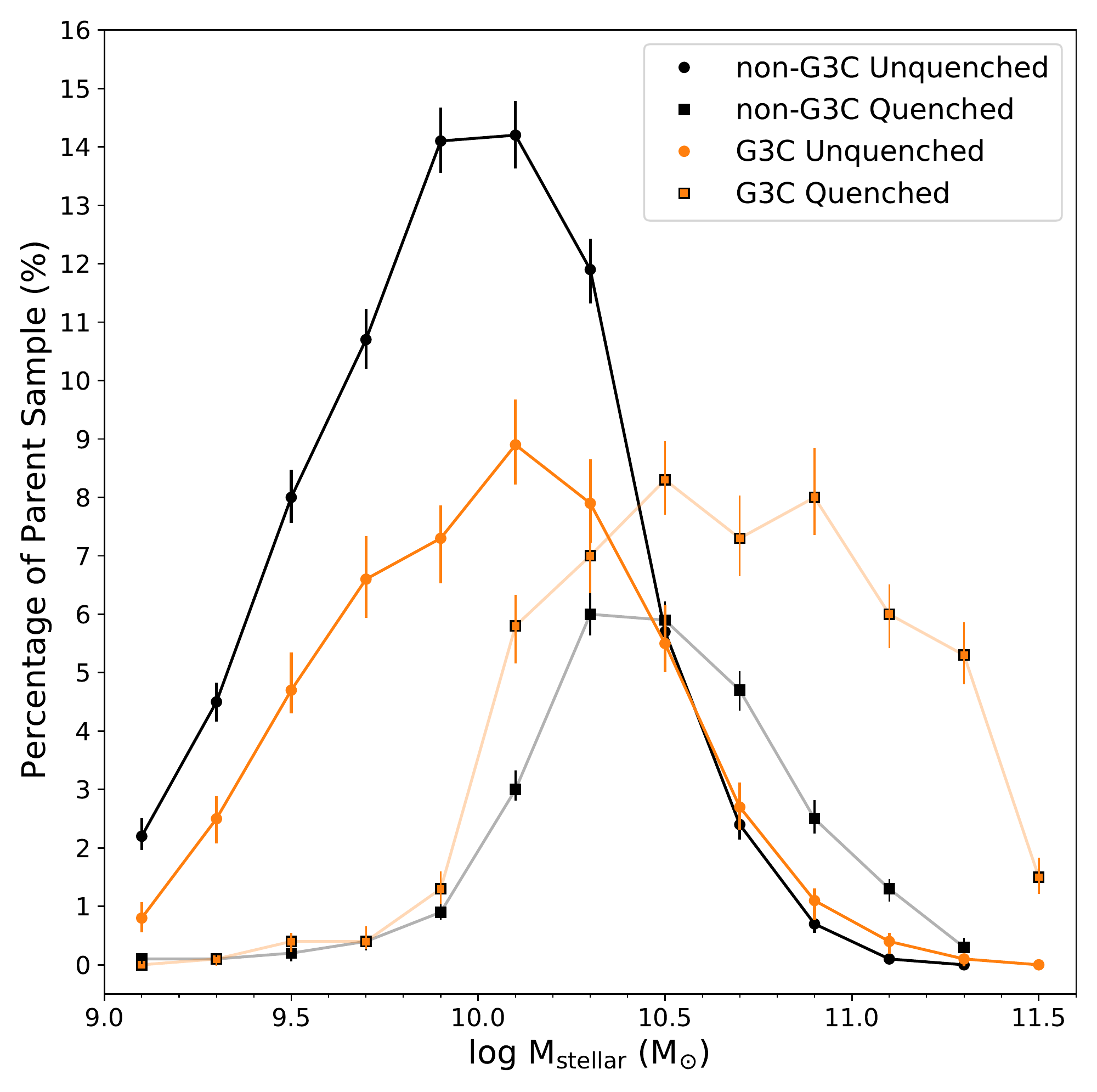}{0.5\textwidth}{(a) non-G3C and G3C comparison}
             \fig{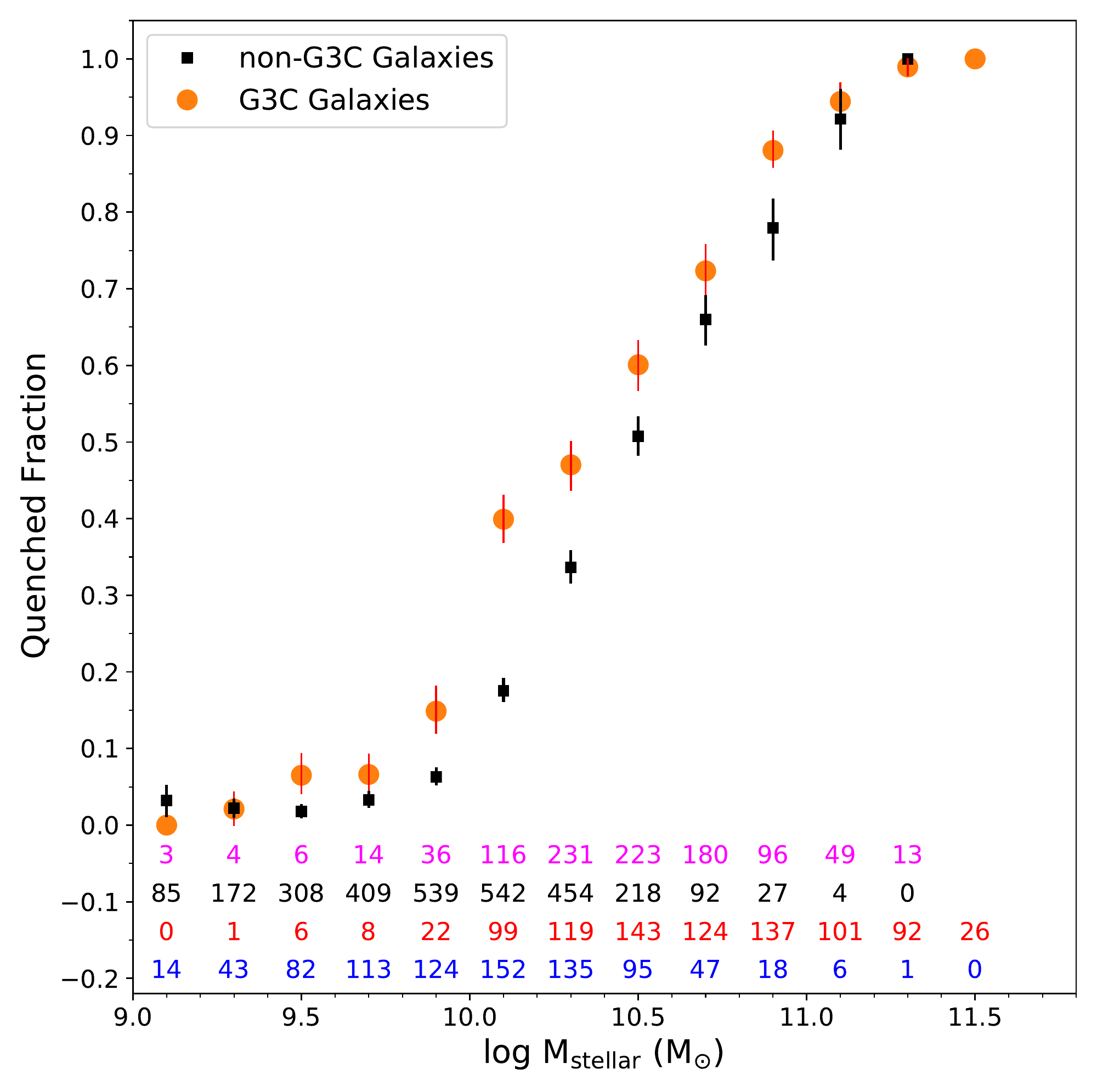}{0.5\textwidth}{(b) quenched fraction of the non-G3C (black) and G3C (orange) samples}}
\caption{(a) The percentage of unquenched (circles) and quenched (squares) systems in each stellar mass bin for the non-G3C (black) and G3C galaxies (orange), respectively. The shape of the distribution of unquenched systems appears similar for both non-G3C and G3C samples, but the quenched distribution for the G3C sample indicates a larger population at high stellar mass. 
(b) The quenched fraction in each mass bin for the non-G3C (black) and G3C galaxies (orange). The numbers at the bottom of the plot reflect the galaxies found in each bin with non-G3C quenched in pink, non-G3C unquenched in black, G3C quenched  in red, and G3C unquenched in blue. Errors are calculated using bootstrap resampling in each bin. \label{fig:ms4}}
\end{figure*}

In the G3C sample (Figure \ref{fig:ms3}b), we clearly see the ``turnover" of SFR at high stellar mass, such that for M$_{\rm stellar}$ $>10^{10.5}$\,M$_\odot$ the SFR of galaxies is broadly decreasing, consistent with what is found in other studies \citep[e.g.][]{Kauf03, Sal07}. This is less evident in the non-G3C sample due to the steep drop-off in high-mass systems.

Here we see why the slope of the SFMS is important; Figures \ref{fig:ms1} and \ref{fig:ms2} have shown that the shape of our star-forming selection in both samples is similar. The difference is in Figure \ref{fig:ms3}b where the greater proportion of massive galaxies with intermediate star formation rates means that if we would fit a relation to the entire star-forming sample (i.e. with no \wise color cut), the slope would be quite different. Any difference would then reflect the proportion of star-forming versus quiescent galaxies, rather than differences in the properties of star-forming galaxies.

\begin{deluxetable}{ccc}[!thb]
\tablecaption{Unquenched and Quenched Fractions for the Ungrouped (non-G3C) and  Grouped (G3C) Samples \label{tab:t3}}
\tablecolumns{3}
\tablenum{3}
\tablewidth{0pt}
\tablehead{
\colhead{\hspace{.5cm} Sample}\hspace{.5cm} &
\colhead{\hspace{.5cm} Unquenched} \hspace{.5cm}&
\colhead{\hspace{.5cm} Quenched} \hspace{.5cm}
}
\startdata
non-G3C &  74.6 ($\pm$1.9)\% & 25.4 ($\pm$0.9)\% \\
G3C &  48.5 ($\pm$2.1)\% & 51.5 ($\pm$2.1\%) \\
\tableline
G3C Mass 1 & 57.8 ($\pm$4.0)\% &  42.2 ($\pm$3.2)\% \\
G3C Mass 2 & 46.3 ($\pm$3.4)\% & 53.7 ($\pm$3.8)\% \\
G3C Mass 3 & 41.0 ($\pm$3.3)\% &  59.0 ($\pm$4.2)\%\\
\enddata
\end{deluxetable}

The overall fraction of quenched systems (see Table \ref{tab:t3}) in the G3C sample (51.5$\pm$2\%) is significantly larger than that of the non-G3C sample (25$\pm$1\%); these differences appear to be largely driven by the increase in galaxies with stellar mass $> 10^{10}$\,M$_\odot$, i.e. high-mass systems, in the G3C sample. We note that the fractions of quenched and unquenched galaxies presented in this section should be interpreted relative to each other and not in absolute terms, due to their dependence on choice of SFMS and quenching separator.

In Figure \ref{fig:ms4}a we consider the stellar mass distribution of quenched and unquenched systems for the non-G3C (black) and G3C (orange) samples. The steep drop-off of unquenched galaxies at M$_{\rm stellar}$ $>10^{10.1}$\,M$_\odot$ is mirrored in both samples and can be explained in terms of the relative paucity of high SFR systems at high stellar mass. In the G3C population, the quenched and unquenched populations are approximately equal, with a transition from one to the other at M$_{\rm stellar}$ $\sim10^{10.3}$\,M$_\odot$. The star-forming, or unquenched, population of the non-G3C sample dominates, while the G3C sample shows a larger fraction of quenched systems at stellar masses of M$_{\rm stellar}$ $>10^{10.3}$\,M$_\odot$. 

The quenching of galaxies with M$_{\rm stellar} > 10^{10.5}$\,M$_\odot$ (corresponding to a galaxy halo mass of M$_{\rm stellar} > 10^{11.7}$\,M$_\odot$) is expected from the break in the to stellar mass - halo mass (SMHM) relation \citep[][]{Beh13}, and corresponds to the mass scale relevant to the viral shock heating of accreting gas. The quenched population is clearly dominated by massive systems and this connection to stellar mass can be understood as being driven by a lack of incoming gas to replenish star formation leading to starvation (so-called mass quenching or secular evolution). In Figure \ref{fig:ms4}a, we see that the rising quenched fraction in the G3C sample peaks at $\sim$M$_{\rm stellar}$ $\sim10^{10.5}$, whereas it peaks in a lower mass bin for the non-G3C sample. This would indicate a very rapid response to viral shock-heating, whereas starvation is expected to lead to a gradual decline. Therefore, although the connection to stellar mass is clear, additional mechanisms may be at work. 

To investigate this further, we extract the number of unquenched and quenched systems in each stellar mass bin and calculate the quenched fraction (i.e. number of quenched galaxies relative to the total number of galaxies in that stellar mass bin). Figure \ref{fig:ms4}b shows how the quenched fraction of G3C galaxies compares to non-G3C galaxies in each stellar mass bin. In the highest mass bins, the quenched fractions of both samples converge to 1 due to a lack of galaxies on the SFMS at high mass. However, as can be seen in Figure \ref{fig:ms4}a, the relative dearth of high mass galaxies in the non-G3C sample overall drives up the quenched fraction somewhat artificially. This highlights a limitation of using a comparison to an ungrouped (or ``field") sample due to the underlying differences in their stellar mass functions. However, it is clear that the quenched fraction in the G3C sample is higher in each mass bin for M$_{\rm stellar}$ $\geq10^{9.5}$\,M$_\odot$, reflecting that star formation in the G3C sample is being impacted at both low and high mass. We note that the poor statistics in the two lowest mass bins of our sample means we are insensitive to any differences in this mass range (see section \ref{sec:A1} of the Appendix).

\begin{figure*}[!thp]
\centering
\subfloat[Group Mass 1]{
\includegraphics[width=0.45\textwidth]{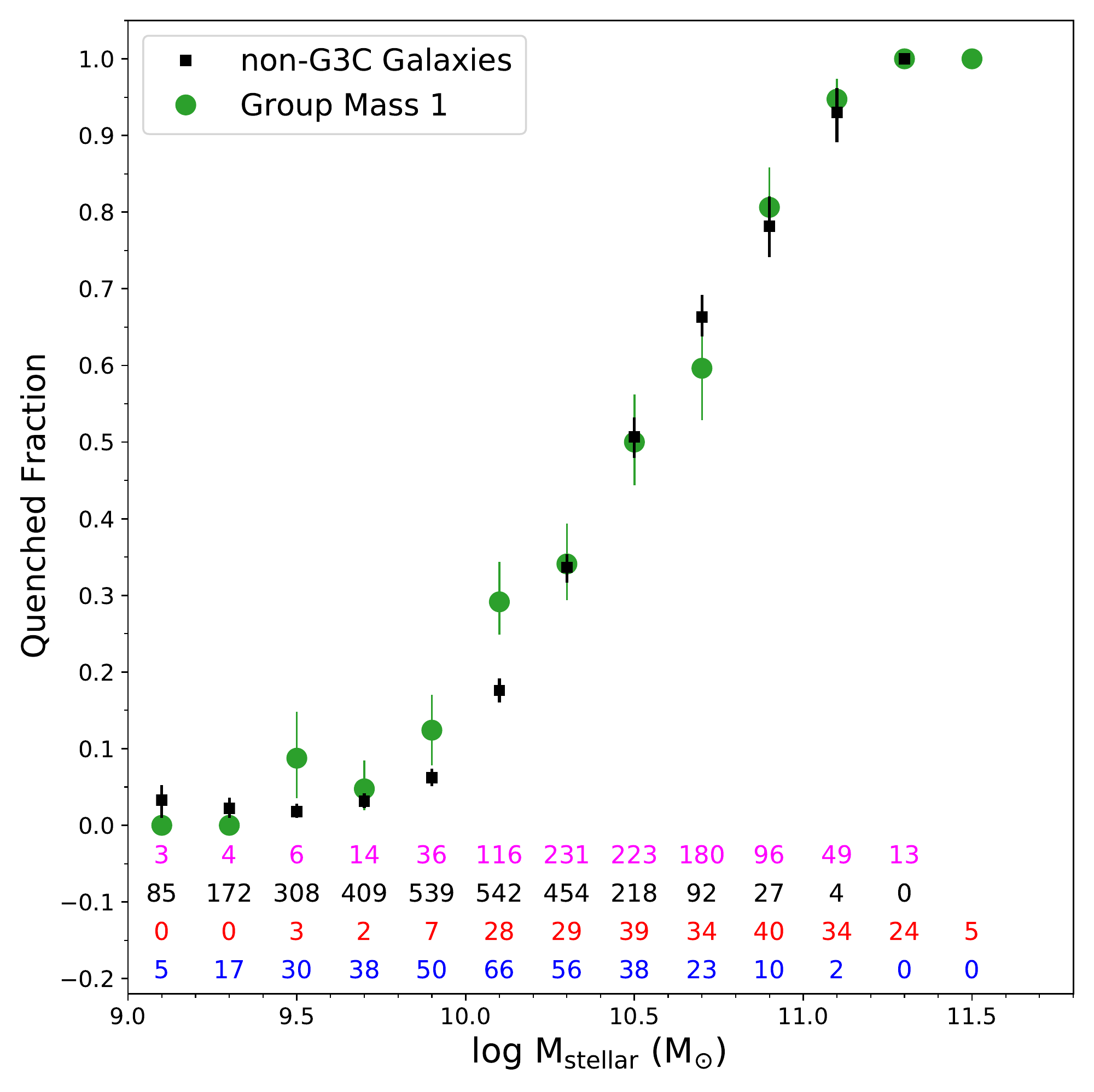}}
\\
\subfloat[Group Mass 2]{
\includegraphics[width=0.45\textwidth]{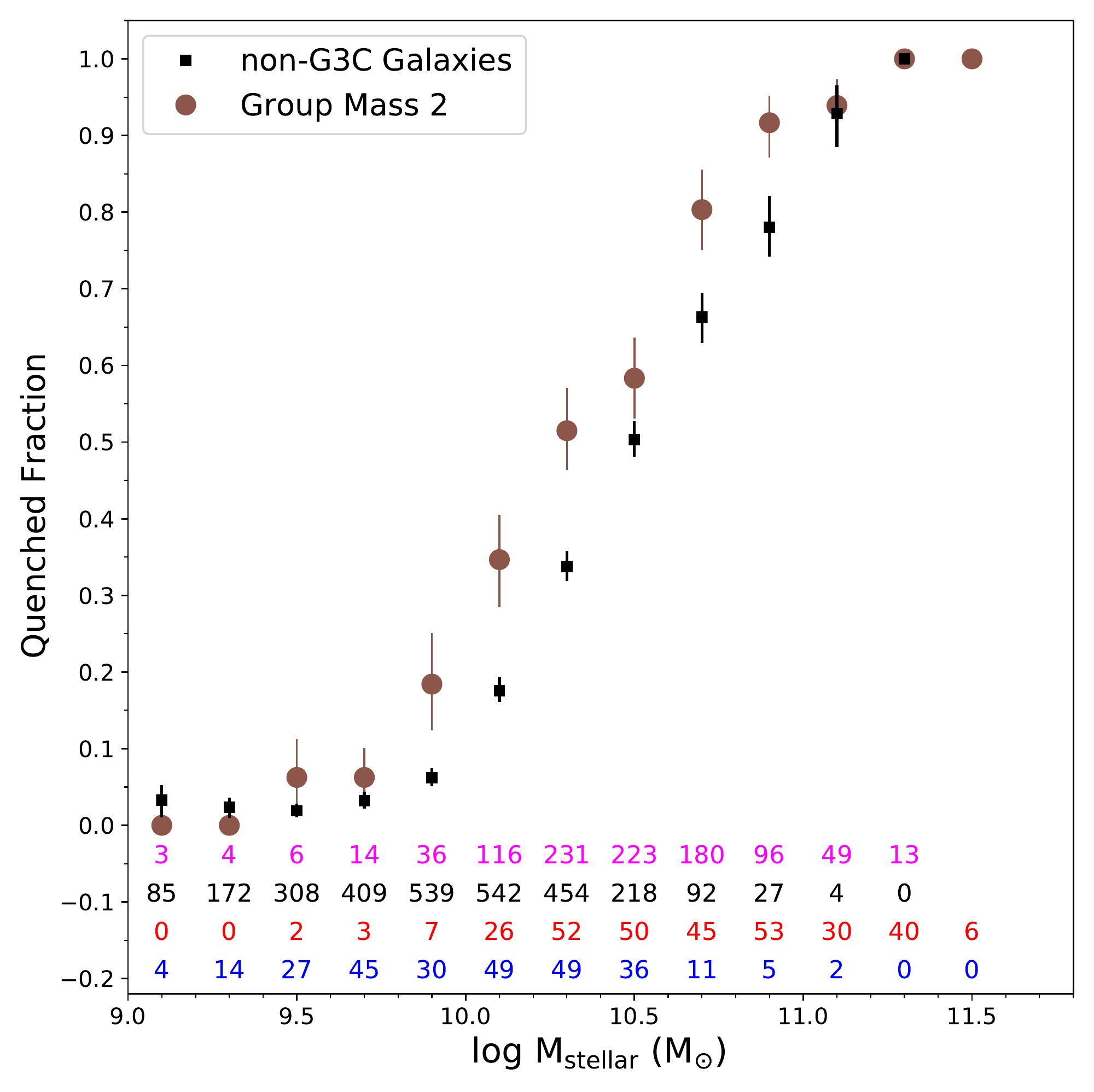}}
\subfloat[Group Mass 3]{
\includegraphics[width=0.45\textwidth]{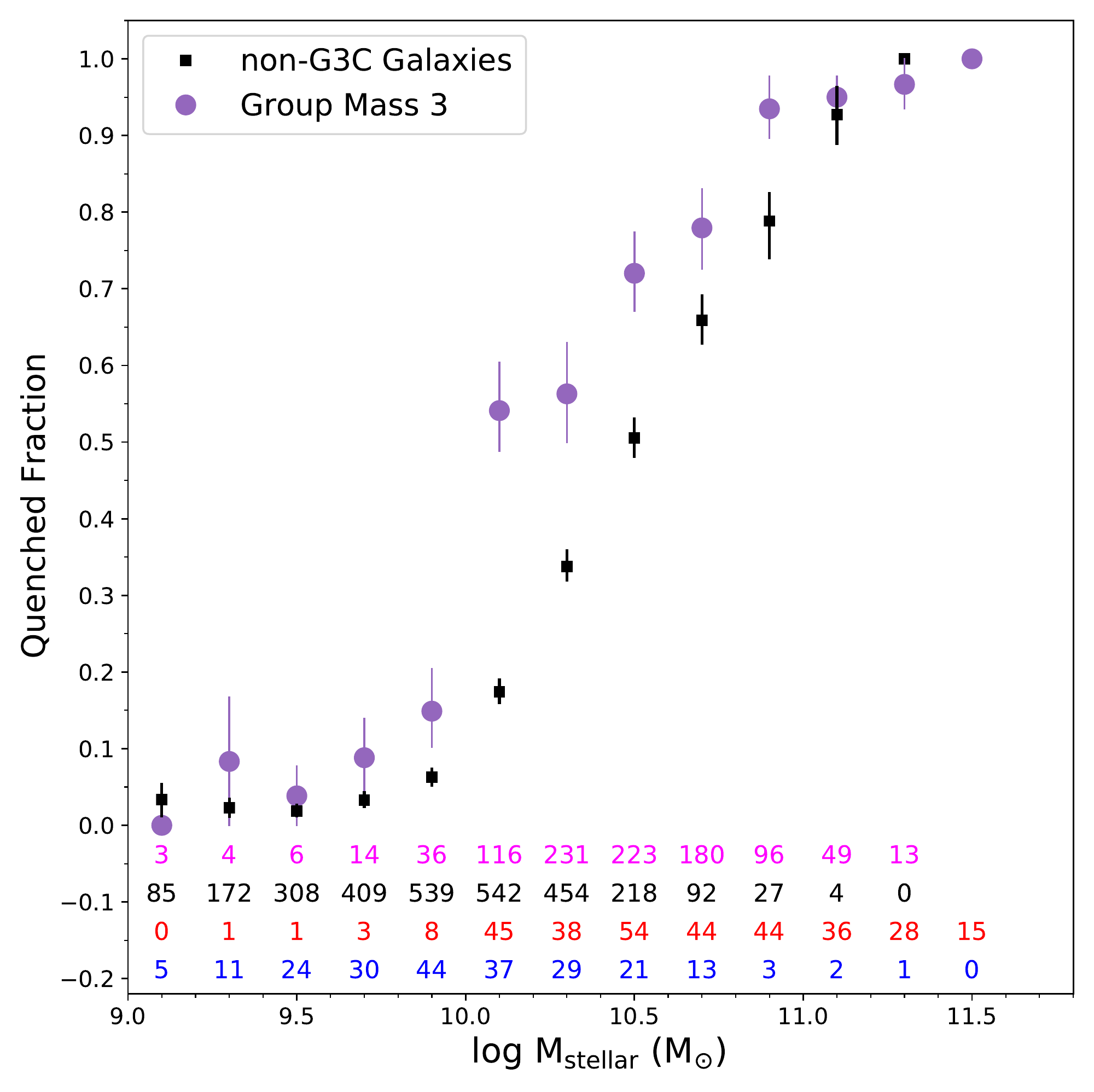}}
\caption{The quenched fraction in each mass bin for the three group mass bins. In each panel, the number of galaxies in each bin is given at the bottom, with the non-G3C quenched galaxies in pink and the non-G3C unquenched in black. For the group mass bins, quenched galaxies are in red and unquenched in blue. As in Figure \ref{fig:ms4}b, errors are calculated using bootstrap resampling in each bin. Notably for Group Mass 2 and 3 (panels (b) and (c), respectively) the quenching mechanisms operating at high stellar mass are more effective i.e. in more massive group halos. \label{fig:ms5}}
\end{figure*}

The quenched fraction at low mass (M$_{\rm stellar} <10^{10.5}$\,M$_\odot$) is expected to reflect mechanisms such as gas stripping, strangulation, harassment etc. acting on the star-forming population i.e. galaxies at low mass are likely more susceptible to environmental quenching in more massive halos \citep[see, for example,][]{Peng10, Peng12, Dav19b, Liu19, Li20}. Our lowest mass bins do not show a clear separation, but this is also where we have the least statistical power (see section \ref{sec:A1}). However, all other mass bins in this range reflect a larger quenched fraction in the G3C sample, indicative of environmental (i.e. external) processes affecting star formation.

Table \ref{tab:t3} provides the corresponding quenched and unquenched populations in each of the different group mass bins; this shows that although the increase of quenched fraction with halo mass is clear, there is a large step from Group Mass 1 to Group Mass 2, and only a relatively small increase to Group Mass 3. This motivates for a larger sample that would allow for increased divisions in halo mass to explore this transition further.

In Figure \ref{fig:ms5} we provide the quenched fractions in each stellar mass bin for the individual group mass samples. Considering the range M$_{\rm stellar} <10^{10.5}$\,M$_\odot$, an elevated quenched fraction is seen at all group masses for M$_{\rm stellar} \gtrsim10^{9.5}$\,M$_\odot$. We note that as we approach M$_{\rm stellar}$ $\sim10^{10.5}$, the quenched fractions appear to increase with Group Mass, which is broadly consistent with \citet{Dav19b}.
Above M$_{\rm stellar} \sim10^{10.5}$\,M$_\odot$, Group Mass 1 shows little difference compared to the non-G3C sample, consistent with its relatively smaller proportion of high mass systems. Both Group Mass 2 and Group Mass 3 show a large quenched fraction, relative to the ungrouped and Group Mass 1 sample, between M$_{\rm stellar} \sim10^{10.5}$\,M$_\odot$ and M$_{\rm stellar} \sim10^{11}$\,M$_\odot$, but with no clear halo mass dependence. 

\begin{figure*}[!thbp]
\gridline{\fig{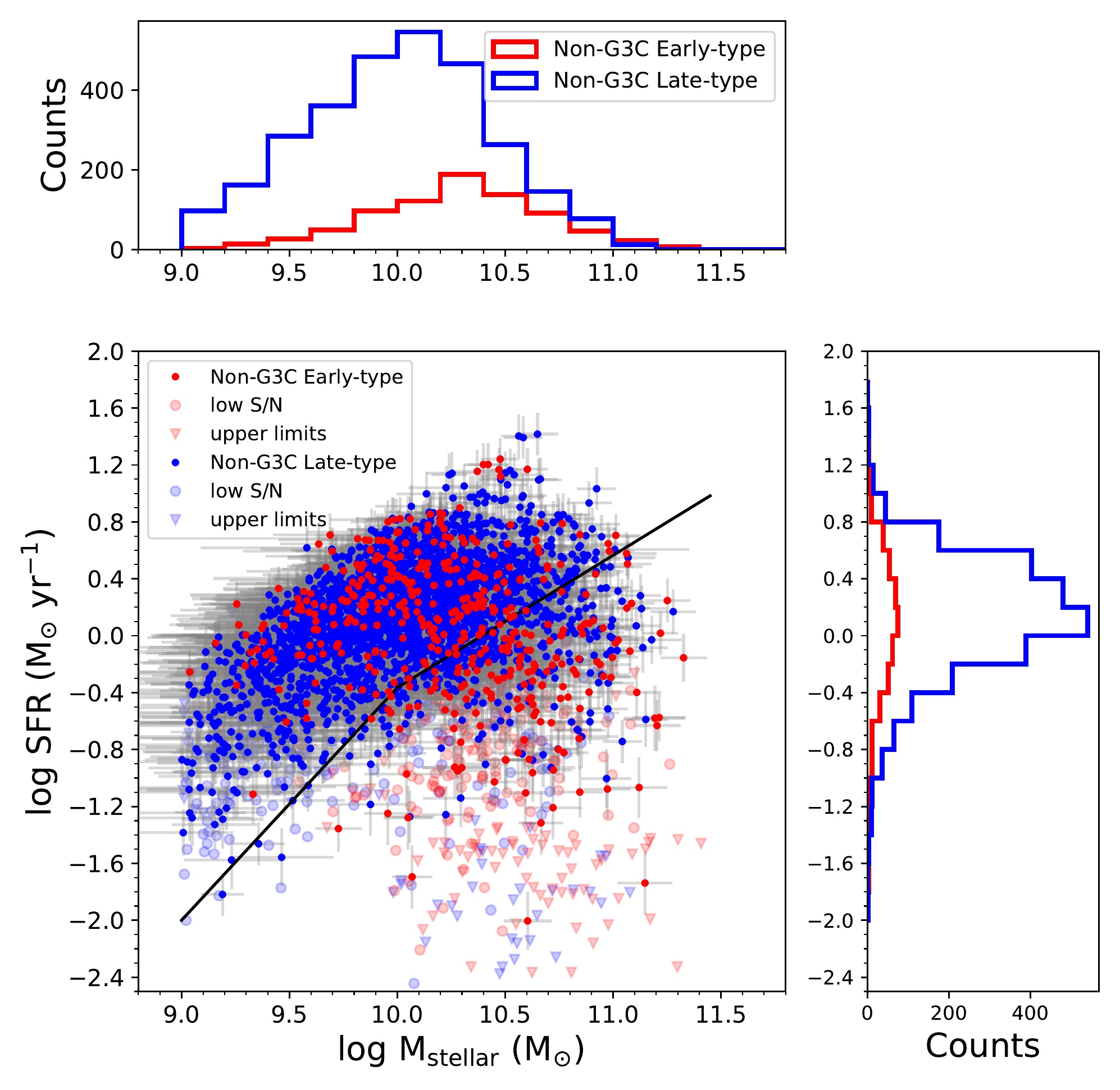}{0.5\textwidth}{(a) non-G3C sample}
          \fig{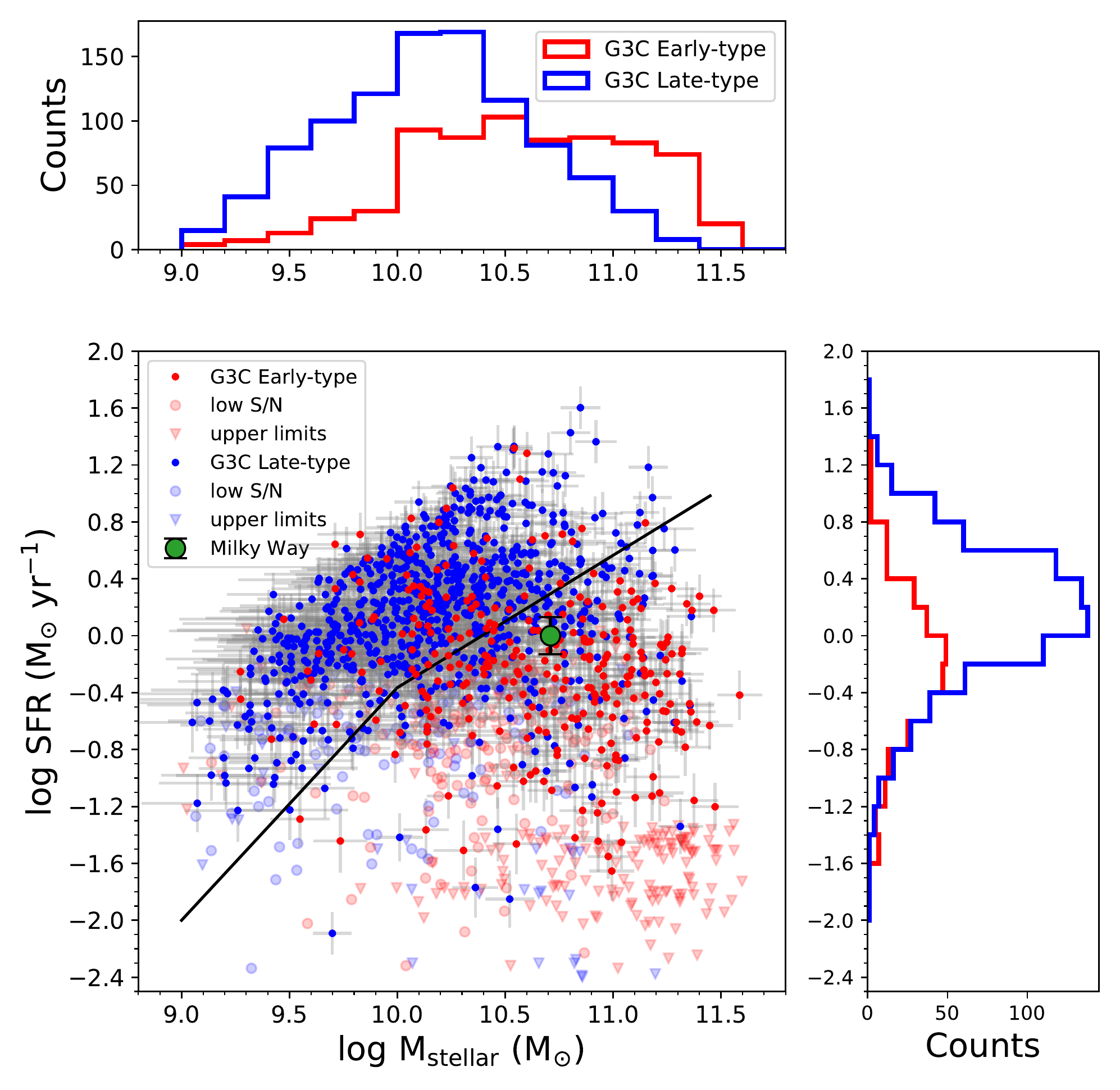}{0.5\textwidth}{(b) G3C sample}}
\caption{The log\, SFR--log\,M$_{\rm stellar}$ distribution for the (a) non-G3C and (b) G3C sample, now color-coded by morphological classification (see also Figure \ref{fig:msA2} in the Appendix for comparison). The G3C sample sees a shift to higher stellar mass driven by an increase in early-types. In panel (b) we include the location of the Milky Way \citep[from][]{Mutch11}. \label{fig:ms6}}
\end{figure*}

\subsubsection{Quenching and Morphology}

We next fold our visual morphology classifications into the SFMS analysis by color-coding the non-G3C and G3C galaxies as early- or late-type (Figure \ref{fig:ms6}). For the non-G3C sample we see the dominance of late-types noted previously, but also that the SFR and stellar mass distributions of the early-types show the greatest difference between the non-G3C and G3C samples. For the G3C sample we observe a shift of the early-type distribution to high stellar mass (also seen in Figure \ref{fig:f4}b). This is accompanied by the (expected) shift to lower SFRs of the early-type distribution, which can be seen in the right side panel of Figure \ref{fig:ms6}b. Considering the late-type stellar mass distributions (blue histograms in the upper panels of Figure \ref{fig:ms6}), we observe a shift to higher masses in the G3C sample (also seen in Figure \ref{fig:f4}). 
To confirm the validity of the observed distribution of late- and early-types using our adopted classification, we use the detailed bulge-to-disk decomposition of Casura et al. (in prep), available for a sub-sample of our galaxies, in Section \ref{ap:ms} of the Appendix (Figure \ref{fig:msA2}).

\begin{figure*}[!thbp]
\gridline{\fig{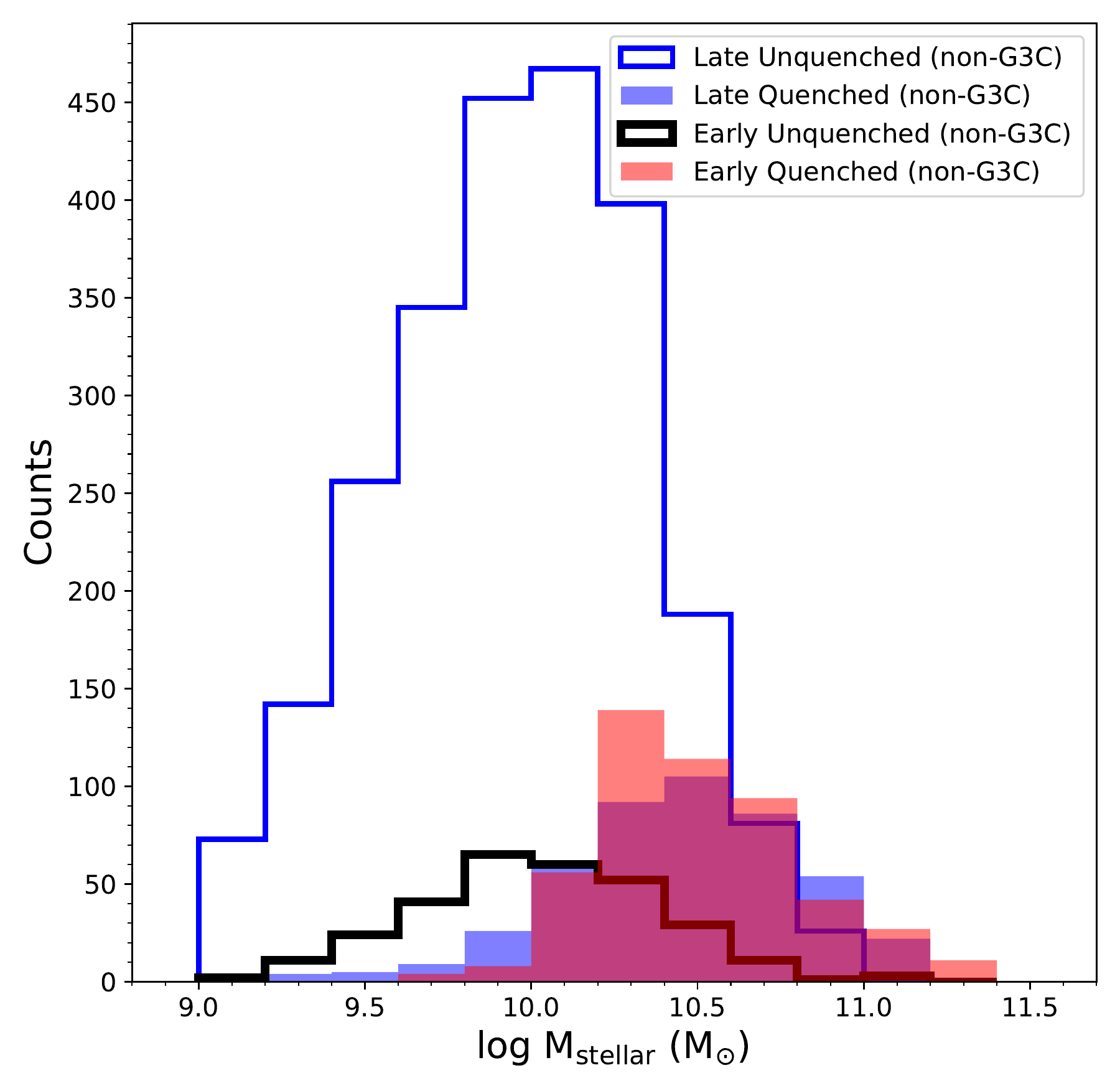}{0.45\textwidth}{(a) non-G3C sample}
          \fig{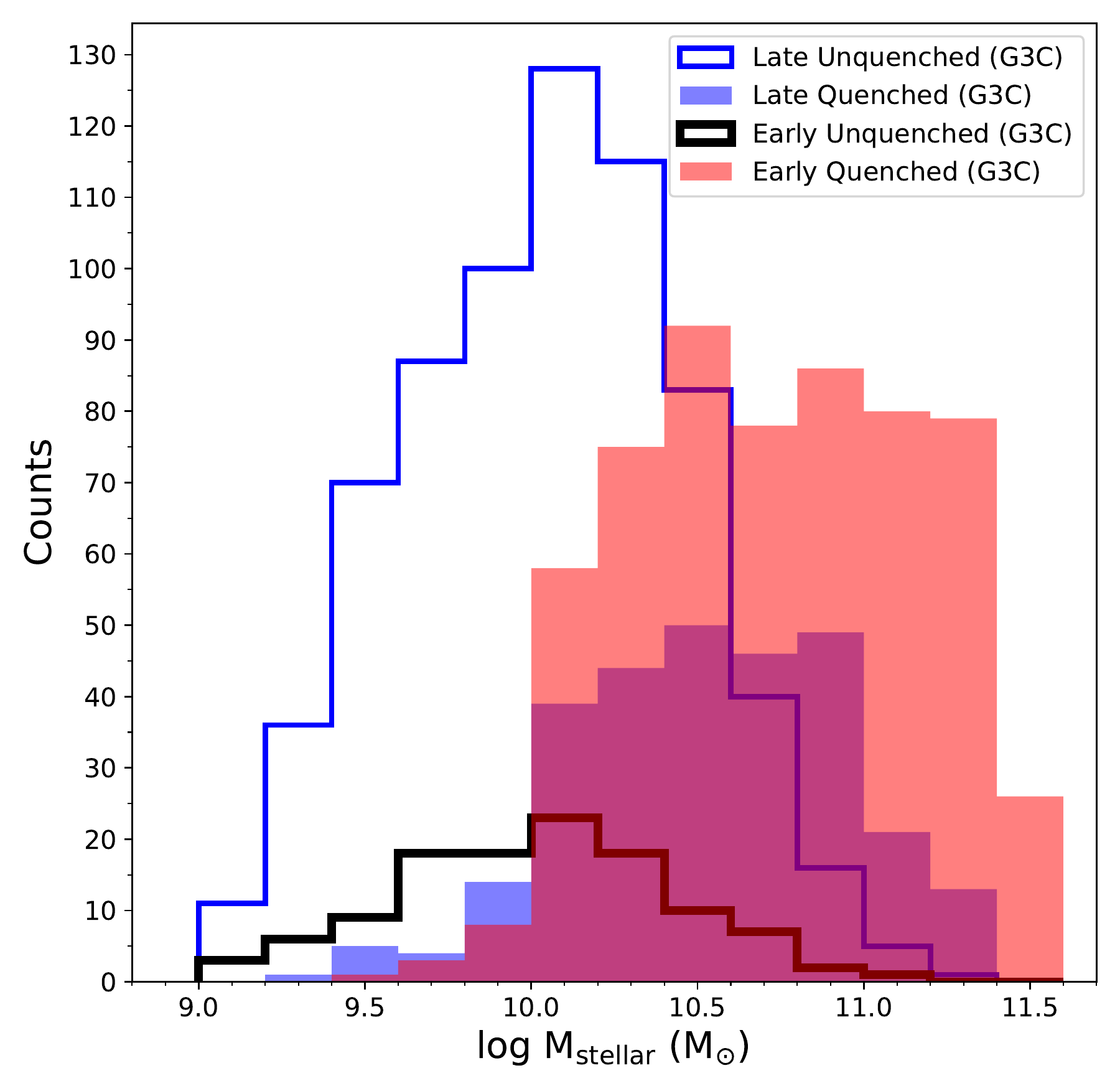}{0.45\textwidth}{(b) G3C sample}}
\caption{The stellar mass distributions of the unquenched (unshaded) and quenched (shaded) systems, sub-divided into early- and late-type. The non-G3C sample (a) is dominated by unquenched late-type systems, whereas the G3C sample (b) shows a large fraction of quenched early-type, and also a large fraction of quenched late-type systems. \label{fig:ms7}}
\end{figure*}

In Figure \ref{fig:ms6}b, we include the position of the Milky Way in log\,SFR--log\,M$_{\rm stellar}$ space, from \citet{Mutch11}, who found it to lie within the ``green valley". In our classification, the Milky Way is a late-type, quenched system i.e. below the SFMS. As noted earlier, we expect this class to include systems in the process of quenching and the Milky Way is a good example of a galaxy with low, but not extinguished star formation. \citet{Mutch11} suggest that both the Milky Way and M31 represent a population of galaxies
in the midst of a transitional process and that the limited availability of cold gas is the main cause for the observed decline of star formation. They posit that both galaxies will quiescently evolve onto the red sequence before they merge in $\sim$5\,Gyr producing a remnant elliptical galaxy, suggesting an alternative pathway to the paradigm where gas-rich mergers of spirals evolve to form the red sequence.

Using the separation into quenched and unquenched systems discussed earlier, we divide the early-type and late-type systems of the non-G3C and G3C sample, accordingly; their stellar mass distributions are shown in Figure \ref{fig:ms7} with their fractions listed in Table \ref{tab:t4}. Here the role of the group environment in mass assembly can clearly be discerned, where groups host higher masses and more bulge-dominated galaxies.

\begin{deluxetable*}{lcccc}[!htb]
\tablecaption{Unquenched and Quenched, Early- and Late-types in the non-G3C and G3C Samples \label{tab:t4}}
\tablecolumns{5}
\tablenum{4}
\tablewidth{0pt}
\tablehead{
\colhead{Sample} &
\colhead{\hspace{.5cm}Unquenched\hspace{.5cm}} &
\colhead{\hspace{.5cm}Quenched\hspace{.5cm}} &
\colhead{\hspace{.5cm}Unquenched\hspace{.5cm}} &
\colhead{\hspace{.5cm}Quenched\hspace{.5cm}}\\
\colhead{} &
\colhead{\hspace{.5cm}Late-type\hspace{.5cm}} &
\colhead{\hspace{.5cm}Late-type\hspace{.5cm}} &
\colhead{\hspace{.5cm}Early-type\hspace{.5cm}} &
\colhead{\hspace{.5cm}Early-type\hspace{.5cm}}
}
\startdata
non-G3C &   65.8($\pm$1.7)\%  & 12.6($\pm$0.6)\%    &  8.1($\pm$0.5)\%  & 13.4($\pm$0.6)\% \\
G3C   &   41.2($\pm$1.9)\%  & 17.0($\pm$1.1)\%    &   6.8($\pm$0.7)\%   &  35.0($\pm$1.7)\%   \\
\tableline
G3C Mass 1 &  50.0($\pm$3.6)\%  &14.5($\pm$1.7)\%   &  7.4($\pm$1.2)\%  &  28.1($\pm$2.5)\% \\
G3C Mass 2 &  38.1($\pm$3.0)\%  & 18.2($\pm$1.9)\%   & 7.5($\pm$1.2)\%   & 36.2($\pm$2.9)\%  \\
G3C Mass 3 &  34.9($\pm$3.0)\%  & 18.4($\pm$2.0)\%   &  5.6($\pm$1.1)\%   & 41.1($\pm$3.3)\%  \\
\enddata
\end{deluxetable*}

For the non-G3C sample, the late-type, unquenched systems clearly dominate. In comparison, the G3C sample (Figure \ref{fig:ms7}b), shows a large proportion of quenched early-types (chiefly at M$_{\rm stellar}> 10^{10}$\,M$_\odot$), but the stellar mass distribution of late, unquenched galaxies appears quite similar to the non-G3C sample with the only difference being the relative proportion (41\% vs. 66\%). As with the non-G3C sample, the fraction of unquenched, early-type systems in the G3C sample is the smallest contributor and comparable to that of the non-G3C distribution. The increase of high-mass, late-type systems in the G3C sample seen in Figures \ref{fig:f4} and \ref{fig:ms6} appears to drive a corresponding increase in the quenched late-type population. 

In the context of this study, therefore, both late-type and early-type systems 
are found below the star-forming main-sequence, but the latter to a greater degree than the former. This is consistent with the fact that at high stellar mass (M$_{\rm stellar}> 10^{10.5}$\,M$_\odot$), where systems turn off the star-forming sequence, early-types are more common than late-types \citep[see also][]{Kel14a, Moff16}.

Bearing in mind that early-types were excluded from the SFMS fit through morphology and color, the fraction of early-types that are unquenched (i.e. within the star-forming sequence) only make up 6.8\% of the total (see Table \ref{tab:t4}. This is very slightly lower than what is found for the non-G3C sample (8.1\%). There is therefore no indication that bulge-dominated galaxies are more prevalent on the SFMS in grouped galaxies, despite their increased numbers at high mass.

\begin{figure*}[!thp]
\centering
\subfloat[Group Mass 1 ($<10^{12.95}$\,M$_\odot$/h)]{
\includegraphics[width=0.45\textwidth]{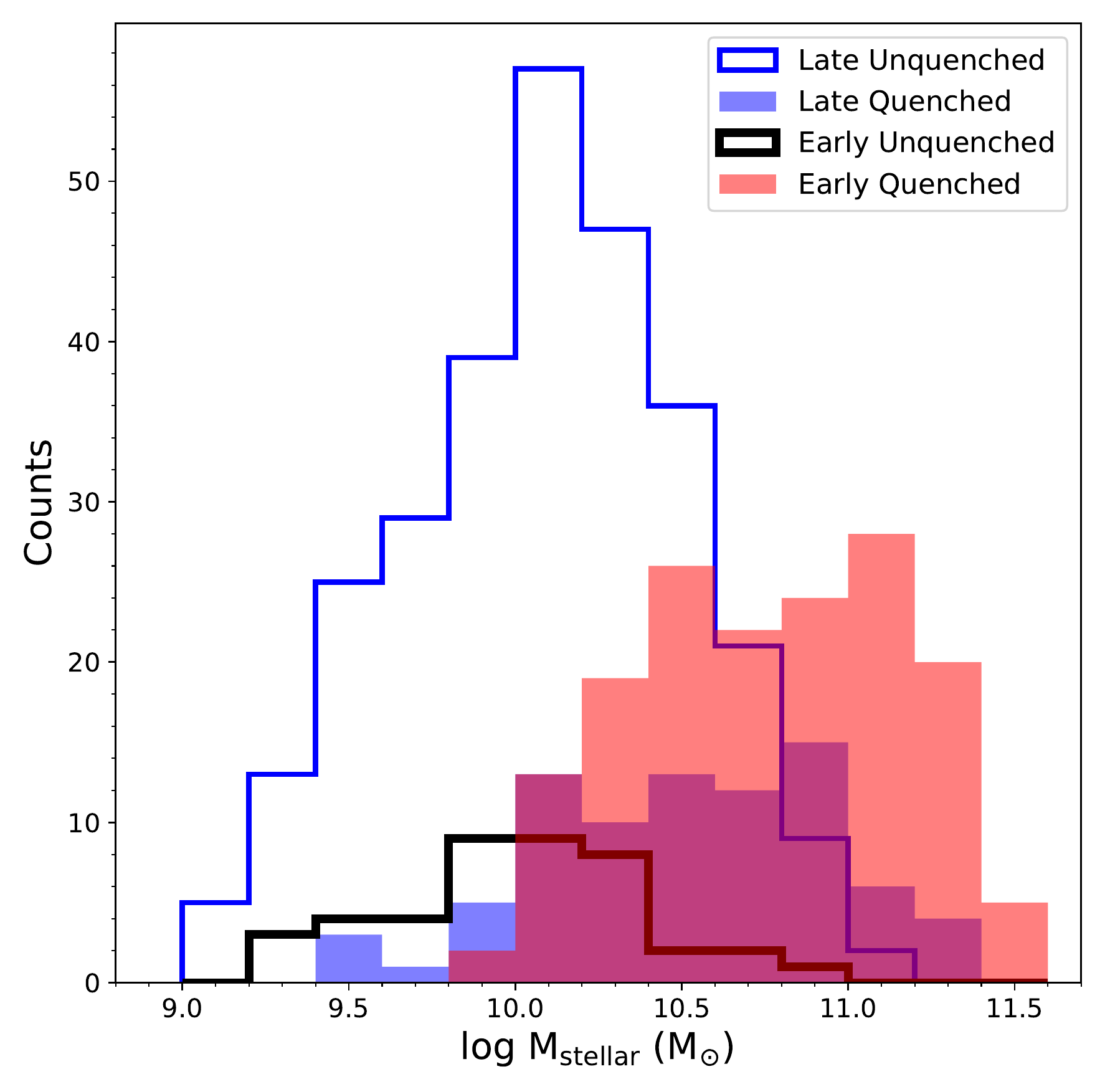}}
\\
\subfloat[Group Mass 2 ($10^{12.95}$ -- $10^{13.4}$\,M$_\odot$/h)]{
\includegraphics[width=0.45\textwidth]{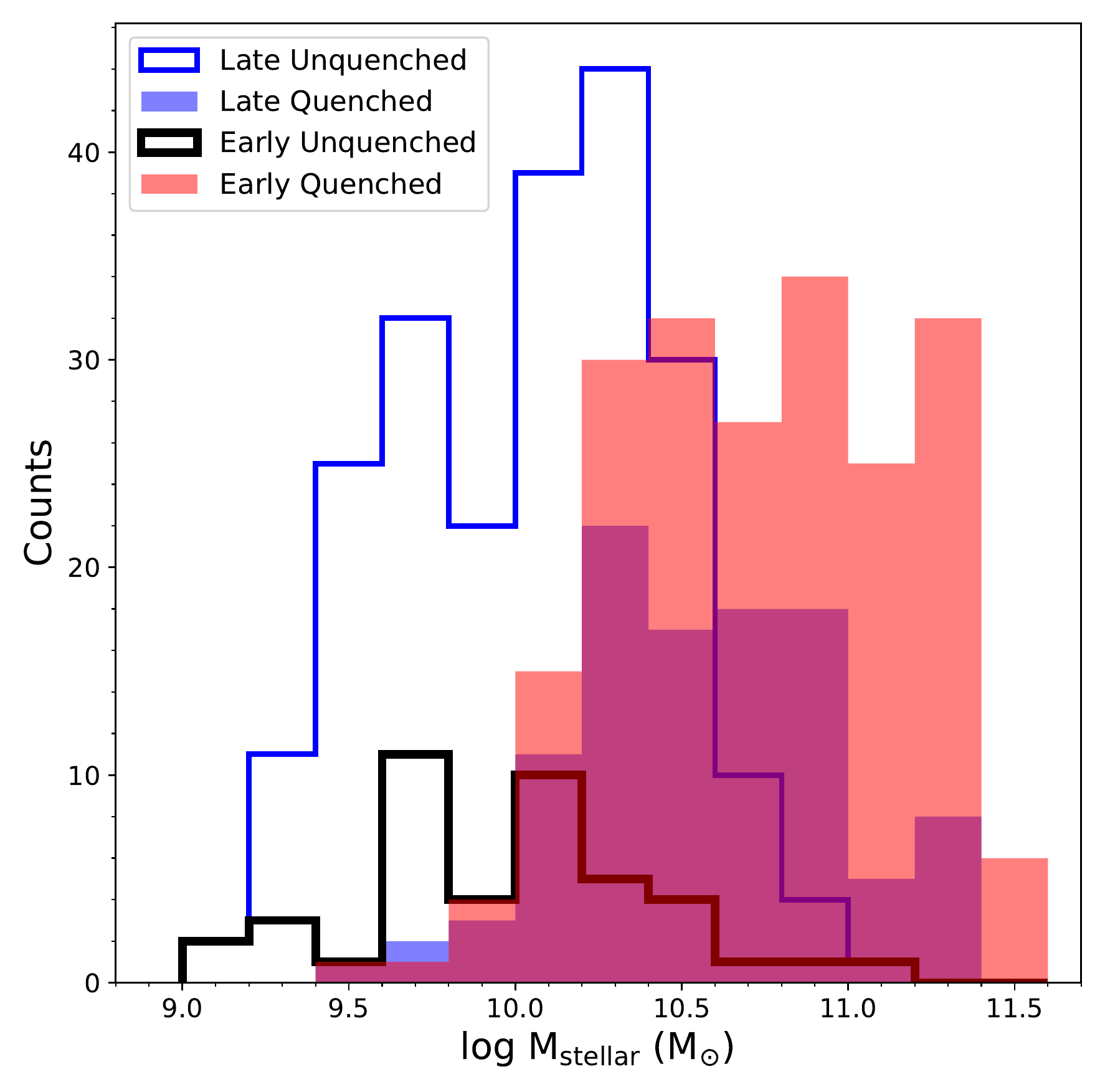}}
\subfloat[Group Mass 3 ($>10^{13.4}$\,M$_\odot$/h)]{
\includegraphics[width=0.45\textwidth]{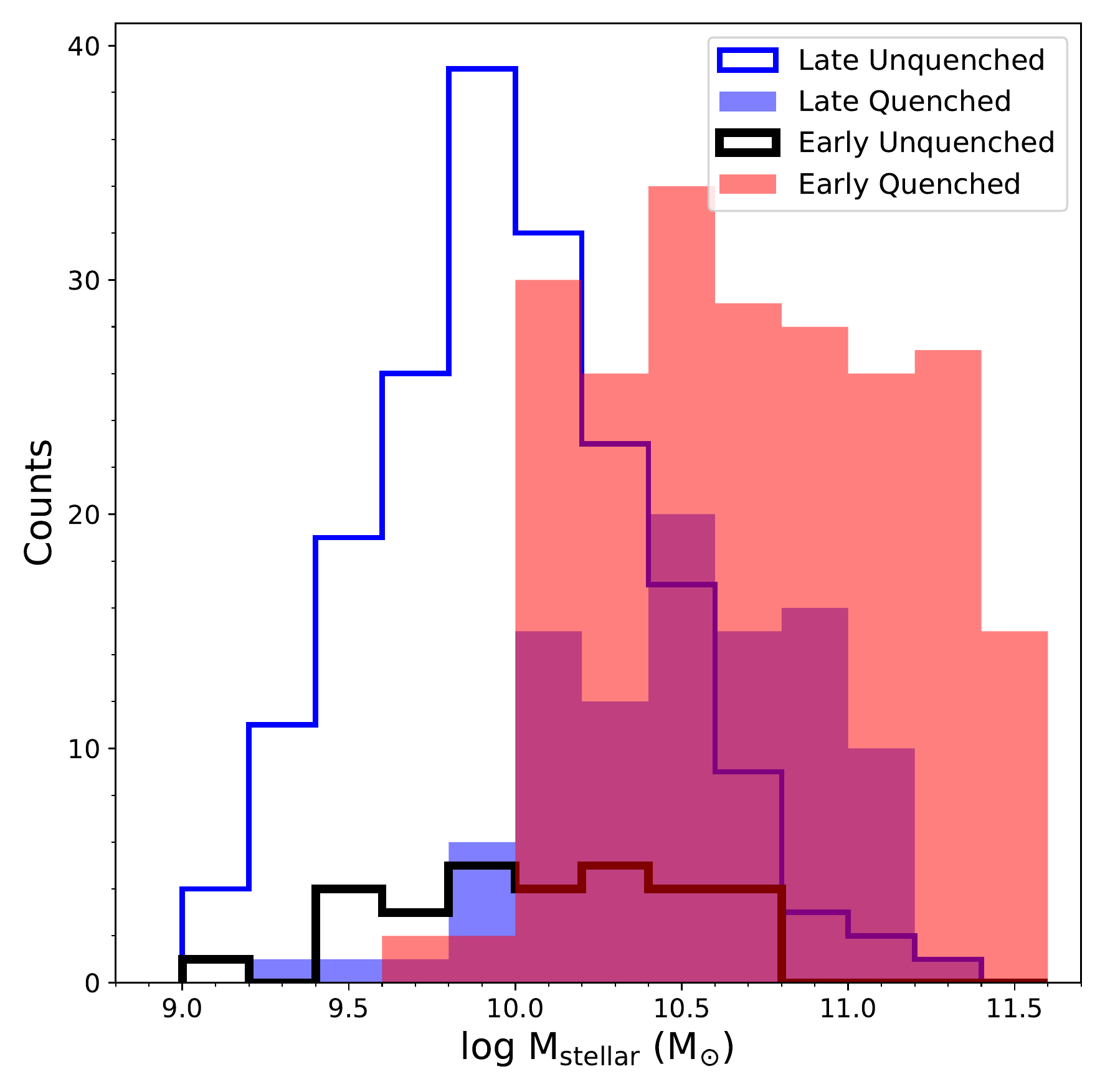}}
\caption{The stellar mass distributions of the unquenched (unshaded) and quenched (shaded) systems, sub-divided into early- and late-type, for the three group mass bins. The increasing quenched fraction, of both early- and late-type systems, with increasing group mass is evident, although with a less dramatic shift from Group Mass 2 to Group Mass 3. An increase in quenched late-types is observed, particularly from Group Mass 1 to Group Mass 2. \label{fig:ms8} }
\end{figure*}

\begin{figure*}[!thp]
\centering
\subfloat[non-G3C and G3C]{
\includegraphics[width=0.45\textwidth]{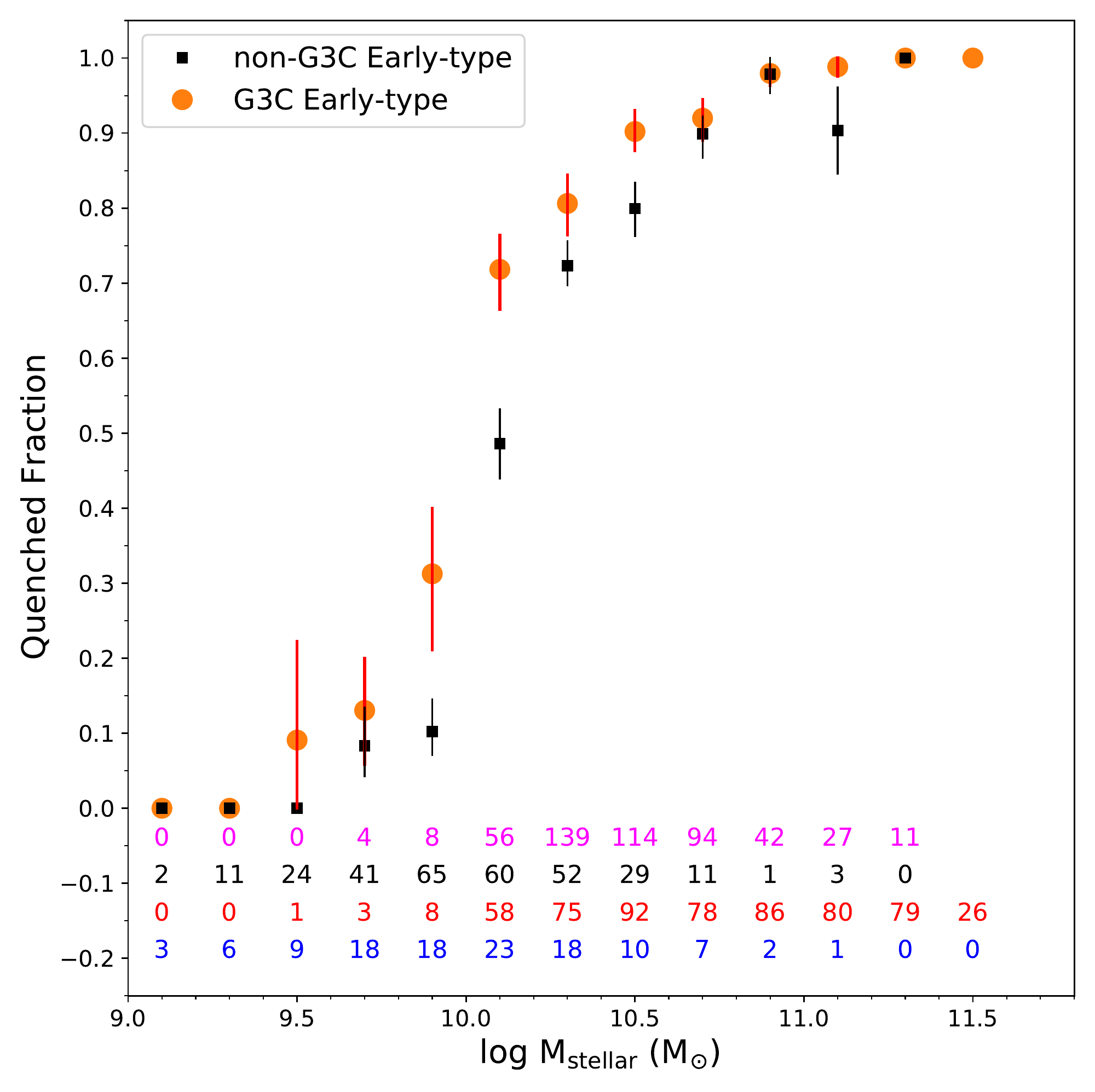}}
\subfloat[Group Mass 1 ($<10^{12.95}$\,M$_\odot$/h)]{
\includegraphics[width=0.45\textwidth]{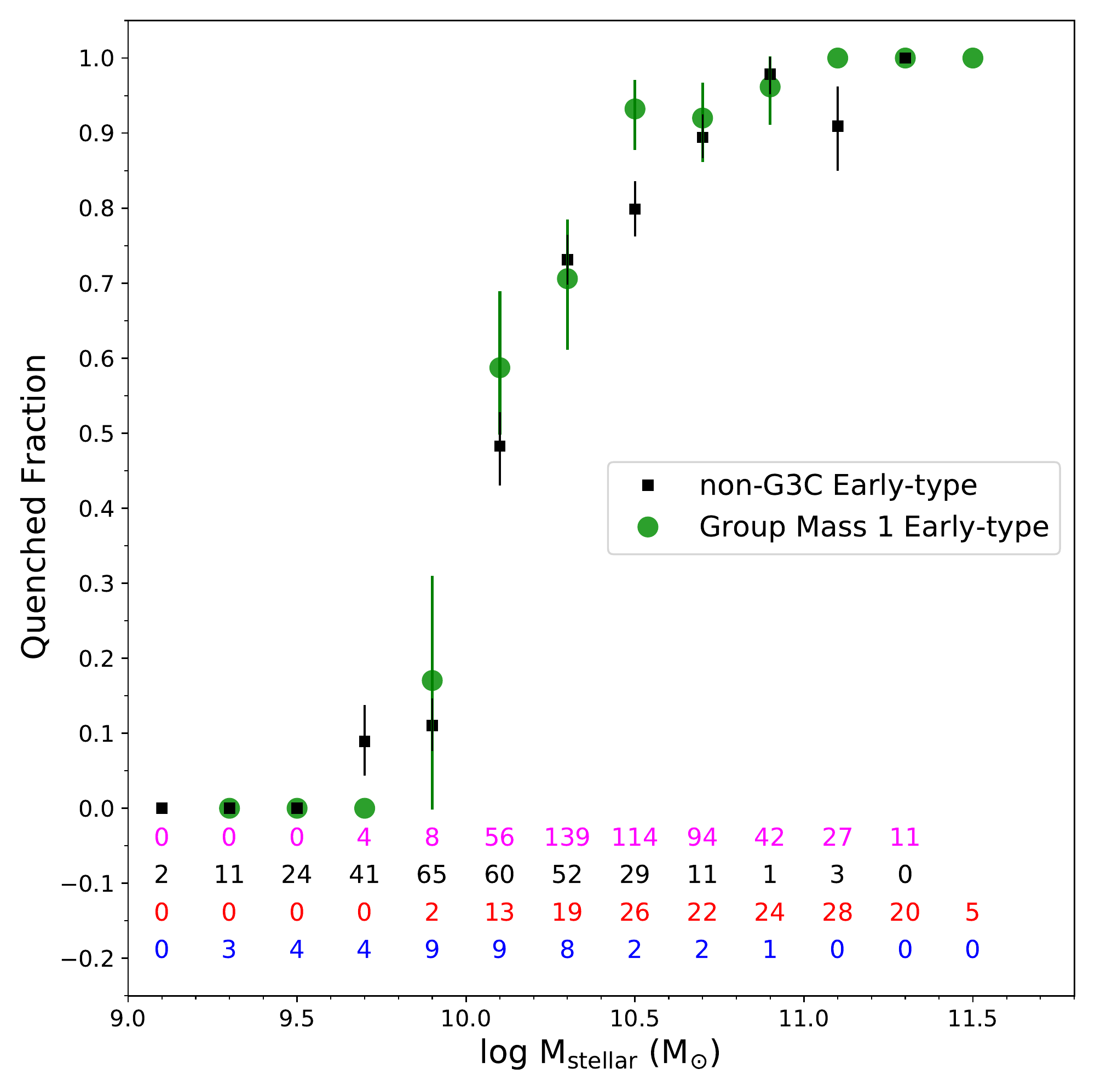}}
\\
\subfloat[Group Mass 2 ($10^{12.95}$ -- $10^{13.4}$\,M$_\odot$/h)]{
\includegraphics[width=0.45\textwidth]{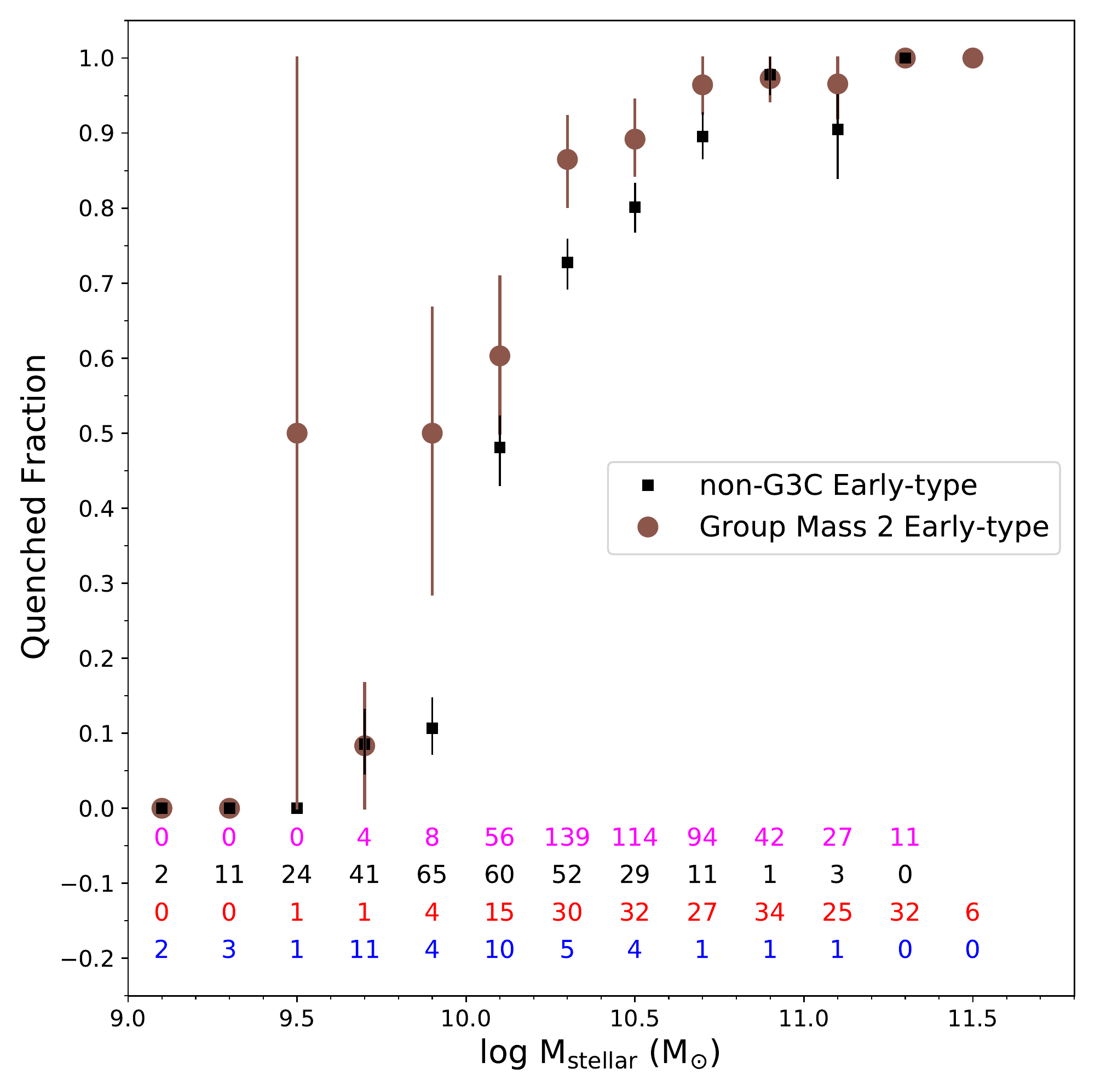}}
\subfloat[Group Mass 3 ($>10^{13.4}$\,M$_\odot$/h)]{
\includegraphics[width=0.45\textwidth]{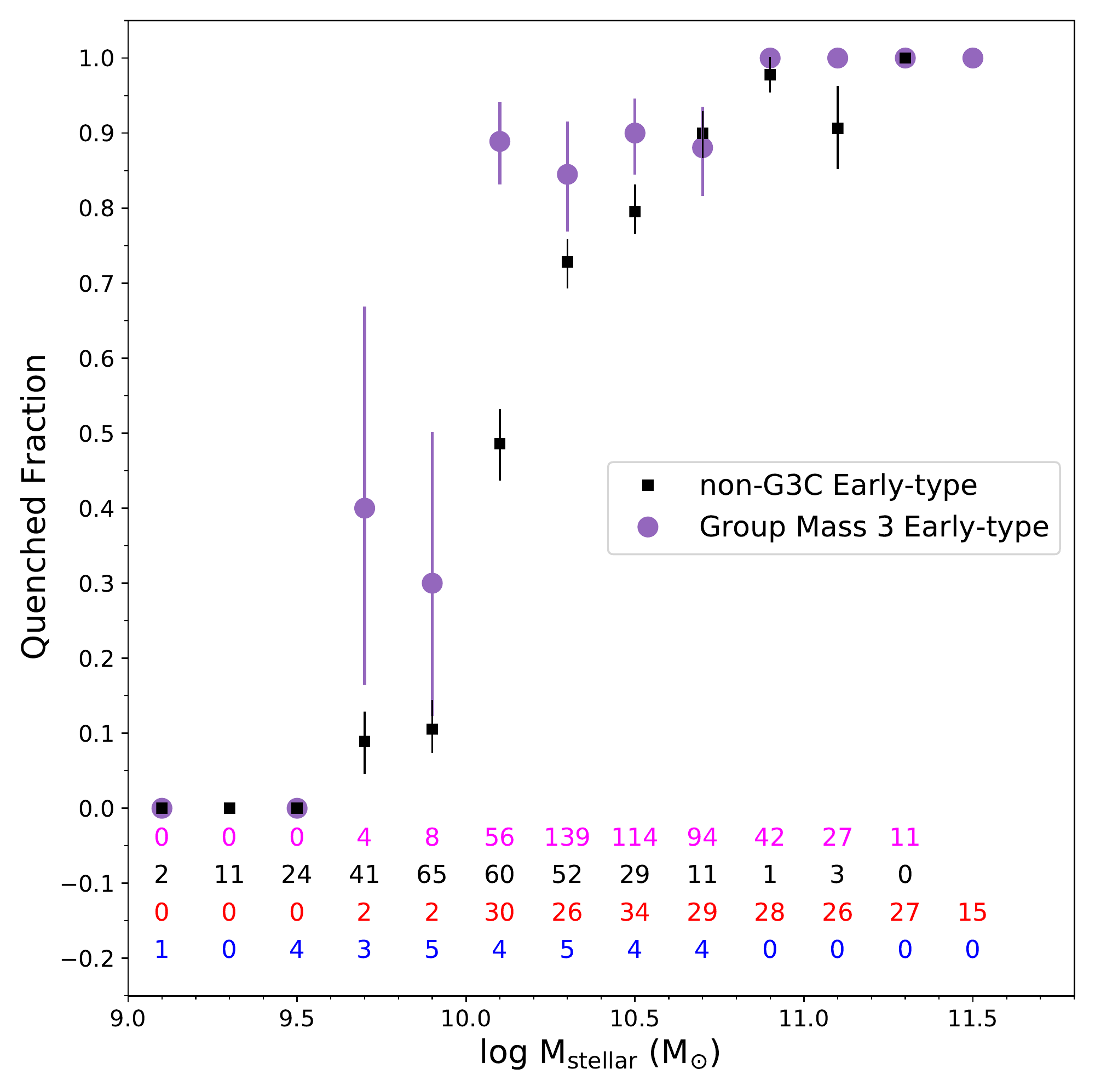}}
\caption{The quenched fraction of early-types per stellar mass bin, comparing the non-G3C sample to a) the G3C sample, b) Group Mass 1, c) Group Mass 2, and d) Group Mass 3 samples, respectively. In each panel, the number of galaxies in each bin is given at the bottom, with the non-G3C quenched galaxies in pink and the non-G3C unquenched in black. For the G3C samples, quenched galaxies are in red and unquenched in blue. As in Figure \ref{fig:ms5}, errors are calculated using bootstrap resampling in each bin. \label{fig:ms9} }
\end{figure*}

Using the quenching separator as before (Figure \ref{fig:ms7}), we provide the breakdown of quenched and unquenched systems, separated by morphology in Figure \ref{fig:ms8}, for the different group mass bins. The relative fractions of unquenched and quenched, late- and early-types (also for the non-G3C and parent G3C samples) is provided in Table \ref{tab:t4}. We observe that the fraction of quenched early-types increases with increasing halo mass, from 28($\pm$3)\%\% to 36($\pm3$)\% to 41($\pm3$)\%. We see a corresponding decrease in unquenched late-types with increasing halo mass. We note that the fraction of quenched early-types relative to the overall quenched population does not appear significantly different within the errors (66.5\% in the lowest halo mass bin increasing to 69\% in the highest halo mass bin), broadly consistent with the results from \citet{Liu19}. 

\begin{figure*}[!thp]
\centering
\subfloat[non-G3C and G3C]{
\includegraphics[width=0.45\textwidth]{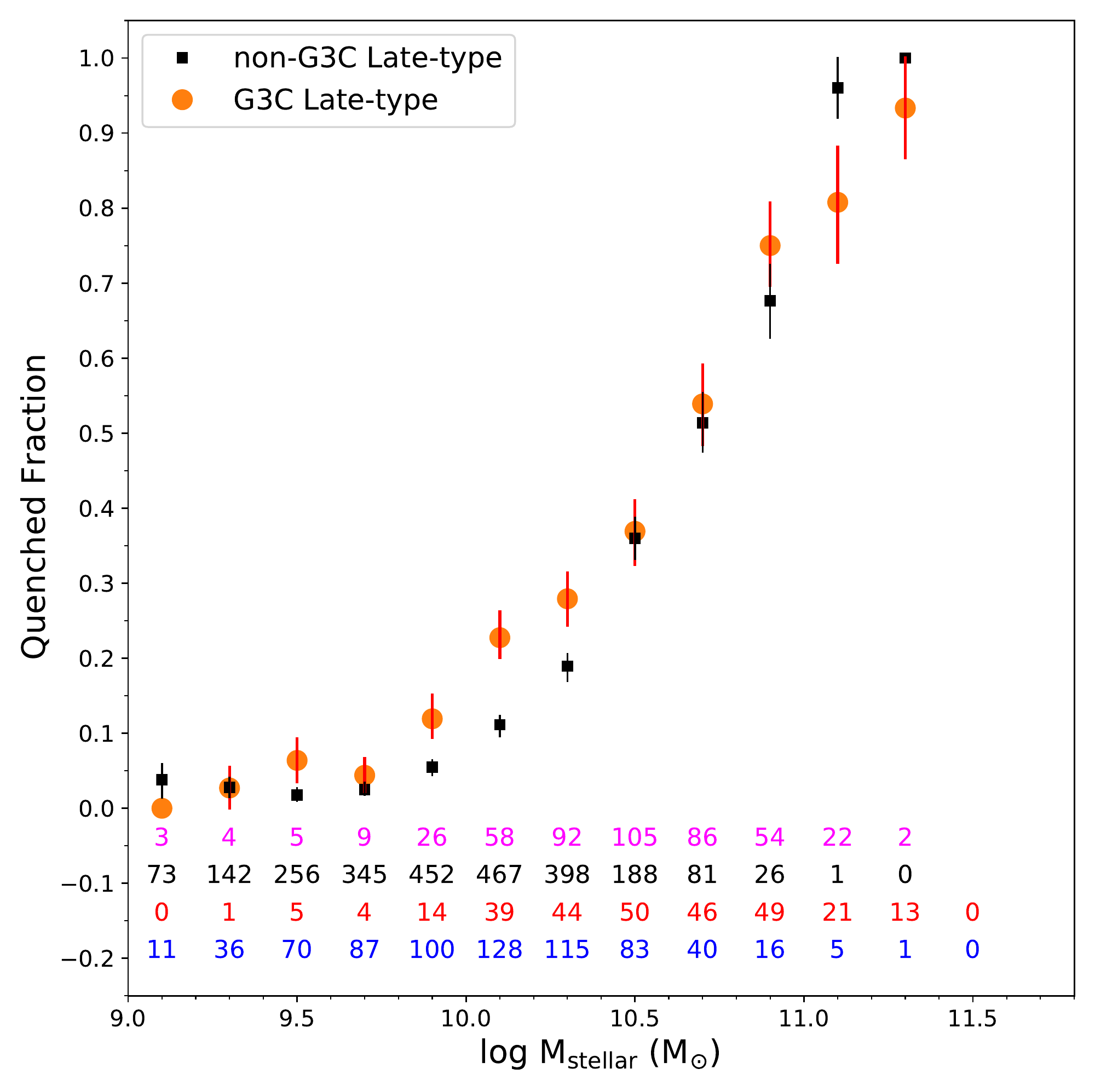}}
\subfloat[Group Mass 1 ($<10^{12.95}$\,M$_\odot$/h)]{
\includegraphics[width=0.45\textwidth]{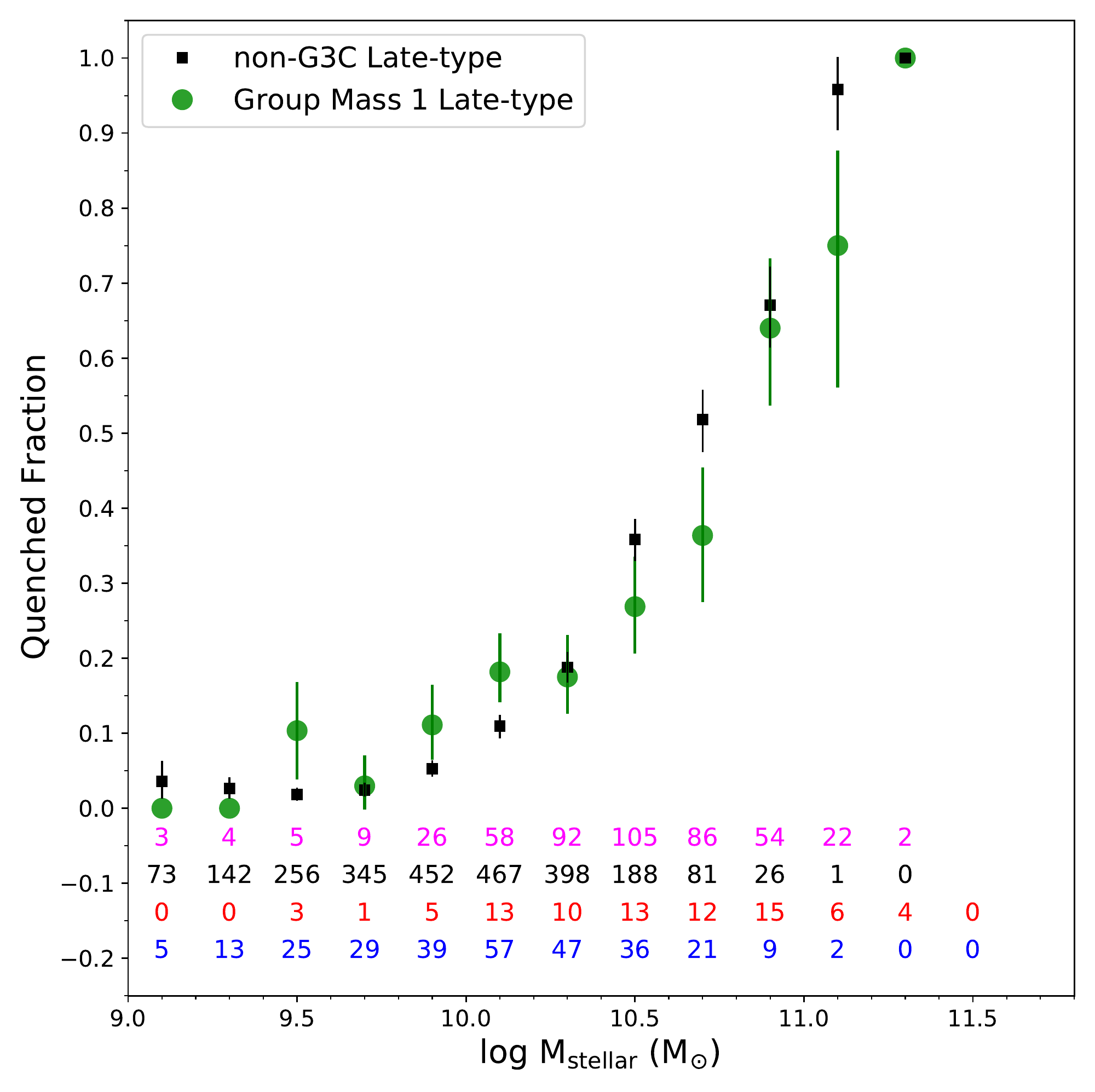}}
\\
\subfloat[Group Mass 2 ($10^{12.95}$ -- $10^{13.4}$\,M$_\odot$/h)]{
\includegraphics[width=0.45\textwidth]{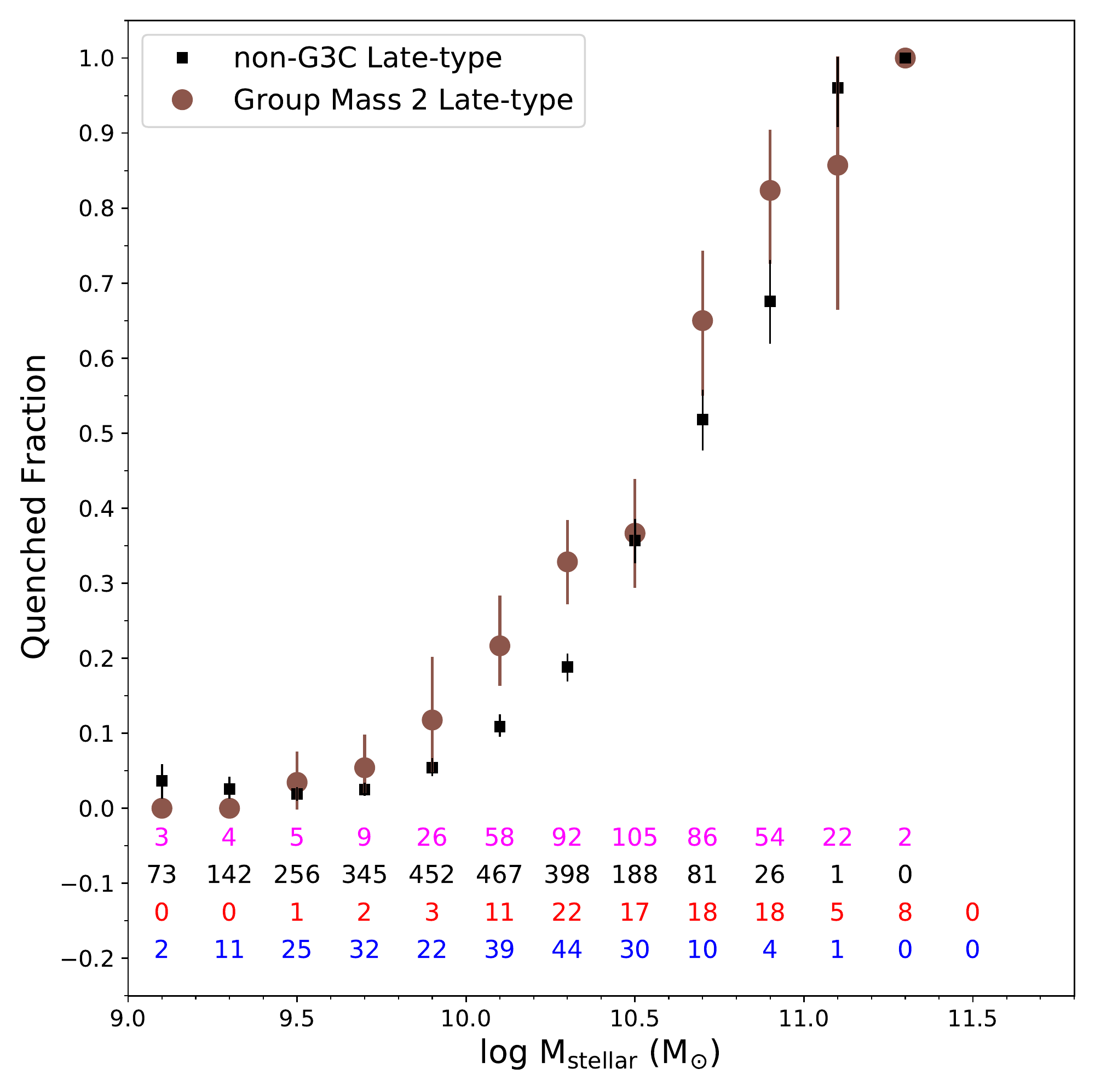}}
\subfloat[Group Mass 3 ($>10^{13.4}$\,M$_\odot$/h)]{
\includegraphics[width=0.45\textwidth]{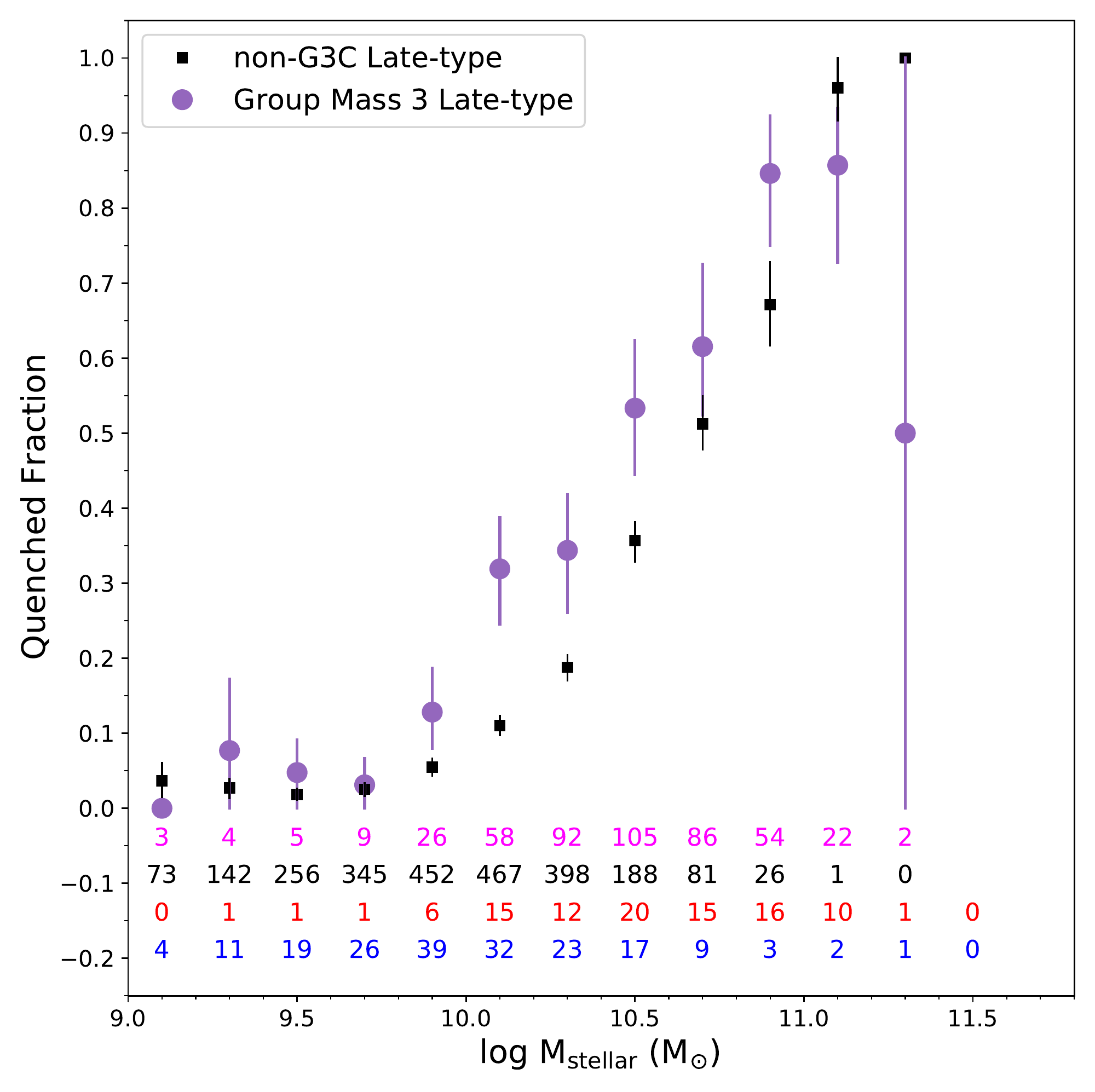}}
\caption{The quenched fraction of late-types per stellar mass bin, comparing the non-G3C sample to a) the G3C sample, b) Group Mass 1, c) Group Mass 2, and d) Group Mass 3 samples, respectively. In each panel, the number of galaxies in each bin is given at the bottom, with the non-G3C quenched galaxies in pink and the non-G3C unquenched in black. For the G3C samples, quenched galaxies are in red and unquenched in blue. As in Figure \ref{fig:ms5}, errors are calculated using bootstrap resampling in each bin. \label{fig:ms10} }
\end{figure*}

Turning now to the unquenched early-types, i.e. early-types on the SFMS, we do not observe an increase of this population with halo mass. In fact, this population appears (tentatively) to decrease with increasing halo mass. The increasing fraction of early-types with increasing halo mass we observe, are therefore almost exclusively below the star-forming sequence, in agreement with previous works \citep[e.g.][]{Bluck14, Cook20}.

The fraction of quenched late-types shows a slight increase from Group Mass 1 to Group Mass 2 (which does not change within the uncertainties to Group Mass 3) with the expected shift to more massive stellar masses compared to the late, unquenched population. In stellar mass distribution (Figure \ref{fig:ms8}), the late-type, quenched systems appear to form an intermediate population between the late, unquenched and early, quenched populations. We expect several pathways of transformation to be operating in groups (including mergers) and this may indicate a pathway of transformation (late, unquenched to early, quenched via late-type, quenched) that operates more efficiently at some halo masses.

Finally, the quenched fraction by morphology, per stellar mass bin, is presented for early-types (Figure \ref{fig:ms9}) and late-types (Figure \ref{fig:ms10}) in each of our samples. Considering the early-types, Figure \ref{fig:ms9}a shows that for M$_{\rm stellar}> 10^{10}$\,M$_\odot$, early-types in groups show a higher quenched fraction, driven largely by galaxies in Group Mass 2 (Figure \ref{fig:ms9}c) and Group Mass 3 (Figure \ref{fig:ms9}d). The statistics in the lowest mass bins are understandably poor given the low numbers of early-types at these stellar masses. 

Considering the late-types (Figure \ref{fig:ms10}), we see that for M$_{\rm stellar}<10^{10.5}$\,M$_\odot$, late-types in groups are preferentially quenched -- this is particularly clear in Group Mass 2 and Group Mass 3. This points to external (environmental) processes acting on late-types and impacting their star formation. Comparison of Figure \ref{fig:ms10}c and Figure \ref{fig:ms10}d indicates that the effect is most noticeable in Group Mass 3 in the $10^{9.7}<$ M$_{\rm stellar}<10^{10.5}$ range.

For masses M$_{\rm stellar}>10^{10.5}$, we curiously observe that in our lowest halo mass bin, the quenched fraction of late-types is lower than what is found for the non-G3C sample. This implies that in these mass bins the late-type galaxies are actually forming more stars and not less. For Group Mass 2 and 3, however, we see increased quenching of late-types with values that are broadly consistent with each other. Once again, larger samples and finer divisions of halo mass will allow for a more definitive investigation.

In this section we have observed the effect of group environment on the star formation properties of galaxies. Group galaxies are quenched (i.e. have moved below the SFMS) relative to the ungrouped sample, at both high mass (mass quenching) and low mass (environmental quenching). Early-types dominate the quenched fraction at high stellar mass (M$_{\rm stellar}>10^{10.5}$), particularly in high mass halos. The quenched fraction at low stellar mass M$_{\rm stellar}<10^{10.5}$ is dominated by late-types and is observed (to varying degrees) in all halo mass bins. We find an intermediate population (in stellar mass) of late-type, quenched systems suggesting an evolutionary pathway via the quenching of disk-dominated galaxies.

\subsection{Compactness \label{sec:comp}}

Although relatively rare in the local universe \citep{Mc09}, compact groups are considered laboratories of extreme (and seemingly rapid) evolution \citep{John07,Walk10}, where interactions and merging pathways dominate \citep[e.g.][]{Bar89, Ver01}. For example, the oldest known compact group, Stephan's Quintet \citep{Steph77}, offers a unique and pristine perspective on shock cooling pathways \citep[e.g.][]{App06, Clu10} and the role of turbulence in the suppression of star formation \citep[e.g.][]{Gu12, App17}. 

\citet{Hick82} identified 100 compact groups in the Palomar Observatory Sky Survey, introducing a set of three criteria: a) richness, b) isolation, and c) compactness. This methodology has been used to construct similar catalogues using surveys such as the COSMOS-UKST Southern Galaxy Catalogue \citep{Iov02}, 2MASS \citep{Diaz12, Diaz15}, and SDSS \citep{Mc09}. Compact groups have also been identified using the FoF algorithm on surveys such as the CfA2 \citep{Bar96}, and SDSS \citep{Sohn16}; however, incompleteness due to fibre collsions particularly disrupts dense, compact structures, such as compact groups. Recent work has advocated for combining the two selection methods, for example, \citet{Diaz18} using SDSS, and \citet{Zh20} who combined redshift information from SDSS, LAMOST, and GAMA.

In this work we take an alternative approach and consider the on-sky ``compactness" of our entire G3C group sample, making use of the convex hull parameters included as part of the G3C catalog.
A convex hull describes the minimum 2D area (or 3D volume) that encloses a collection of points, such that the surface is only permitted to bend inwards. Therefore, for each galaxy group, an independent smooth surface is constructed, in comoving space, that contains all the galaxies in the group. This hull, therefore, has a 2D surface area and contains a 3D volume. It can also be projected onto the RA-Dec plane to reflect the projected area of the group (in units of (Mpc/h)$^2$) -- for the G3C sample this is referred to as the ``d2radec" parameter.

\begin{figure*}[!thp]
\centering
\subfloat[Loose Groups]{
\includegraphics[width=0.45\textwidth]{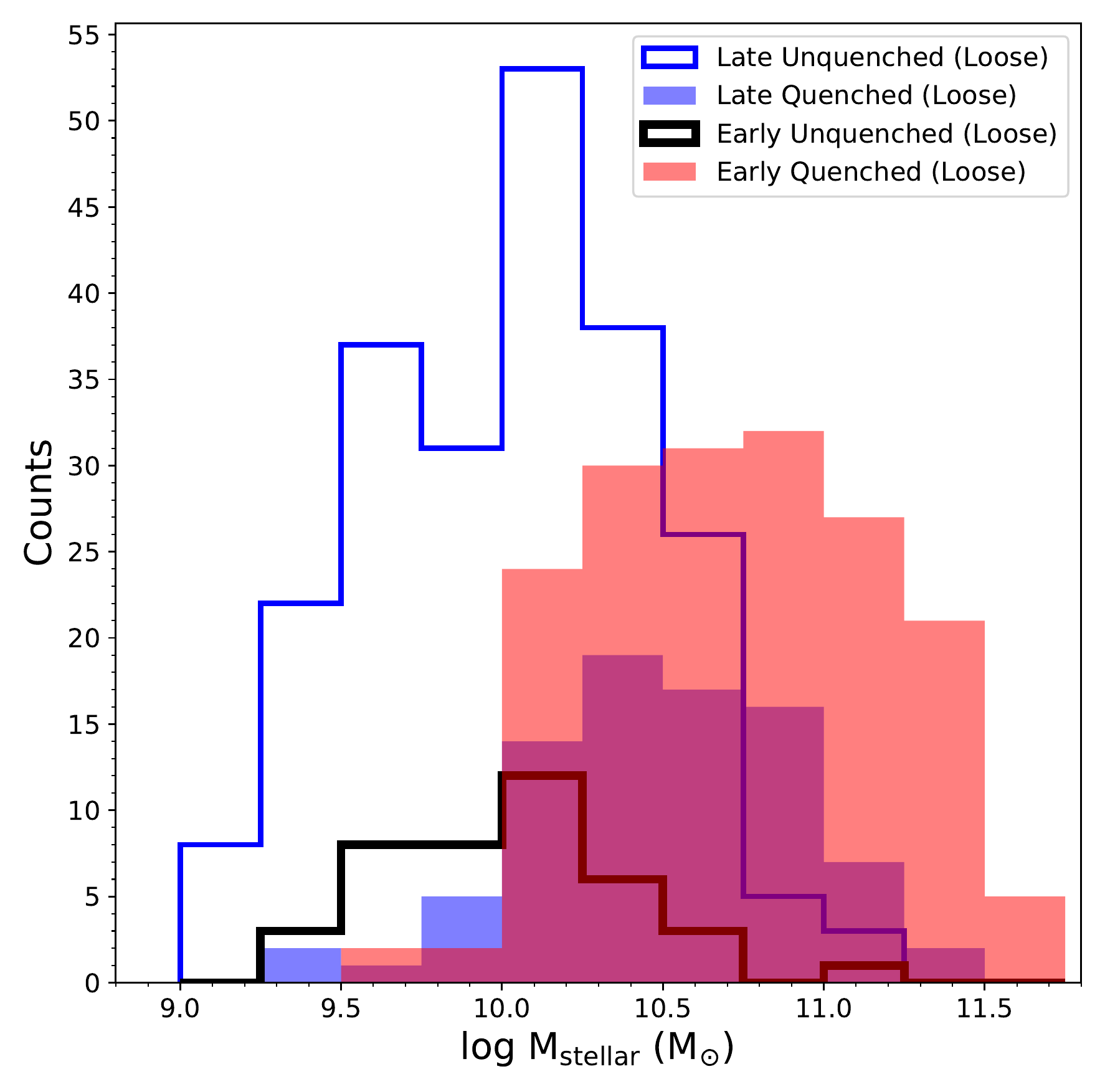}}
\\
\subfloat[Nominal Groups]{
\includegraphics[width=0.45\textwidth]{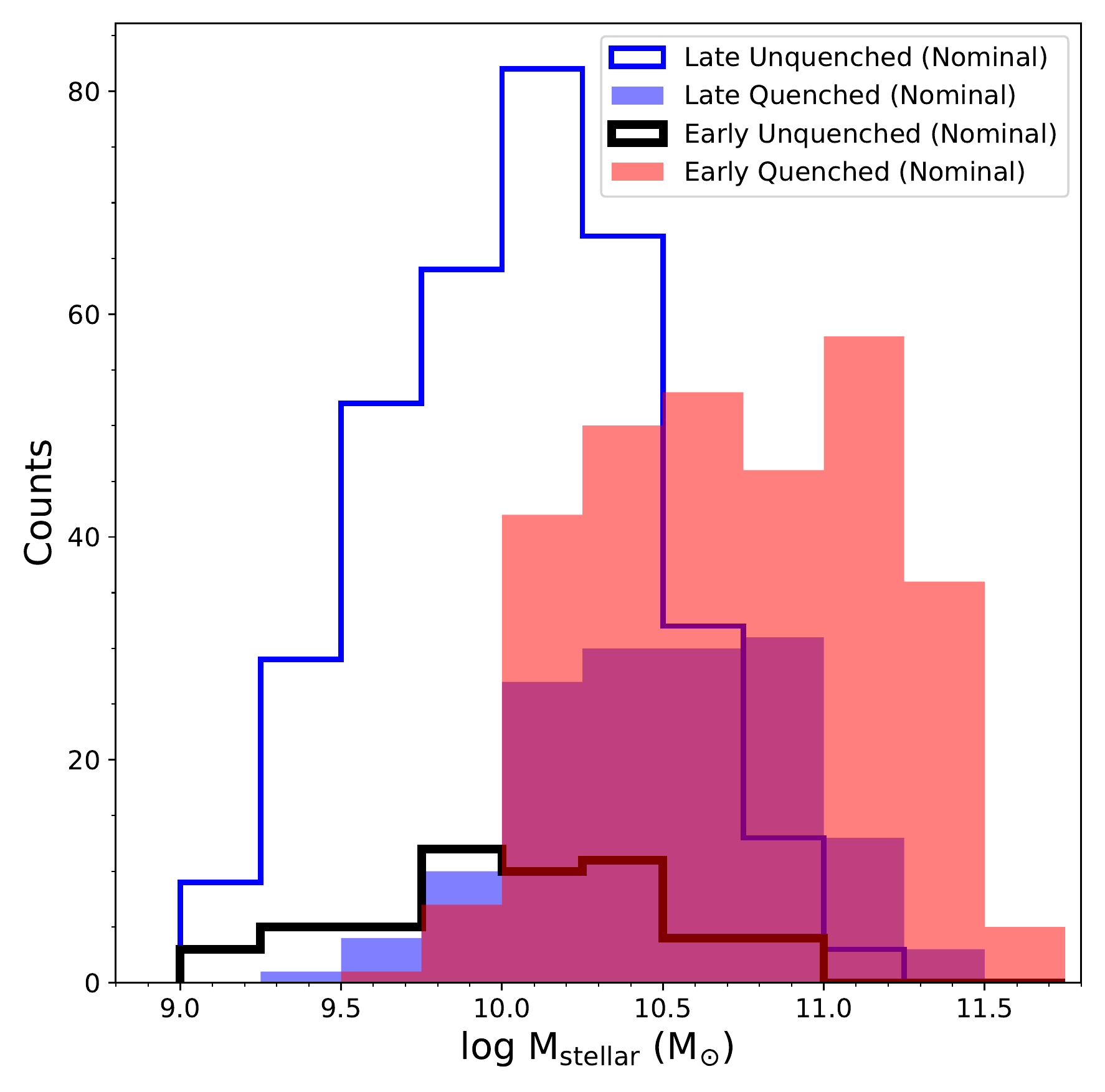}}
\subfloat[Compact Groups]{
\includegraphics[width=0.45\textwidth]{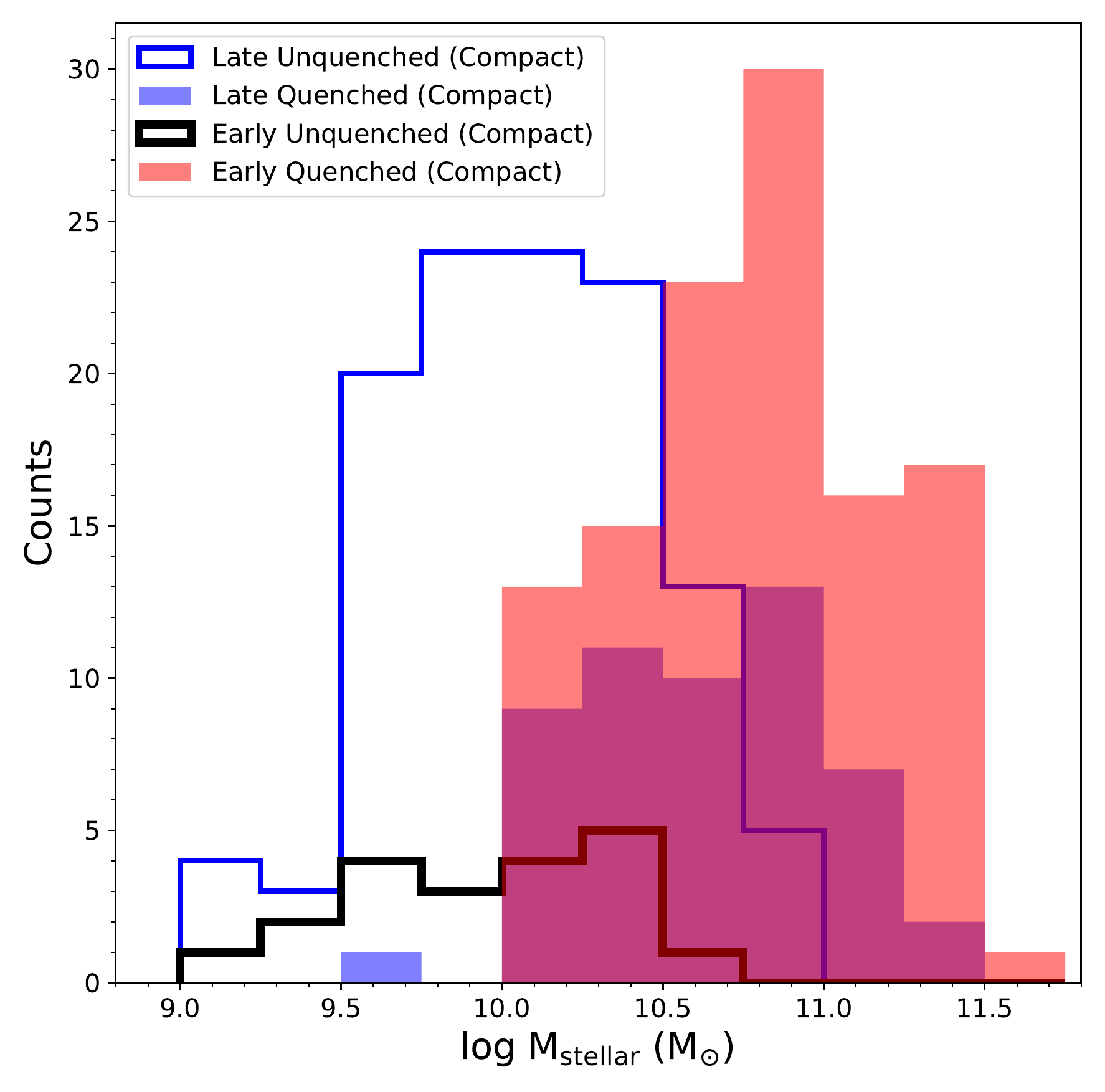}}
\caption{The stellar mass distributions of the unquenched (unshaded) and quenched (shaded) systems, sub-divided into early- and late-type, for the loose, nominal and compact group delineations described in the text. The loose and nominal stellar mass distributions of unquenched late-types and quenched early-types appear fairly similar, however, the compact systems are strongly peaked at M$_{\rm stellar} \sim 10^{10.5}$\,M$_\odot$ in both the late unquenched and early quenched populations. \label{fig:comp1} }
\end{figure*}

We investigate the distribution of the d2radec parameter as a function of group membership and group halo mass in section \ref{ap:comp} of the Appendix. Unsurprisingly, when comparing d2radec and dynamical mass (see Figure \ref{fig:A6}), an increase in halo mass broadly correlates to an increase in d2radec; i.e., more massive groups tend to have a larger projected area on the sky (i.e. in the RA-Dec plane) than smaller groups. We apply a linear fit to this distribution, which is described by:  
\begin{multline}
{\rm log_{10}}\, {\rm d2radec}\, ({\rm Mpc/h)^2} = 1.01\, {\rm log_{10}}\, {\rm Mass}_{\rm dyn.}\, ({\rm M}_{\odot}/{\rm h})\\ - 14.87, 
 \end{multline}
with an intrinsic spread of $\sigma=0.59$, as shown in Figure \ref{fig:A6} of section \ref{ap:comp} of the Appendix. We use this to define a compactness criterion where groups that lie in the top quartile defined by this relation are designated ``loose", whereas those in the bottom quartile are designated ``compact". We note that the largest point of difference to traditional compact groups is the richness criterion that requires that at least 4 member galaxies are of similar brightness (within 3 magnitudes of the brightest member), and therefore mass. Our definition is purposefully more general and can therefore not be considered true compact groups \citep[in the][sense]{Hick82}, and should not be compared as such. We refer to groups that lie between the top and bottom quartile as ``nominal", i.e. they are as compact on the sky as we would expect from their group halo mass. The numbers of groups and the corresponding number of galaxies in each division is given in Table \ref{tab:t5}.

\begin{deluxetable}{lcc}[!tb]
\tablecaption{Division of G3C Sample according to Compactness Criterion\label{tab:t5}}
\tablecolumns{3}
\tablenum{5}
\tablewidth{0pt}
\tablehead{
\colhead{Designation} &
\colhead{No. of Groups } &
\colhead{No. of Galaxies} 
}
\startdata
Loose & 125 & 793  \\
Nominal & 251  & 1299  \\
Compact & 118 &  469\\
\enddata

\end{deluxetable}

In order to examine the SFR properties, we use the quenching separator as before and show in Figure \ref{fig:comp1} the relative proportions of early- and late-types that are forming stars at an expected rate (unquenched), and those that have fallen below that threshold (quenched). Given the low numbers in our loose and compact samples, we increase our stellar mass bin size to 0.25 dex compared to previous sections.

\begin{figure*}[!thp]
\centering
\subfloat[Loose vs Nominal Groups]{
\includegraphics[width=0.45\textwidth]{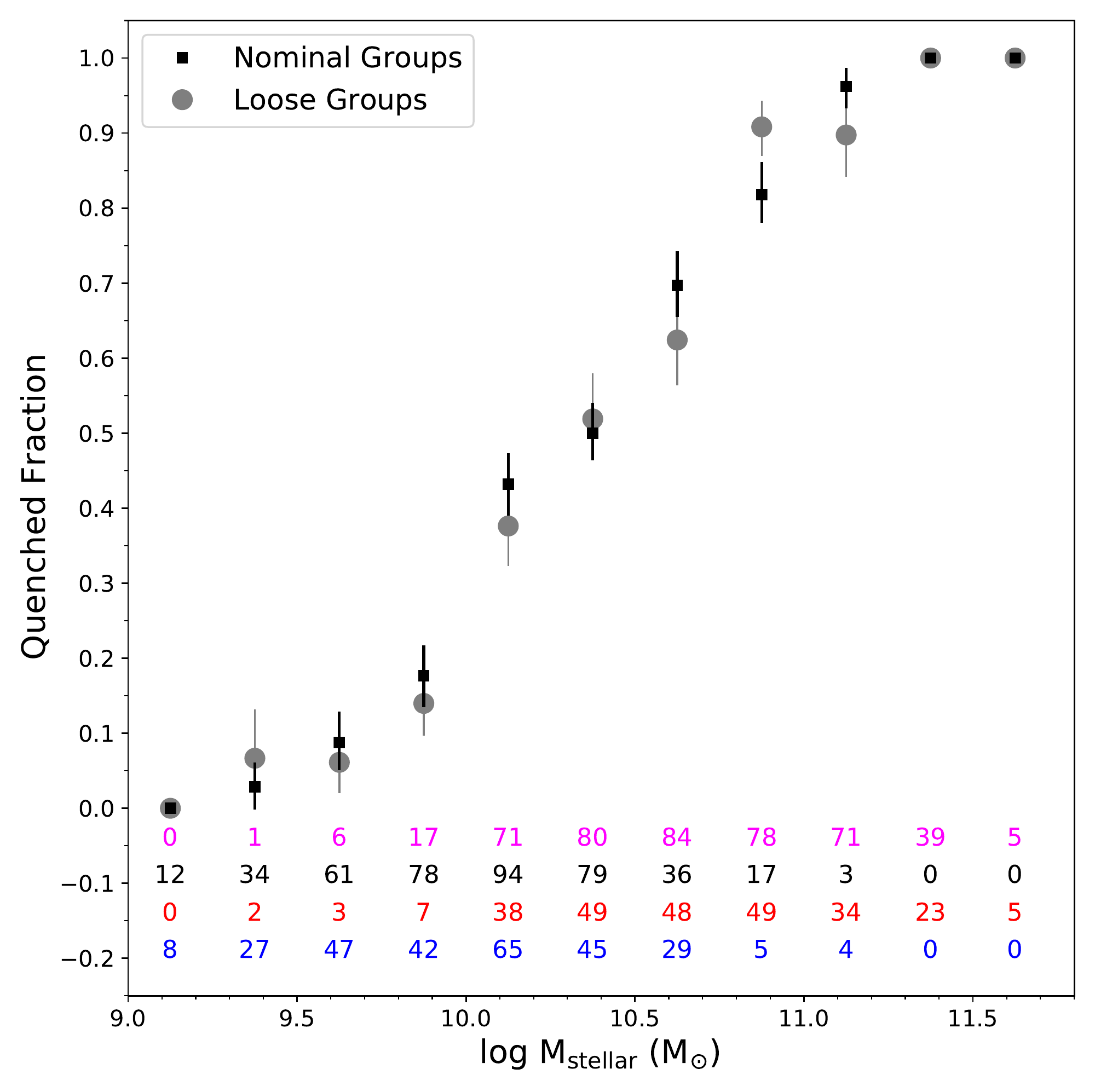}}
\subfloat[Compact vs Nominal Groups]{
\includegraphics[width=0.45\textwidth]{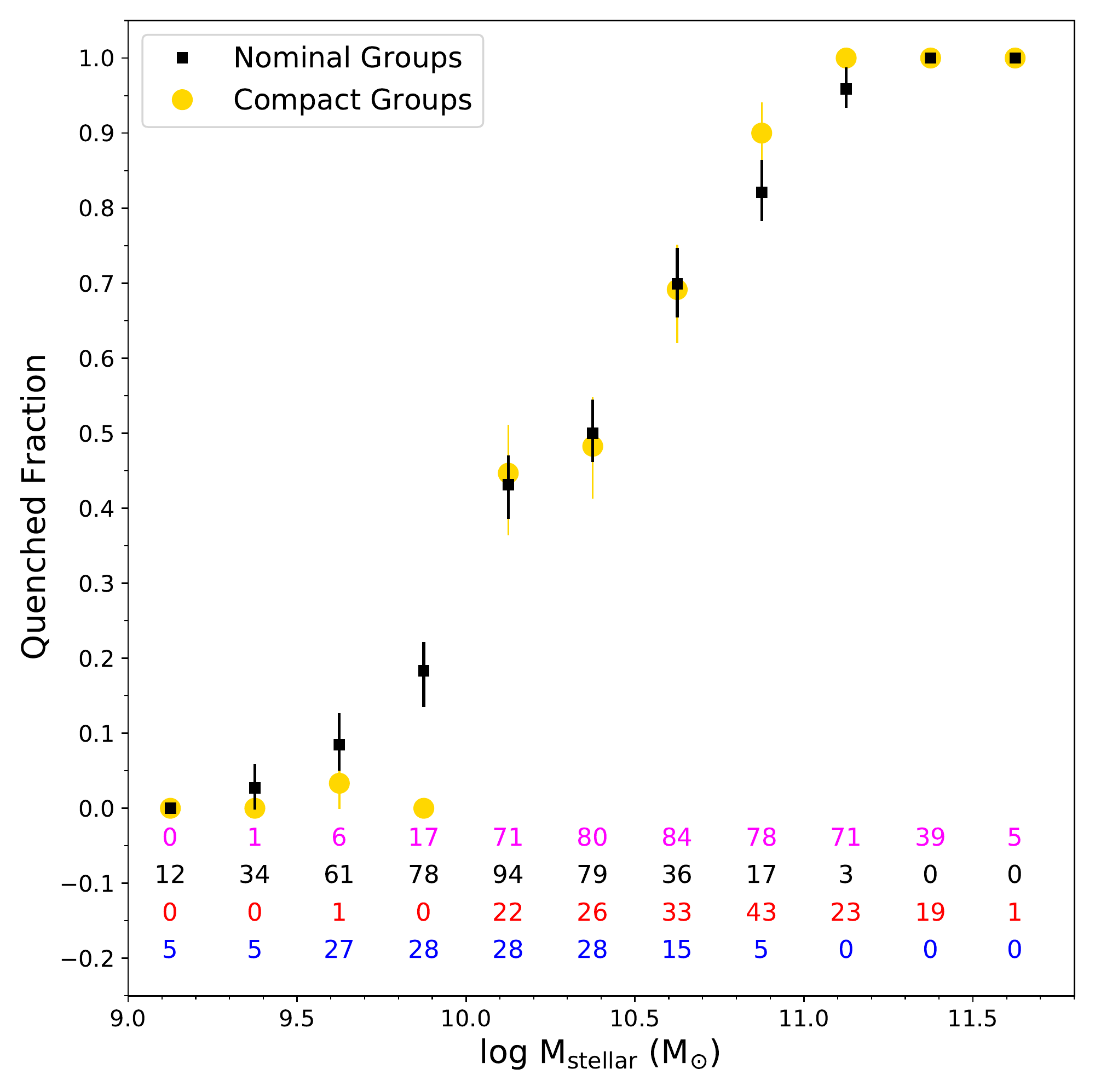}}
\caption{The quenched fraction for (a) loose vs nominal and (b) compact vs nominal groups, as defined by the compactness criterion outlined in the text. In each panel, the number of galaxies in each bin is given at the bottom, with the nominal quenched galaxies in pink, and nominal unquenched galaxies in black. The red text indicates the number of quenched galaxies in either loose (panel a) or compact (panel b) groups. The numbers of unquenched loose galaxies (panel a) and unquenched compact galaxies (panel b) are given in blue. As previously indicated, errors are calculated using bootstrap resampling in each bin. From this it appears that quenching efficiency is less impacted by compactness, as compared to morphological transformation. \label{fig:comp2} }
\end{figure*}

Comparing the loose and nominal groups, the stellar mass distributions of late-type, unquenched and early-type, quenched systems appear similar. The compact group galaxies, however, clearly have a larger proportion of early-type, quenched systems, relative to their late, unquenched population. This is reflected in Table \ref{tab:t6} where those populations have roughly equal numbers, compared to the loose and nominal groups, albeit with large uncertainties due to the small sample size.

\begin{deluxetable*}{lcccc}[!tbh]
\tablecaption{Unquenched and Quenched, Early- and Late-types in the Loose, Nominal and Compact Samples \label{tab:t6}}
\tablecolumns{5}
\tablenum{6}
\tablewidth{0pt}
\tablehead{
\colhead{Sample} &
\colhead{Unquenched} &
\colhead{Quenched} &
\colhead{Unquenched} &
\colhead{Quenched}\\
\colhead{} &
\colhead{Late-type} &
\colhead{Late-type} &
\colhead{Early-type} &
\colhead{Early-type}
}
\startdata
Loose &  42.8($\pm$3.4)\%  & 15.9($\pm$1.9)\%    & 7.9($\pm$1.3)\%  & 33.4($\pm$2.9)\% \\
Nominal   &   41.2($\pm$2.6)\% &   17.5($\pm$1.6)\%   &  6.3($\pm$0.9)\%   & 35.0($\pm$2.4)\%   \\
Compact &  38.2($\pm$4.2)\%  & 17.4($\pm$2.6)\%   & 6.6($\pm$1.5)\%  &  37.8($\pm$4.1)\% \\
\enddata
\end{deluxetable*}

Considering only the high stellar mass population (M$_{\rm stellar}>10^{10.5}$\,M$_\odot$), both early- and late-type, the compact groups have relatively more galaxies at high stellar mass  (45\%), relative to the loose (37\%) and nominal groups (39\%). Considering the whole sample, they have an increased proportion of early-types, both quenched and unquenched (44\%) compared to the loose (41\%) and nominal (41\%) groups. This could indicate increased tidal interactions in the more compact systems, promoting the growth of bulges (via inside-out growth) and overall stellar mass. This is consistent with simulations \citep[e.g.][]{Brass09} and observations \citep[e.g.][]{Deng08,Walk10} indicating increased fractions of early-type galaxies in traditional compact groups, compared to typical groups and ungrouped galaxies. However, within the errors, we do not see any significant differences in the overall fractions of quenched systems when comparing the loose, nominal and compact populations. 

We investigate this further in Figure \ref{fig:comp2} where we show the quenched fractions for nominal, loose and compact groups in each stellar mass bin. Although our samples of loose and compact group galaxies are small, we see that for M$_{\rm stellar} < 10^{10.5}$\,M$_\odot$, neither loose nor compact groups show evidence for increased quenching. However, for M$_{\rm stellar} > 10^{10.5}$\,M$_\odot$, we see that there may be differences compared to nominal groups, but this requires further investigation.

\section{Discussion \label{sec:disc}}

Our aim in this study is to examine the mid-infrared (\textit{WISE}) and morphological properties of aggregate samples of galaxies: ungrouped, grouped, differing in halo mass, and varying in compactness. We have shown that \wise is well-suited to examine the mass and star formation properties of galaxies as a function of environment, in the $z<0.1$ universe.

The bimodality in \wise W2$-$W3 color of the G3C sample (Figure \ref{fig:f3}) appears to be largely driven by the morphological composition of this population. In the non-G3C sample, by comparison, we observe a dearth of galaxies at low W2$-$W3 (the mid-IR blue end), which can be attributed to a lack of high-mass, early-type galaxies. Within the G3C sample, increasing halo mass corresponds to a larger proportion of early-type galaxies at high stellar mass (Figure \ref{fig:f5}). These trends can be largely understood in the context of the morphology-density relation, driven by the variation of the stellar mass function with halo mass \citep[e.g.][]{Yang09, Al15} combined with the stellar mass function of morphological types \citep[e.g.][]{Moff16}, which produces a change in the mix of disk- and spheroid-dominated morphologies in the denser group environment \citep[see also][]{Groot17}, and is consistent with the findings of, for example, \citet{Bluck14, Liu19}. In Figure \ref{fig:new}, however, we see tentative evidence for a changing early-type fraction in stellar mass bins $10^{10} \lesssim$ M$_{\rm stellar} \lesssim10^{11}$\,M$_\odot$, as well as an overall increased fraction of early-types when considering M$_{\rm stellar} > 10^{10}$\,M$_\odot$ in increasing halo mass bins. This suggests that stellar mass is not solely driving the morphology-density relation \citep[see e.g.][]{Bam09}.

Compared to the SFMS determined for ungrouped galaxies (Figure \ref{fig:ms1}), we find that a similarly constructed SFMS in grouped galaxies (Figure \ref{fig:ms2}) shows a slightly steeper slope, suggesting increased star formation as a function of mass in this population. This may be connected to the behaviour we observe in Figure \ref{fig:ms10}b where the quenched fraction of late-types in our lowest halo mass bin, show a lower quenched fraction compared to the ungrouped sample in the stellar mass range M$_{\rm stellar} \geq 10^{10.5}$\,M$_\odot$.

Considering systems that have moved below the SFMS, we find an increase in quenched fraction with increasing halo mass (Figure \ref{fig:ms5} and Table \ref{tab:t3}), consistent with the increase in passive fraction observed in other studies \citep[e.g.][]{Brinch04, Bluck14, Dav19b}. The quenched fraction is dominated by early-types at high mass (Figure \ref{fig:ms7} and Table \ref{tab:t4}), in line with previous findings \citep[e.g.][]{Bluck14, Liu19} and consistent with a varying stellar mass function combined with the mass-morphology mapping. The overall increase in the fraction of quenched, early-types with increasing halo mass is clearly aligned to an increase in high stellar mass systems, in line with secular processes. 

However, the pathways that lead to this behaviour are not as clear. We find evidence of an increase in the quenched population at M$_{\rm stellar} < 10^{10.5}$\,M$_\odot$ (Figure \ref{fig:ms5}), which appears to be driven by the quenching of late-types at these stellar masses, particularly in more massive halos (Figure \ref{fig:ms10}), which we expect from environmental (i.e. external) quenching mechanisms operating in this regime \citep[see also][]{Peng10, Dav19b, Liu19, Li20}. 

Although we see a clear separation in the stellar mass distributions of unquenched, disk-dominated galaxies and quenched, bulge-dominated galaxies across halo mass, the population of quenched, late-types (Figure \ref{fig:ms8}) at intermediate stellar mass \citep[where we expect stellar mass quenching to become important;][]{Peng10} could be the product of environmental processes and may in the future transition to early-type morphologies as a result of mergers \citep[see also][]{Mutch11}. This population increases from Group Mass 1 to Group Mass 2, but does not change significantly when moving to the largest group mass bin (see Table \ref{tab:t4}). This suggests that the processes forming this population do not necessarily become more efficient in larger mass group halos, possibly due to other mechanisms becoming more dominant. Larger samples would be needed to track the behavior of this population as a function of group halo mass, with the inclusion of cold and hot gas measures providing a more definitive picture of their evolutionary state.

Turning to the SFMS itself, the lack of unquenched early-types i.e. early-types on the sequence (Figure \ref{fig:ms6} and \ref{fig:ms7}) is in agreement with recent work from \citet{WangB20}, using SDSS-IV MaNGA, who find that the SFMS is dominated by spirals with small bulges. The lack of increase (in fact, decrease) of the early-type, unquenched population with increasing halo mass (Figure \ref{fig:ms8}, Table \ref{tab:t4}), when the overall number of early-types is increasing, suggests that these galaxies are predominantly found below the SFMS, in line with previous studies \citep[e.g.][]{Bluck14, Cook20}. However, this does not imply a physical causation between the presence of a bulge and the quenching of star formation \citep[e.g.][]{Bluck14, Lilly16, WangE18, Cook19}.

When considering group compactness, our results suggest that relatively compact groups have a tendency to host a larger fraction of high-mass and early-type systems, but only reflect a small  increase in overall quenched fraction compared to loose and nominal groups (Figure \ref{fig:comp1} and Table \ref{tab:t6}). They do, however, have the smallest fraction of unquenched late-types and the largest fraction of quenched early-types, suggesting either more rapid evolution in these environments or, alternatively, reflecting their earlier collapse and therefore more advanced evolution.

Taken together, our results emphasise the need to control for halo and stellar mass when investigating environmental effects \citep[for example, the impact on \HI mass content;][]{Hess13}, as the observed changes in quenched fraction can be an expected consequence of an increase in early-type fraction, itself a consequence of the changing stellar mass distribution with increasing halo mass. 

Although it is tempting to connect the changing properties we observe across environments to mechanisms operating as a function of density, such as interactions and mergers, the differing formation histories of galaxies makes direct comparisons infeasible. For example, galaxies in relatively large group halo masses are expected to have experienced more mergers, leading to an increased proportion of early-types at high stellar mass \citep[e.g.][]{Rod16, Deel17}. However, the observed merger fraction in the local universe does not appear to be more than $\sim$5\% \citep[e.g.][]{Darg10, Robot14}. In addition, halos in dense environments form earlier than halos of the same mass in less dense environments, and are therefore subject to ``assembly bias" \citep[e.g.][]{Sh04, Cro07, Wil13}. 

Observationally, we have the added conundrum that only by identifying the real progenitors of galaxies in our samples can we study how galaxies have been transformed by environment. This ``progenitor bias" means that changes in populations can be driven by changes in membership, rather than through changes in individual members \citep[e.g.][]{Dokk01, Car13, Cor19}. In fact, it has been shown that observed correlations between galaxy structure and the quenching of galaxies can be explained as a consequence of the size–mass relation for star-forming galaxies \citep{Lilly16}. And we note that in the work of \citet{Cor19}, current numerical simulations, such as EAGLE \citep{Sch15}, indicate that pre-processing has a limited effect on the structural properties of galaxies \citep[but, see also][]{DeL12}.

Observationally, recent studies have found growing evidence for a transitional or ``characteristic" mass associated with environmental processes \citep{WangE18, WangE20, Li20}. The characteristic stellar mass (M$_\ast,_{\rm ch}$) \citep{Li20} is a function of the halo mass of the group and the mass of the central; it implies that the quenching of galaxies is not driven by a simple central-satellite dichotomy, but rather by the interactions between internal and external processes \citep{WangE20, Li20}. In this paradigm, galaxies above M$_\ast,_{\rm ch}$ quench by internal (secular) processes and not environmental processes. Conversely, galaxies below M$_\ast,_{\rm ch}$ are more likely to quench due to external, environmental processes. In addition, it is chiefly the galaxies above the characteristic mass that are building stellar mass by mergers and tidal interactions \citep[but see also ][]{Josh20}. \citet{Li20} also investigates how M$_\ast,_{\rm ch}$ relates to quenched fraction as a function of bulge-to-total ratio and location within the halo; their findings lend support to the physically-motivated dichotomy that arises from M$_\ast,_{\rm ch}$.

Forthcoming \HI interferometric surveys will likely play a key role in progressing our understanding of environmental processes; by linking the detailed spatial information of cold gas to environmental measures and multiwavelength data, morphological transformation and pre-processing can be identified and studied as it takes place. Further to that, linking these studies to the relative location of dense structures within the cosmic web will enable a more comprehensive view of the role of secular versus environmental processes.

\section{Conclusions}

Making use of the high fidelity group properties and visual morphologies of the GAMA survey, we show that \wise can be used effectively to investigate the colors, star formation, and stellar mass of galaxies in different environments (measured as halo/dynamical mass) to $z<0.1$. We summarise our findings as:

\begin{itemize}

\item{The G3C population (i.e. galaxies within groups; 4$\leq$ Nfof $\leq$ 20) clearly show different \wise color, stellar mass, and morphological composition compared to the ungrouped (non-G3C) sample. The changing stellar mass function with increased group halo mass is also evident. We find tentative evidence of an overall increasing early-type fraction with increasing halo mass (when considering systems with M$_{\rm stellar} >10^{10}$\,M$_\odot$) and also in stellar mass bins between $10^{10}\lesssim$M$_{\rm stellar} \lesssim10^{11}$\,M$_\odot$ that may suggest that the morphology-density relation is not purely a consequence of the mass-morphology relation and a varying stellar mass function.}

\item{We determine a SFMS (using late-type, ungrouped galaxies, with \wise colors of W2$-$W3$>3$) of the form: 
\begin{multline} 
{\rm log_{10}}\, {\rm SFR} ({\rm M}_{\odot}/{\rm yr^{-1}}) = 0.93\,{\rm log_{10}}\, {\rm M}_{\rm stellar} ({\rm M}_{\odot})\\ - 9.08,
\end{multline}
and use this to define a quenching separator, delineating star-forming galaxies on the sequence, from those that are transitioning to being passive systems.}

\item{Using the quenching separator we show that with increasing halo mass, there is an accompanied increase in high mass (M$_{\rm stellar} >10^{10.5}$\,M$_\odot$) early-type systems that have moved off the star-forming sequence, in line with mass quenching.}

\item{We also find evidence of an increase in the quenched fraction of galaxies in groups in the mass range M$_{\rm stellar} <10^{10.5}$\,M$_\odot$, indicative of environmental quenching processes acting on late-type systems. This effect appears more significant in our most massive halo, but requires a larger sample for more robust statistics.}


\item{We observe quenched, late-type galaxies to form an intermediate population in stellar mass between the late, unquenched and early, quenched samples consistent with an evolutionary pathway where disk galaxies experience declining star formation on the way to the red sequence; i.e. in addition to the paradigm of gas-rich disk galaxies merging to form the red sequence.}

\item{Galaxies in groups that are compact, compared to the aggregate relation of compactness versus halo mass, have a higher proportion of early-type and high mass systems overall, but reflect a similar fraction of quenched galaxies compared to nominally compact groups.}

\end{itemize}

Expanding to a larger $z<0.1$ sample, with similarly high fidelity group measures, will mean better statistics for dividing by (a) stellar mass, (b) halo mass, and (c) star-forming/ transitioning/ passive systems. The addition of neutral gas content will be essential for examining the efficiency of quenching mechanisms \citep[e.g.][]{Cook19, Cook20, Jan20} and in the future, the SKA \HI Pathfinders will be pioneers in this phase space. For galaxy groups, measuring the baryon content locked in hot gas (e.g. from eROSITA) will allow for detailed studies of the baryon cycle as a function of environment. 

However, the robust determination of environment metrics through highly complete redshift information will be the limiting factor to extending this kind of study to larger volumes with the statistical fidelity needed to study the detailed pathways by which the cosmic web forms and galaxies are built.  

In the interim, our objective has been to show that \wise is well-suited to galaxy evolution studies in the local ($z<0.1$) universe. We have shown that environment is correlated with mass. And also that at fixed mass, environment correlates with the relative number of quenched vs. unquenched galaxies. And moreover, again at fixed mass, environment correlates with the properties of star-forming galaxies.  This demonstrates the difficulty in disentangling the different ways in which galaxies are shaped by their environments, and shows the need for very careful analysis of large galaxy samples. 

\newpage
\appendix

\section{Photometric Data Properties \label{sec:A1}}

\begin{figure*}[!hp]
\gridline{\fig{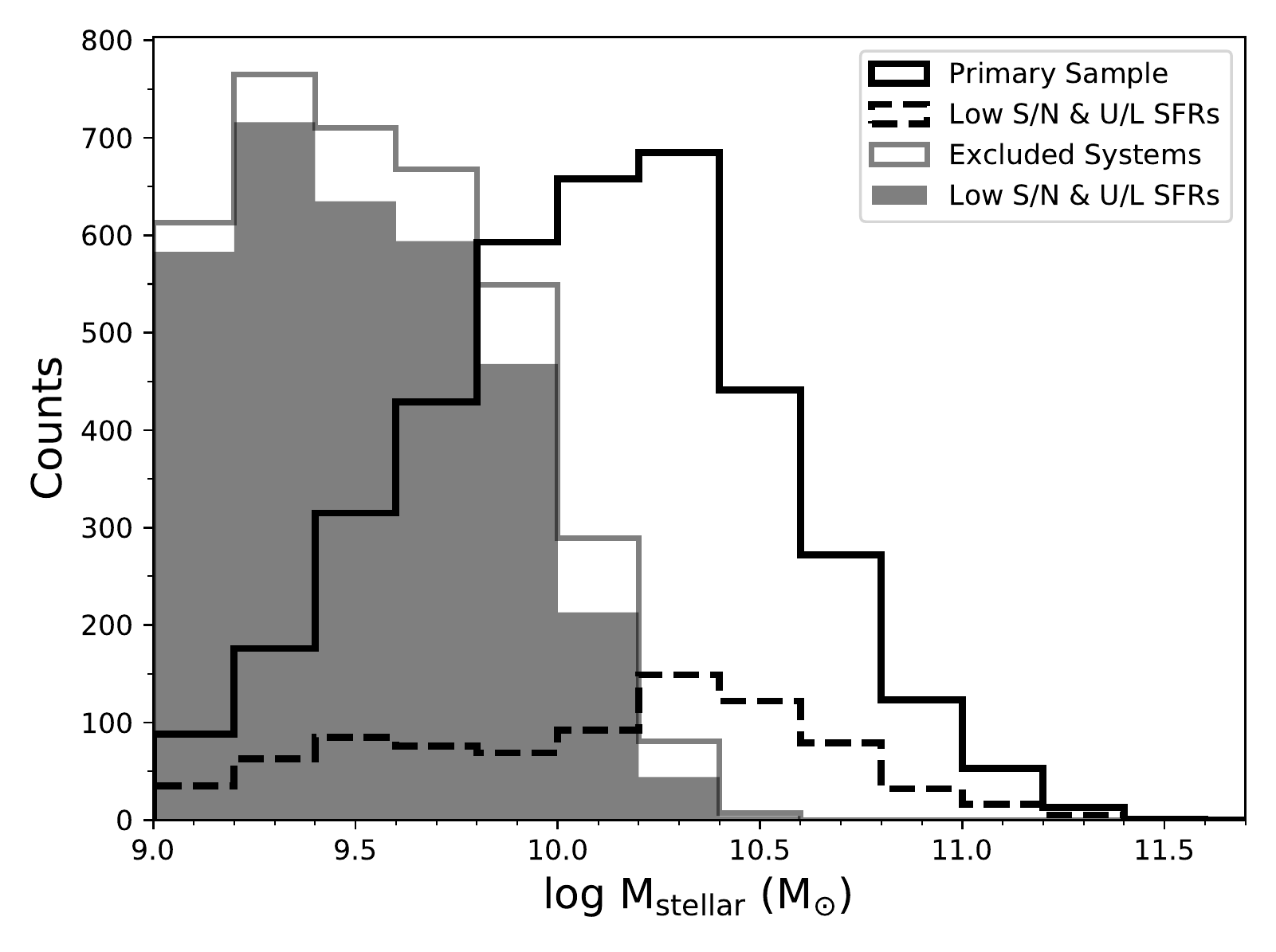}{0.5\textwidth}{(a) non-G3C Sample}
              \fig{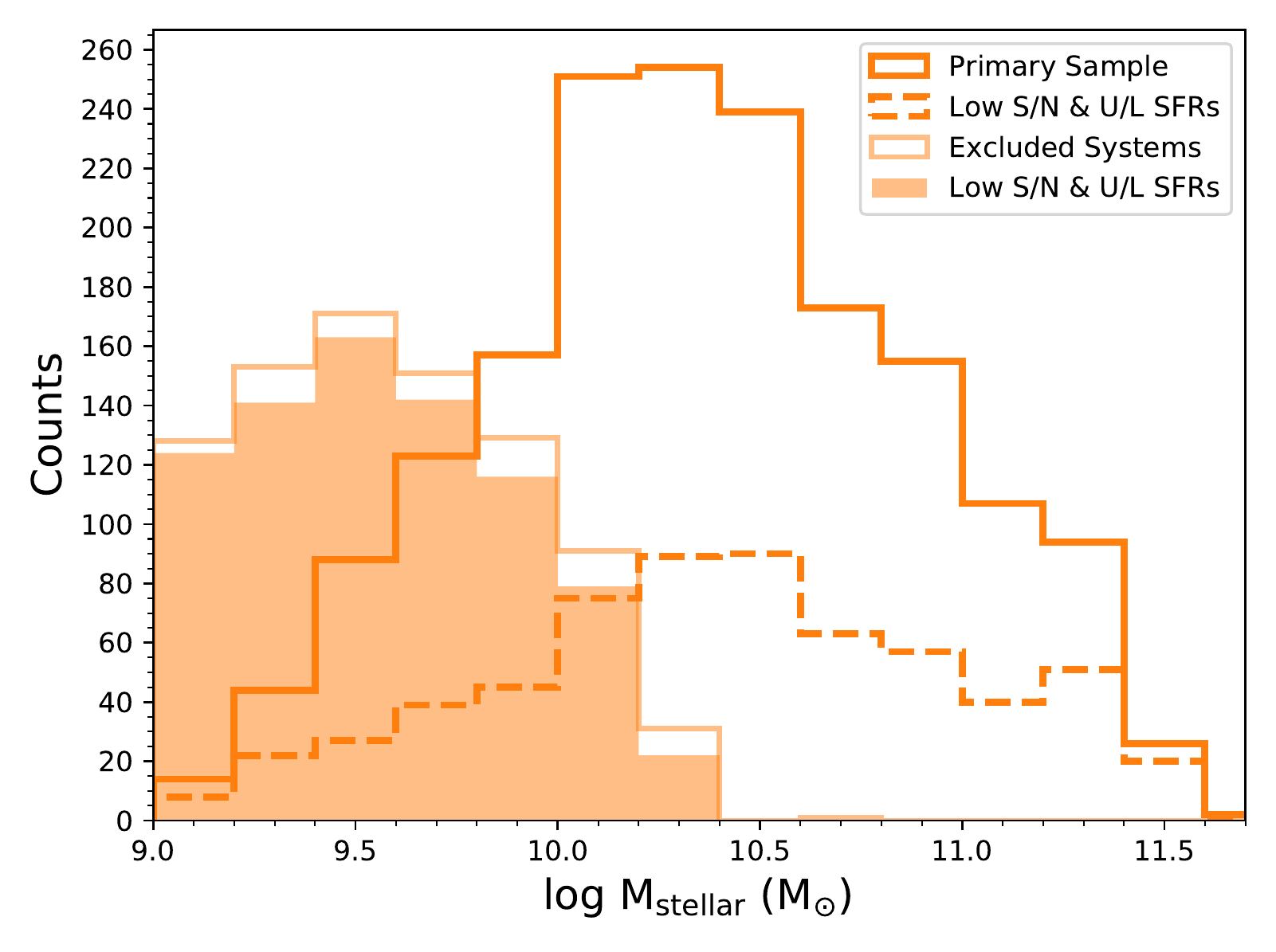}{0.5\textwidth}{(b) G3C Sample}}
\caption{The distribution in \wise stellar mass of the Primary Sample (dark line) and those systems excluded on the basis of the error in stellar mass being larger than 0.5 dex (Excluded Systems; light line) as discussed in the text. We note that by definition the stellar masses of the Excluded Systems, and therefore their true distribution, are highly uncertain. The dashed lines reflects the distribution of sources within the Primary Sample that have low S/N and upper limit SFRs. Similarly, the shaded histogram shows that the low S/N and upper limit SFRs dominate the Excluded Systems. \label{fig:AA1}}
\end{figure*}

As detailed in section \ref{photcut}, we have applied a signal to noise (S/N) photometric quality cut in W1$-$W2 color (S/N$>$5), which corresponds to a stellar mass error less than 0.5 dex, to limit the amount of cross-contamination in our stellar mass bins. In this section, we refer to the systems selected in this way as the ``Primary Sample" -- this is the sample used in this study. To investigate how this selection impacts our sample, we show in Figure \ref{fig:AA1} the stellar mass distributions of the Primary Sample and the so-called ``Excluded Systems" in the non-G3C (a) and G3C (b) samples, respectively. We caution that the stellar masses of the Excluded Systems are by definition highly uncertain, but shown here for illustrative purposes. It is evident that the low S/N W1$-$W2 sources dominate the low mass end (M$_{\rm stellar}<10^{10}$ M$_\odot$) in both samples. This is understandable given that these systems lack an abundant old stellar population that gives rise to the near-infrared light we are using to trace stellar mass.

Figure \ref{fig:AA1} additionally shows the distribution of low S/N and upper limit SFRs within the Primary Sample (dashed histogram) and Excluded Systems (shaded region). This shows that the Excluded Systems are dominated by galaxies with less well-determined SFRs -- this is due to the low SFRs of these low mass galaxies, which rapidly drop beyond the sensitivity of the \wise W3 band. Adding these sources to our Primary Sample would therefore further increase the noise within each mass bin, in addition to the noise across mass bins introduced by relaxing the stellar mass robustness requirement.

Although this reduces the number of galaxies at the lowest stellar masses (particularly for the non-G3C sample) and will limit our statistical power in these mass bins, excluding them does not influence our analysis as we expect little to no evolution to $z=0.1$ and the galaxies at these stellar masses we can use (located at lower redshifts) can be assumed to be representative of those excluded. Additionally, since we are comparing the non-G3C and G3C samples, the main results of this work are unaffected by any biases introduced due to this selection, as it affects the same population in both.  However, to investigate this further  we repeat our quenched fraction analysis for the non-G3C and G3C samples, as presented in section \ref{sec:MS}, but now including the all systems (i.e. no S/N requirement in W1$-$W2); this is shown in Figure \ref{fig:AA1}. We find that our results are highly consistent with what is found in Figure \ref{fig:ms4}b, examining in particular the low mass end where any differences would manifest.

\begin{figure}[!thbp]
\begin{center}
\includegraphics[width=9cm]{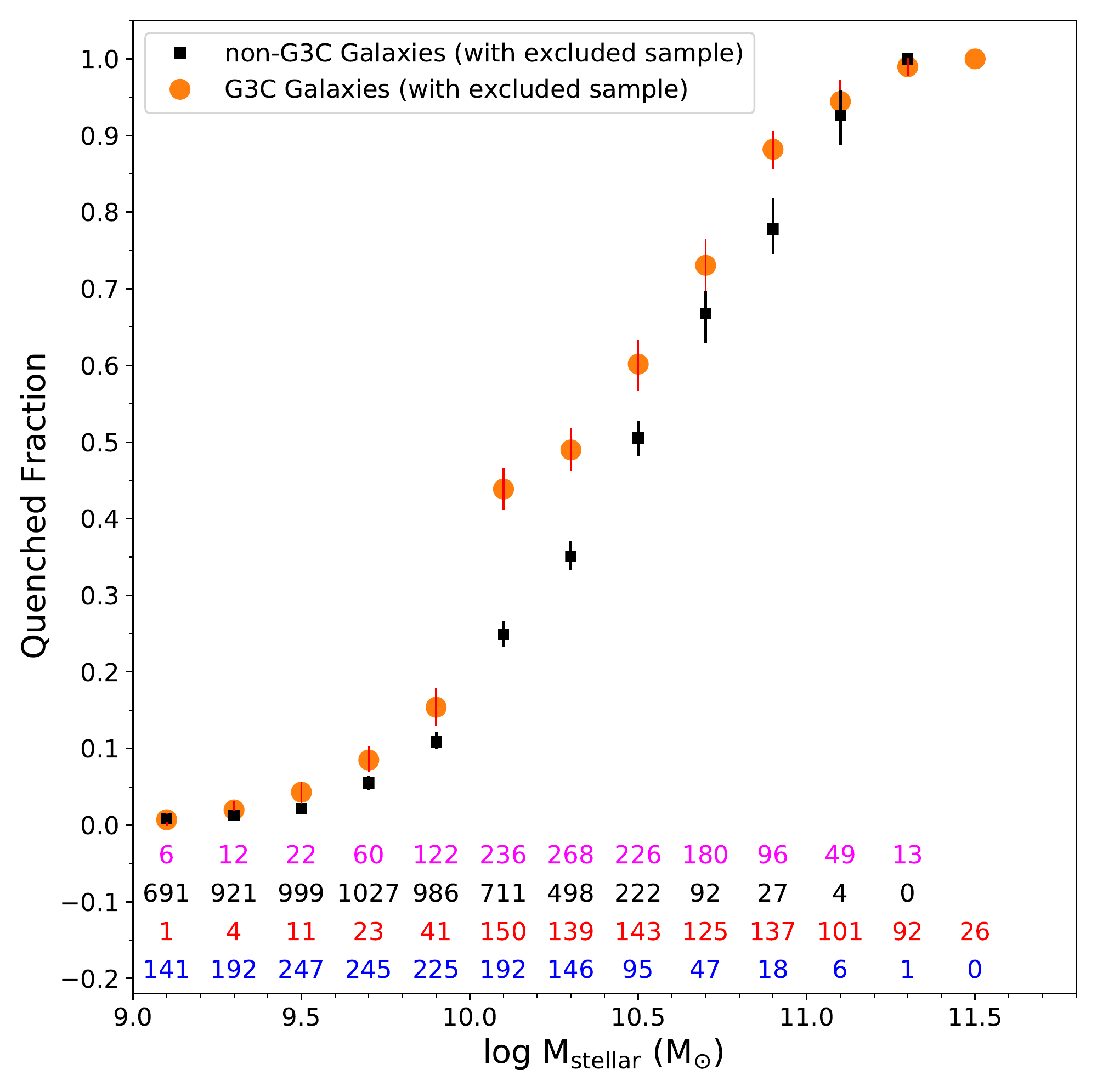}
\caption{Replicating Figure \ref{fig:ms4}b, we calculate quenched fraction in each mass bin for the non-G3C (black) and G3C galaxies (orange), but including the Excluded Systems with stellar mass being larger than 0.5 dex. The numbers at the bottom of the plot reflect the galaxies found in each bin with non-G3C quenched in pink, non-G3C unquenched in black, G3C quenched  in red, and G3C unquenched in blue. Errors are calculated using bootstrap resampling in each bin.}
\label{fig:AA2}
\end{center}
\end{figure}

\section{Mass Properties of the non-G3C and G3C Samples \label{sec:A2}}

\begin{figure}[!hb]
\begin{center}
\includegraphics[width=8.5cm]{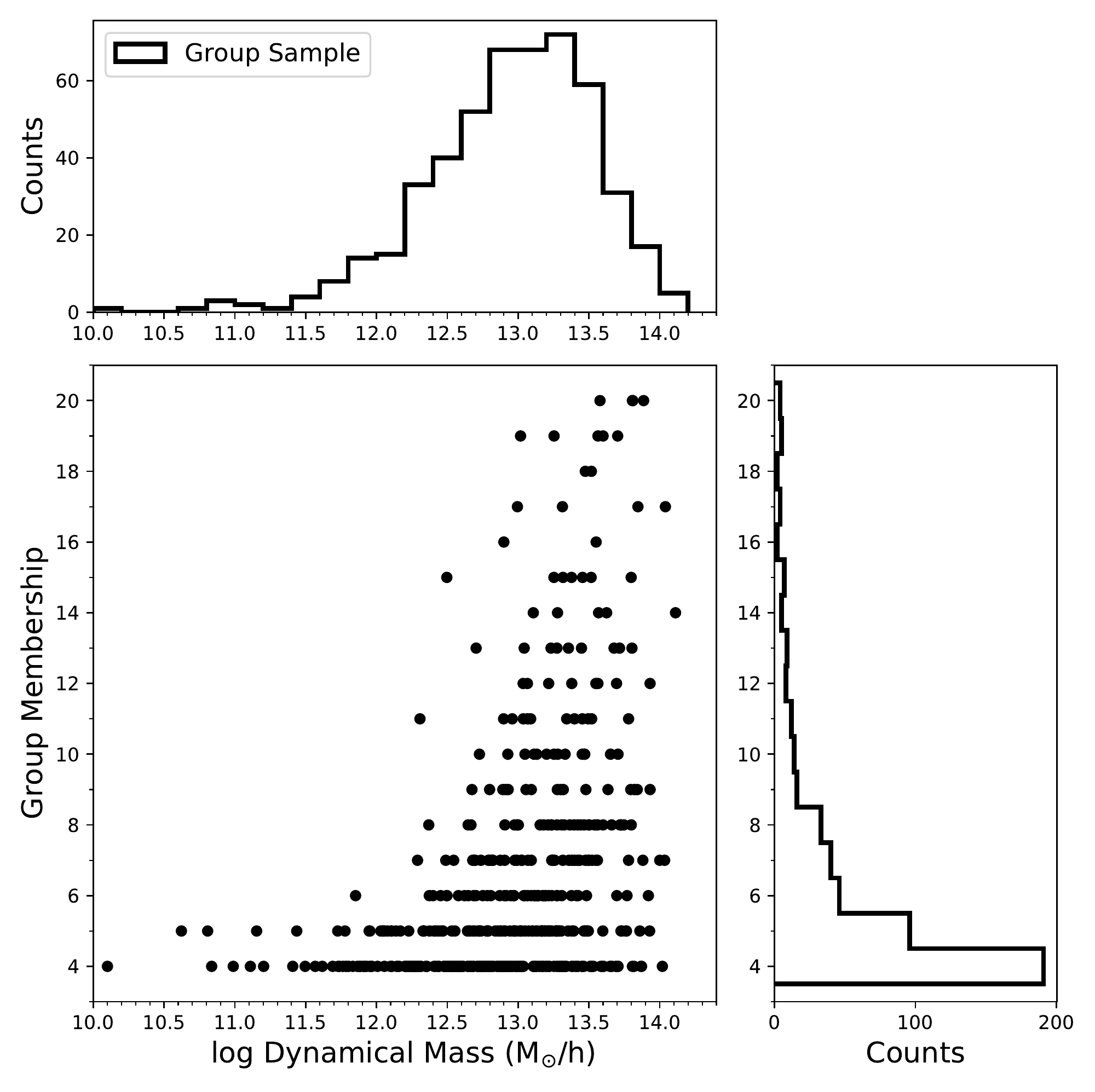}
\caption{The distribution of dynamical (halo) mass and group membership in our $z<0.1$ sample.}
\label{fig:A1}
\end{center}
\end{figure}

We include the distribution of dynamical (halo) mass and group membership of our G3C sample in Figure \ref{fig:A1}. It is evident that groups of membership 4 dominate our sample and show the broadest range in dynamical mass, consistent with the G3C analysis presented in \citet{Robot11}. Groups with membership $10<$ Nfof $\leq20$  are largely limited to having halo masses of log Dynamical Mass $>12.5$ M$_{\sun}$/h. As shown in \citet{Robot11},  the accuracy of the halo mass derived from dynamical mass estimates after applying a scaling factor, are median unbiased for Nfof $\geq4$. However, the standard deviation of the distribution increases strongly as a function of decreasing multiplicity, as given by Equation (20) in \citet{Robot11}. For our sample, this ranges from 0.45 -- 0.74 dex and is largely driven by uncertainties in the derived velocity dispersions for low group membership. It should therefore be borne in mind that the halo masses derived from the dynamical estimates are susceptible to scatter, particularly for the low group memberships that dominate our sample. However, our halo mass bins are chosen to be purposefully large to lessen the impact of this scatter.

\section{The Star Formation -- Stellar Mass Diagram in Halo Mass and Bulge-to-Disk Decomposition \label{ap:ms}}

\begin{figure*}[!htbp]
\gridline{\fig{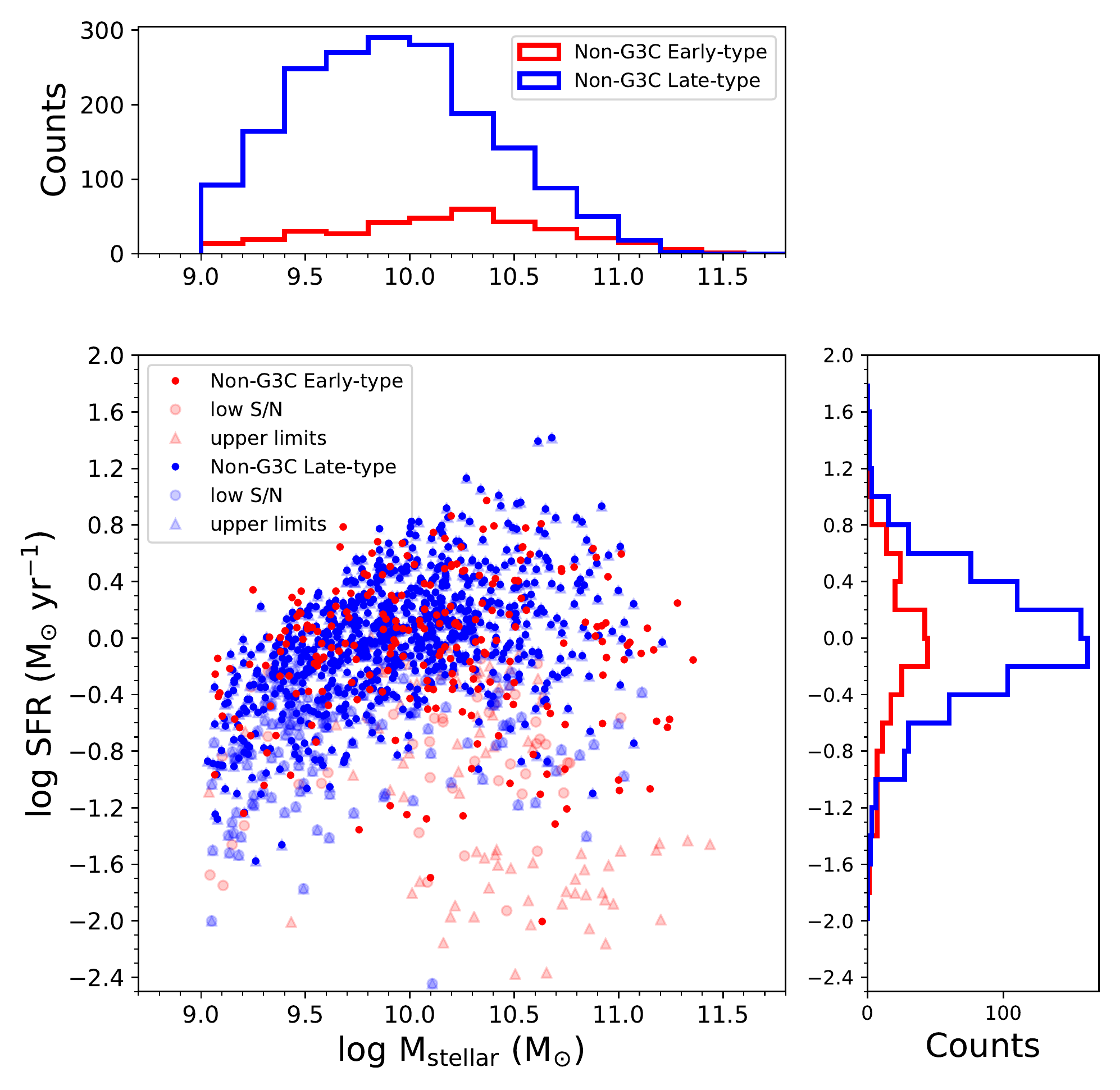}{0.5\textwidth}{(a) non-G3C Sample}
              \fig{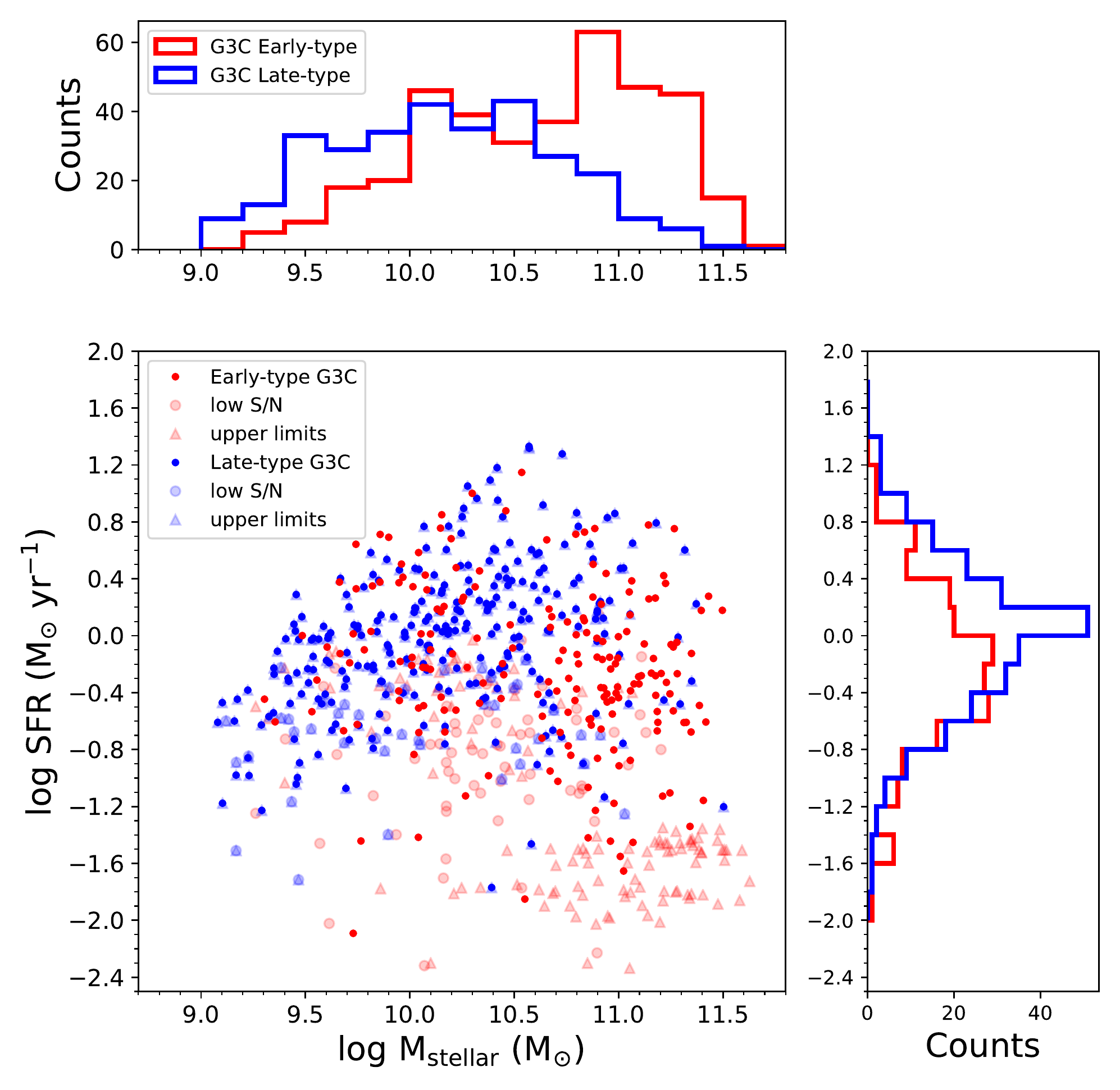}{0.5\textwidth}{(b) G3C Sample}}
\caption{The log SFR--log M$_{\rm stellar}$ distribution for the (a) non-G3C and (b) G3C sample using the bulge-to-total decompositions of Casura et al. (in prep.), available  for galaxies in our sample to $z<0.08$. To make this comparison we have grouped galaxies with a single-fit S{\'e}rsic index of $>2.5$ with those having a bulge-to-total ratio (from a double component fit) of B/T$>0.5$ as the ``Early-type" sample. ``Late-type" systems here, therefore, either have a S{\'e}rsic index $\leq2.5$ or B/T$\leq0.5$. Although this is a subset of our full sample, we see a similar distribution to that of Figure \ref{fig:ms6}, indicating that the visual morphology classifications and detailed decompositions broadly agree. \label{fig:msA2}}
\end{figure*}

In Figure \ref{fig:ms8} we presented the log SFR-- log M$_{\rm stellar}$ distribution for the (a) non-G3C and (b) G3C samples, making use of our visual morphological classifications to separate early- and late-type. Detailed bulge-to-disk comparisons have been performed on the KiDS $g, r,$ and $i$-band imaging \citep{deJ17} for $z<0.08$ galaxies in the GAMA II equatorial survey regions using ProFit \citep{Robot17} and are provided as part of the BDDecompv03 DMU (Casura et al., in prep.). We make use of these measurements to investigate the consistency of our results, albeit with a redshift subset of our sample. In Figure \ref{fig:msA2} we have grouped galaxies with a single-fit S{\'e}rsic index of $>2.5$ with those having a bulge-to-total ratio (from a double component fit) of B/T$>0.5$ as the ``Early-type" sample. Similarly, ``Late-type" systems in this context either have S{\'e}rsic index $\leq2.5$ or B/T$\leq0.5$. Compared to Figure \ref{fig:ms6}, we find good agreement with the morphological composition and distribution of the non-G3C and G3C samples, even with the smaller subset of systems.

\section{Compactness \label{ap:comp}}

\begin{figure}[!thb]
\begin{center}
\includegraphics[width=8.5cm]{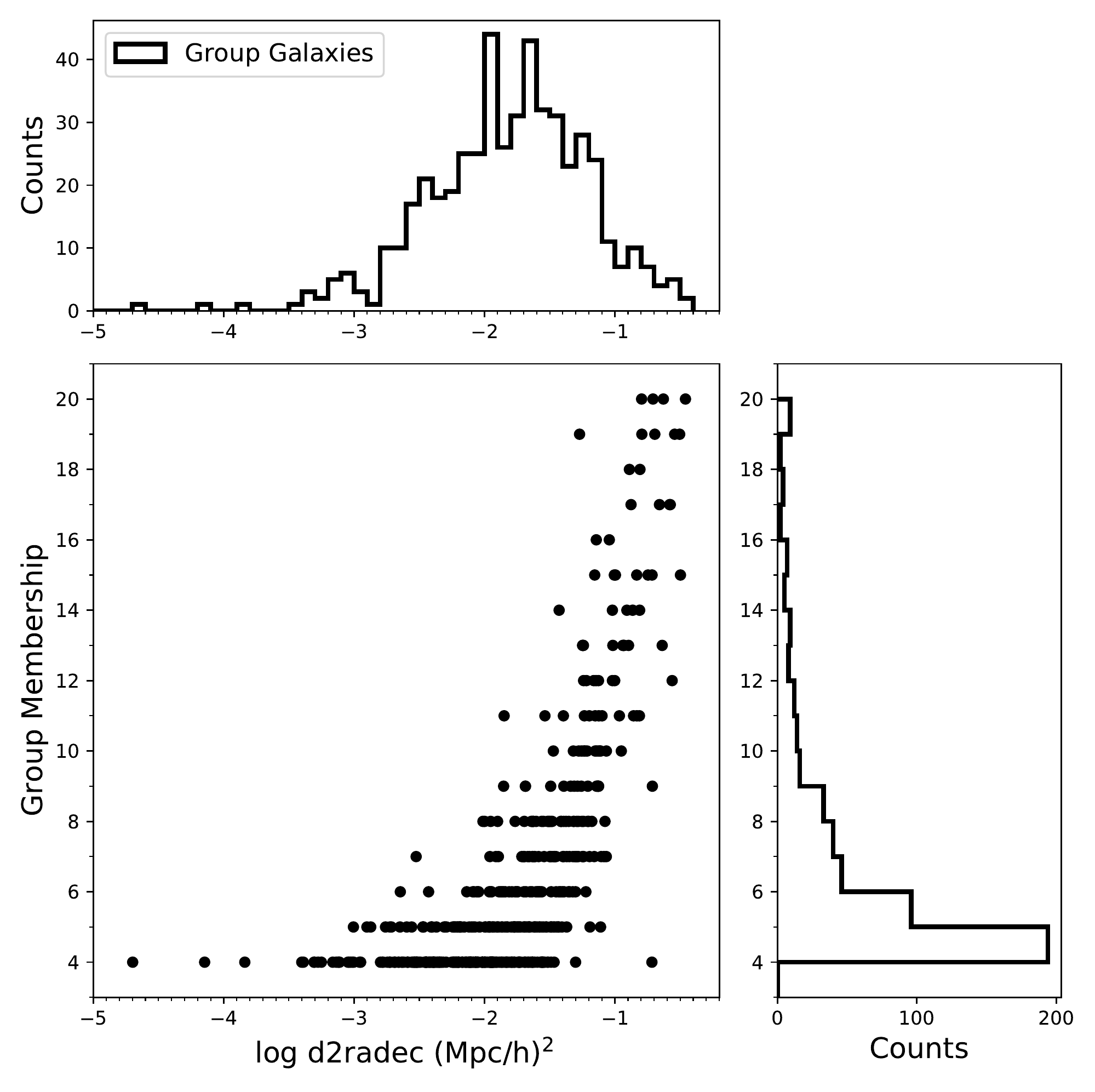}
\caption{Group Membership of the G3C sample as a function of the convex hull parameter, d2radec. As group membership increases, the d2radec  parameter tends to increase, reflecting a larger projection of the group on the sky.}
\label{fig:A5}
\end{center}
\end{figure}

In this section we investigate the convex hull parameter, d2radec, introduced in Section \ref{sec:comp} of the paper. In Figure \ref{fig:A5} we plot group membership as a function of d2radec, which shows that as group membership increases, the projected area of the group on the sky tends to be larger. We note that groups of membership 4 show a broad range in the d2radec parameter, which is consistent with the large spread in halo mass observed in Figure \ref{fig:A1}.

\begin{figure}[!hb]
\begin{center}
\includegraphics[width=8cm]{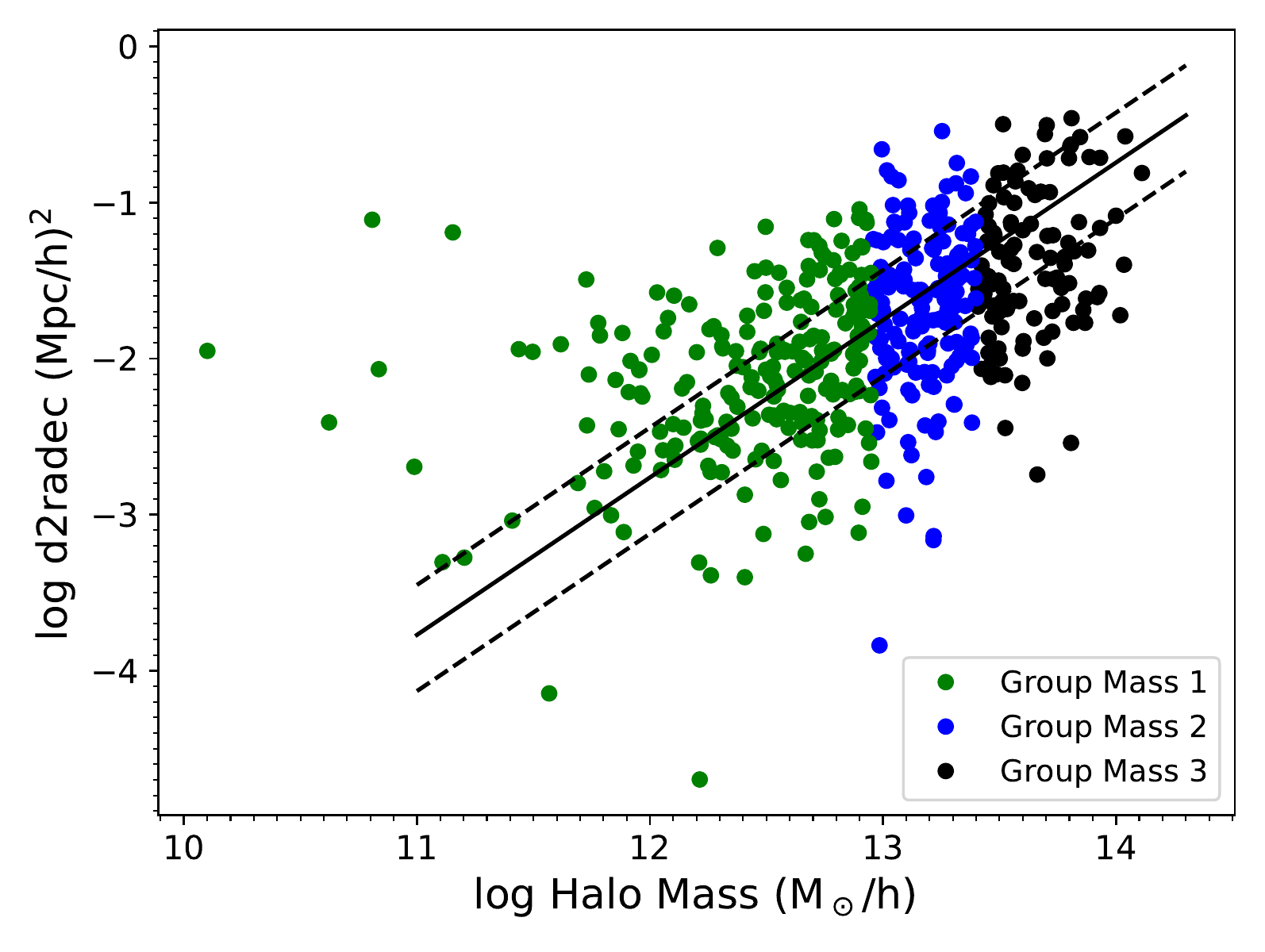}
\caption{The d2radec parameter as a function of dynamical (halo) mass, color-coded by group mass bin (Table \ref{tab:t2}). The solid line in indicates the best fit, with dashed lines indicating the division into top and bottom quartile.}
\label{fig:A6}
\end{center}
\end{figure}

In Figure \ref{fig:A6} we examine the d2radec parameter as a function of dynamical mass (that goes as the square of the velocity dispersion), which shows a clear correlation. We determine a best-fit to the distribution, given by: 

\begin{equation}
{\rm log_{10}}\, {\rm d2radec}\, ({\rm Mpc/h)^2} = 1.01\,{\rm log_{10}}\, {\rm Mass}_{\rm dyn.}\, ({\rm M}_{\odot}/{\rm h}) - 14.87, 
 \end{equation}
 
with an intrinsic spread of $\sigma=0.59$. The dashed lines in Figure \ref{fig:A6} reflect the separation, using this relation, between top and second quartile, and third and fourth quartile, respectively. We consider groups that lie in the second and third quartile to be ``nominally" compact i.e. they have a projected area on the sky that is consistent with this relation. Alternatively, those groups lying in the top quartile are considered to be ``loose" for their dynamical mass, and those in the lower quartile are considered ``compact" on the sky given their dynamical mass.

\acknowledgments

We thank the anonymous referee for helpful comments and suggestions that have improved the content and clarity of this paper. MC is a recipient of an Australian Research Council Future Fellowship (project number FT170100273) funded by the Australian Government. THJ acknowledges support from the National Research Foundation (South Africa).  
GAMA is a joint European-Australasian project based around a spectroscopic campaign using the Anglo-Australian Telescope. The GAMA input catalog is based on data taken from the Sloan Digital Sky Survey and the UKIRT Infrared Deep Sky Survey. Complementary imaging of the GAMA regions is being obtained by a number of independent survey programmes including GALEX MIS, VST KiDS, VISTA VIKING, WISE, Herschel-ATLAS, GMRT and ASKAP providing UV to radio coverage. GAMA is funded by the STFC (UK), the ARC (Australia), the AAO, and the participating institutions. The GAMA website is http://www.gama-survey.org/. Based on observations made with ESO Telescopes at the La Silla Paranal Observatory under programme ID 177.A-3016.
This publication makes use of data products from the Wide-field Infrared Survey Explorer, which is a joint project of the University of California, Los Angeles, and the Jet Propulsion Laboratory/California Institute of Technology, and NEOWISE, which is a project of the Jet Propulsion Laboratory/California Institute of Technology. \wise and NEOWISE are funded by the National Aeronautics and Space Administration. 
This research has made use of {\tt python (https://www.python.org)} and python packages: {\tt astropy} \citep{Ast18, Ast13}, {\tt matplotlib} http://matplotlib.org/ \citep{Hun07}, {\tt NumPy} http://www.numpy.org/ \citep{Walt11}, and {\tt SciPy} https://www.scipy.org/ \citep{Virt20}

\vspace{5mm}
\facilities{WISE, AAT, VST}

\end{document}